\documentclass[onecolumn]{aastex7}
\usepackage{booktabs}
\usepackage{mathtools}
\usepackage{comment}
\usepackage{bm}
\usepackage{overpic}

\def    \apjl  		{\rm {ApJL}}

\def	\K		{\,{\rm K}}
\def	\mum	{\,{\mu \rm{m}}}

\def \bea {\begin{eqnarray}}
	\def \ena {\end{eqnarray}}

\def	\ba	{\boldsymbol{a}}

\def	\bB	{\boldsymbol{B}}

\def	\bJ	{\boldsymbol{J}} 
\def	\bk	{\boldsymbol{k}}

\def	\cm	{\,{\rm cm}}

\def	\d	{{\rm d}}

\def	\D	{{\rm D}}

\def	\erg	{\,{\rm erg}}

\def	\g	{\,{\rm g}}
\def	\gas	{\,{\rm gas}}
\def	\G	{{\rm G}}

\def	\H	{{\rm H}}

\def	\N	{{\rm N}}
\def	\nH	{n_{\rm H}}

\def	\pc	{\,{\rm pc}}

\def	\s	{\,{\rm s}}

\def	\B 	{\rm B}

\def \St {{\rm St}}

\def	\V	{{\rm V}}

\def	\rad	{{\rm rad}}
\def	\yr	{{\rm yr}}

\newcommand{\paperi}{Paper~I}
\def    \Bv     	{\boldsymbol{B}}

\def    \kv     	{\boldsymbol{k}}

\newcommand{\bOmega}{\boldsymbol{\Omega}}
\newcommand{\bmu}{\boldsymbol{\mu}}

\makeatletter
\newcommand*{\rom}[1]{\expandafter\@slowromancap\romannumeral #1@}
\makeatother

\providecommand{\Lsun}{L_{\odot}}

\providecommand{\NHsn}{N_{\mathrm{\rm H}}}
\providecommand{\Td}{T_{\mathrm{d}}}
\providecommand{\Tgas}{T_{\mathrm{gas}}}

\providecommand{\Smax}{S_{\mathrm{max}}}
\providecommand{\Omdisr}{\Omega_{\mathrm{disr}}}
\providecommand{\Omth}{\omega_{\mathrm{T}}}
\providecommand{\aali}{a_{\mathrm{align}}}
\providecommand{\adisr}{a_{\mathrm{disr}}}
\providecommand{\afast}{a_{\rm align}^{\rm fast}}
\providecommand{\fHiISRF}{f_{\mathrm{hiJ}}^{\mathrm{ISRF}}}   % ISRF equilibrium high-J fraction (t=0)
\providecommand{\fLoISRF}{f_{\mathrm{loJ}}^{\mathrm{ISRF}}}   % = 1 - \fHiISRF
\providecommand{\Rhigh}{R_{\mathrm{hiJ}}}
\providecommand{\Rlow}{R_{\mathrm{loJ}}}
\providecommand{\fhif}{f_{\mathrm{hiJ}}^{\mathrm{fast}}}       % fast-alignment high-J fraction (t>0)
\providecommand{\fDisrJ}{f_{\mathrm{disr\text{-}J}}^{\mathrm{eff}}}
\providecommand{\fHiISRFeff}{f_{\mathrm{align,hiJ}}^{\mathrm{ISRF,eff}}}  % \fHiISRF × \Rhigh  (t=0 baseline)
\providecommand{\fLoISRFeff}{f_{\mathrm{align,loJ}}^{\mathrm{ISRF,eff}}}  % \fLoISRF × \Rlow   (t=0 baseline)
\providecommand{\fHiFasteff}{f_{\mathrm{align,hiJ}}^{\mathrm{fast,eff}}}  % \fhif × \Rhigh     (t>0 fast high-J)
\providecommand{\fLoFasteff}{f_{\mathrm{align,loJ}}^{\mathrm{fast,eff}}}  % (1-\fhif) × \Rlow  (t>0 fast low-J)

\providecommand{\pext}{p_{\mathrm{\rm ext}}}
\providecommand{\Iem}{I_{\mathrm{\rm em}}}
\providecommand{\Ipol}{I_{\mathrm{pol}}}
\providecommand{\Gamrat}{\Gamma_{\mathrm{RAT}}}
\providecommand{\tdamp}{\tau_{\mathrm{damp}}}
\providecommand{\tLar}{\tau_{\mathrm{Lar}}}
\providecommand{\tkRAT}{\tau_{k}}
\providecommand{\Ncl}{N_{\mathrm{cl}}}
\providecommand{\RVsn}{R_{V}}
\providecommand{\Av}{A_{V}}
\providecommand{\PAH}{\mathrm{PAH}}
\providecommand{\IemPAH}{I_{\mathrm{\rm em}}^{\PAH}}
\providecommand{\IemAstro}{I_{\mathrm{\rm em}}^{\mathrm{astro}}}
\providecommand{\invlam}{1/\lambda}

\newcommand{\BRAT}{B\text{-}\mathrm{RAT}}
\newcommand{\kRAT}{k\text{-}\mathrm{RAT}}

\graphicspath{
  {./}
  {./figures/}
}

\begin{document}

%	\title{Time-Domain Dust Astrophysics. II. TransRAT: A General Framework for Dust Evolution under Cosmic Transients}
	\title{Time-Domain Dust Astrophysics. II. TransRAT: Time-Dependent Grain Alignment and Disruption by Cosmic Transients and Their Observational Signatures}
		
	\author{Thiem Hoang}
	\affiliation{Korea Astronomy and Space Science Institute, Daejeon 34055, Republic of Korea}
	\affiliation{Department of Astronomy and Space Science, University of Science and Technology, 217 Gajeong-ro, Yuseong-gu, Daejeon, 34113, Republic of Korea}

	\email{thiemhoang@kasi.re.kr}

\begin{abstract}
Cosmic transients enhance the local radiation field by orders of magnitude on short timescales, rendering steady-state treatments of grain alignment and disruption by RAdiative Torques (RATs) inadequate. We introduce {\it TransRAT}, a self-consistent time-domain framework that follows the coupled dynamical response of dust to transient irradiation and predicts its observable signatures. For numerical demonstration, we apply {\it TransRAT} to a Type~IIP supernova illuminating a one-zone molecular cloud. Radiative heating raises the grain temperature, modifying the magnetic susceptibility and Larmor precession, while the enhanced radiation field accelerates radiative precession and can switch the alignment axis from the magnetic field ($\BRAT$) to the radiation direction ($\kRAT$). Simultaneously, strong RATs align grains faster than gas randomization, extending {\it fast alignment} to smaller grain sizes, and disrupt large grains through RAT disruption (RAT-D), irreversibly modifying the grain size distribution. As the transient fades, the alignment axis returns to $\Bv$ at a rate controlled by the magnetic susceptibility, whereas the enhanced aligned-grain population and RAT-D-modified grain size distribution can persist long after the radiation has faded. We refer to these long-lived departures from the pre-transient state as \emph{physical memory effects}. The time-dependent evolution of grain alignment and disruption produces distinctive observational signatures: a flare in extinction and thermal dust polarization followed by a RAT-D-induced dip; a blueward shift of the polarization peak $\lambda_{\max}$; a polarization-angle rotation equal to the projected $\Bv$--$\kv$ separation ($45^{\circ}$ for our adopted geometry) when RAT-D removes large high-$J$ aligned grains and leaves low-$J$ grains aligned with $\kRAT$, followed by angle reverberation as alignment returns to $\Bv$; and a non-monotonic evolution of $R_V$. The persistent observational manifestations of the transient irradiation, including enhanced polarization, blueshifted $\lambda_{\max}$, modified $R_V$, and polarization-angle evolution, constitute \emph{fossil imprints} of past cosmic transients. {\it TransRAT}  therefore establishes a physical and computational framework for time-domain dust astrophysics, connecting the real-time evolution of dust during cosmic transients with the long-lived fossil imprints that encode their irradiation history.

\end{abstract}


\section{Introduction}
\label{sec:intro}

Dust absorbs, scatters, and re-emits starlight, while aligned nonspherical grains polarize both transmitted starlight and thermal dust emission. Dust-induced polarization therefore provides a powerful probe of grain properties (size, shape, compositions), grain alignment, and magnetic fields across a wide range of environments, from the diffuse interstellar medium (ISM) and molecular clouds (MCs) to dense cores and protostellar systems \citep{PattleFissel.2019,Maury:2019jv}. In the RAdiative Torque (RAT) paradigm, anisotropic radiation exerts torques on irregular grains, spinning them up or down, inducing precession around the radiation direction, and aligning their angular momenta with a preferred axis in space \citep{LazHoang.2007,HoangLaz.2008,Hoangetal.2022}. In typical of interstellar conditions, Larmor precession of the grain magnetic moment around the ambient magnetic field is faster than radiative precession, so the magnetic field becomes the preferred alignment axis (aka. $\BRAT$). However, sufficiently strong radiation can make radiative precession dominant and the radiation direction becomes the alignment axis, aka. $\kRAT$ \citep{LazHoang.2007}. The competition between these two alignment regimes depends sensitively on the radiation field strength, magnetic field strength, and grain magnetic susceptibility \citep{Hoang.2025}. Moreover, sufficiently rapid rotation induced by RATs in strong radiation field can also lead to centrifugal disruption of grains through RAT disruption (RAT-D; \citealt{Hoangetal.2019}).

%% ── P3: The fundamental gap---the steady-state assumption ──────────────────
Most applications of RAT theory to date have focused on the spatial variation of grain alignment and dust polarization-—from the diffuse ISM to MCs, dense cores, and protostellar regions—-while assuming a stationary or slowly-varying radiation field produced by stars (see reviews by, e.g., \citealt{Andersson.2015,LAH.2015,TramHoang.2022}). This steady-state approximation may break down dramatically in the vicinity of \textit{cosmic transients}. Novae, supernovae (SNe), gamma-ray bursts (GRBs), tidal disruption events (TDEs), fast blue optical transients (FBOTs), changing-look active galactic nuclei \citep{Villar.2017,Eftekhari.2022}, and young stars undergoing episodic accretion \citep{Audard.2014} can change their luminosities by many orders of magnitude on timescales comparable to, or shorter than, the characteristic timescales governing grain rotation and alignment \citep{LazHoang.2021}. Dust in such environments is therefore driven far from the steady states usually assumed in grain alignment calculations. This raises a fundamental question: {\it How do grain alignment and dust polarization evolve with time under rapidly varying irradiation?}

%% ── P4: Four fundamental physical effects ──────────────────────────

Time-dependent illumination triggers several coupled physical processes that are usually treated separately in steady-state models. Radiative heating changes the grain temperature and hence the magnetic susceptibility, $\chi$, through Curie's law \citep{Hoang.2026}, thereby modifying the Larmor precession rate. As the relative rates of Larmor and radiative precession evolve, the preferred alignment axis can switch between $\Bv$ and $\kv$. At the same time, rapid RAT-driven spin-up can change the population of aligned grains through fast alignment and transitions between low-$J$ and high-$J$ states, while RAT-D can irreversibly modify the grain size distribution (GSD). Strong radiation may additionally deform grains toward increasingly oblate shapes with larger axial ratios \citep{Reissl.2024}. These processes operate on different characteristic timescales and can leave persistent imprints after the transient has faded. Consequently, dust extinction, emission, and polarization depend not only on the instantaneous radiation field but also on its prior history.

%% ── P5: Transients commonly occur in dusty environments ──────────────────────

This time-dependent regime is astrophysically important because many cosmic transients occur within or near dusty environments. Massive-star explosions and long GRBs are associated with star-forming regions, and molecular structures are observed around numerous supernova remnants \citep{Zhou.2023}. Runaway massive stars and dense clumps can also place substantial amounts of dusty material within a few parsecs of a SN explosion \citep{Dincel.2026,slane.2015supernova-c88}. Type~Ia SNe, although not preferentially associated with star-forming regions, commonly occur within the diffuse ISM, allowing their intense radiation to influence dust over a potentially large surrounding volume. In all of these cases, grains that were initially aligned under the ambient interstellar radiation field (ISRF) can suddenly be exposed to an intense, rapidly evolving radiation pulse, driving their rotational and alignment states far from their pre-transient conditions.

%% ── P6: Connections to fundamental astrophysical questions ───────────────────
%Because the radiation front outruns the ejecta-driven shock, the dust response can probe the magnetic field and grain properties before shock compression and processing. The timing of $\BRAT\leftrightarrow\kRAT$ switching is especially sensitive to the magnetic susceptibility of dust grains \citep{Hoang.2026}.

%% ── P7: Prior work and the remaining gap ─────────────────────────────

Previous calculations of dust exposed to SN and GRB radiation incorporated time-dependent grain alignment and rotational disruption but assumed that grains remained in pre-existing high-$J$, $\BRAT$-aligned states \citep{Hoang.2017,Gianghoang.2020,Hoanggiang.2020}. Such treatments neglect the time-dependent exchange between low-$J$ and high-$J$ populations and do not allow the alignment axis itself to evolve. They therefore cannot capture the coupled response that arises when radiative precession competes with temperature-dependent Larmor precession, nor can they consistently predict the resulting time-dependent polarimetric signatures. A general framework that evolves these processes simultaneously and connects the grain dynamics directly to observable extinction, emission, and polarization has thus been lacking.

%% ── P8: Introducing TransRAT ────────────────────────────────────

In this paper, we develop {\it TransRAT} (Transient RAdiative Torque), a general framework for modeling the time-dependent response of dust to transient irradiation. {\it TransRAT} jointly evolves the radiation field, grain temperature and sublimation, RAT-driven rotation and disruption, magnetic susceptibility, low-$J$/high-$J$ populations, and the preferred alignment axis, and uses these quantities to compute the resulting extinction, thermal emission, and Stokes parameters. The framework therefore provides a self-consistent connection between the evolving radiation source, the nonequilibrium rotational dynamics of dust grains, and their observable photometric and polarimetric signatures. Because these processes operate on different characteristic timescales, the dust response can retain a physical memory of earlier irradiation and produce fossil imprints that persist after the transient has faded. {\it TransRAT} thus provides a foundation for {\it time-domain dust astrophysics}, in which the evolving and history-dependent response of dust can be used to probe grain properties and magnetic fields in the environments of cosmic transients and to reconstruct the irradiation history experienced by the dust. We apply {\it TransRAT} to a Type~IIP SN illuminating a nearby dense molecular cloud, idealized here as a homogeneous, one-zone system. \citet{HoangPaperI.2026} (hereafter \paperi) presented the principal time-dependent polarimetric signatures predicted by this model and discussed their observational strategy. The present paper provides the full theoretical formulation and modeling methodology, validates the numerical implementation, explores the underlying grain dynamics in greater detail, and presents a more comprehensive set of extinction and polarization predictions.

%% ── P9: Paper structure ──────────────────────────────────────────

The paper is organized as follows. Section~\ref{sec:overview} provides a qualitative overview of RAT physics in steady and transient radiation fields and introduces the time-dependent radiation field and relevant dust properties. Section~\ref{sec:RAT} presents the RAT theory and the dynamical equations governing the time-dependent alignment size, preferred alignment axis, and rotational disruption. Section~\ref{sec:TransRATmodel} develops the {\it TransRAT} framework, while Section~\ref{sec:observables} presents the general formalism for calculating extinction, emission, and polarization. Section~\ref{sec:results} presents numerical results for the Type~IIP SN application. Section~\ref{sec:discussion} discusses the physical implications and observational prospects with current and future facilities. We summarize our main findings in Section~\ref{sec:summary}.

\section{Overview of RAT physics in Steady and Transient Radiation Fields}
\label{sec:overview}

We first provide a qualitative overview of grain alignment and rotational disruption by RATs under steady radiation fields and then describe how this phenomenology extends to the time dependent radiation fields of cosmic transients.

\subsection{Grain alignment in steady radiation fields: weak and strong radiation regimes}

\begin{figure}
	\centering
	\includegraphics*[width=0.6\textwidth]{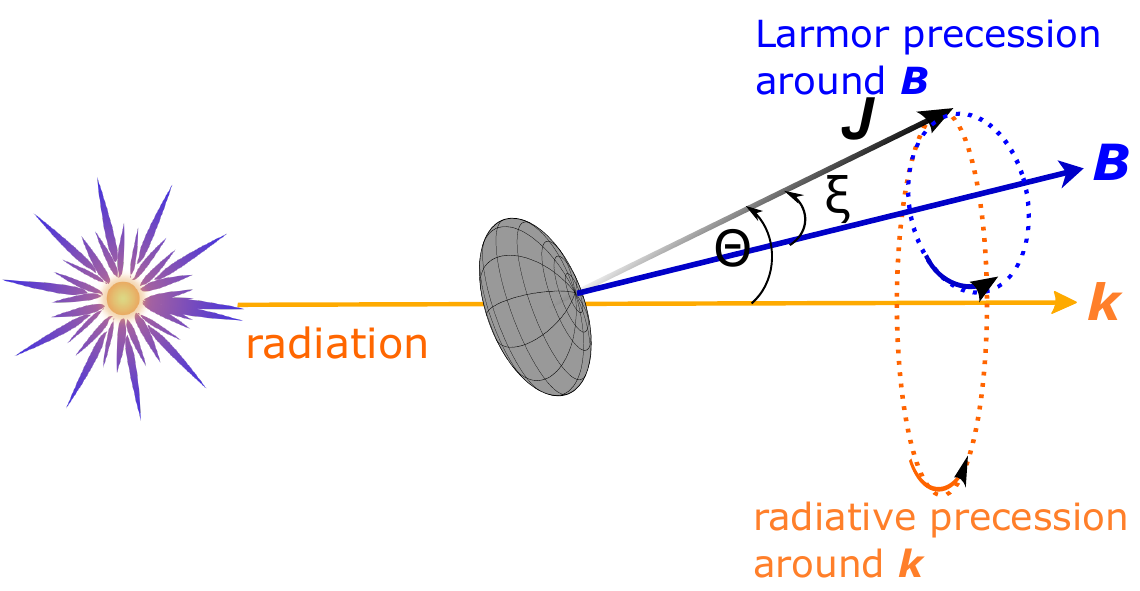}
	\caption{Illustration of grain precession processes and alignment axes in a radiation field. An oblate grain with angular momentum $\bJ$ aligned along its shortest principal axis undergoes Larmor precession about the magnetic field $\bB$ and radiative precession about the radiation direction $\bk$; the faster precession determines the $\Bv$ or $\kv$ alignment axis.}
	\label{fig:BRAT_kRAT}
\end{figure}

The RAT theory describing rotational dynamics of non-spherical grains in a steady radiation field encompasses four coupled effects (see e.g., \citealt{Hoang.2025}). First, RATs induce radiative precession of grains about the radiation propagation direction. Second, RATs modify the grain rotation rate, either accelerating grains to suprathermal rotation or decelerating them toward thermal rotation. Third, RATs drive grain alignment with either the magnetic field ($\BRAT$) or the radiation direction ($\kRAT$), with grains occupying low-$J$ or high-$J$ attractor points. The competition between radiative and Larmor precession determines whether grains undergo $\BRAT$ or $\kRAT$ alignment, as illustrated in Figure~\ref{fig:BRAT_kRAT}: the grain angular momentum precesses about both $\Bv$ and $\kv$, and the faster precession selects the alignment axis. Alignment-axis switching between $\Bv$ and $\kv$ occurs more readily for grains at low-$J$ attractors, whereas grains already aligned at high-$J$ attractors require stronger radiation because of their larger angular momenta. In the collision-dominated regime, gas collisions gradually transfer grains from low-$J$ to high-$J$ attractors, increasing alignment degree and reach perfect alignment over approximately $10$--$100$ gas-randomization times, so-called \textit{slow alignment} \citep{HoangLaz.2008,HoangLaz.2016,Hoang.2025}. Under strong radiation, by contrast, grains can undergo \textit{fast alignment} at either high-$J$ or low-$J$ attractors on a timescale comparable to the radiative precession time $\tau_k$, which is shorter than the gas-damping time $\tau_{\rm gas}$ \citep{LazHoang.2007,LazHoang.2019}. Grains driven to low-$J$ attractors may also remain confined there by strong RATs despite randomizing gas collisions, an effect known as RAT trapping \citep{Hoang.2025}. Finally, RAT-D fragments a grain when its angular velocity exceeds the critical value determined by the grain's material strength. The disruption efficiency depends on both the attractor state (high-$J$ or low-$J$) and the alignment axis ($\BRAT$ or $\kRAT$), as quantified in Sections~\ref{subsec:ratd} and \ref{sec:components}. In addition, rapid rotation by RATs can also deform grain shape due to centrifugal force, leading to more elongated oblate shapes \citep{Reissl.2024}. The resulting partition of the grain population in the size--$J$ plane is illustrated in Figure~\ref{fig:BRAT_kRAT_transient}.

%The left panel of Figure~\ref{fig:BRAT_kRAT} illustrates the alignment of an irregular, nonspherical grain under the combined influence of anisotropic radiation and an ambient magnetic field. The grain angular momentum undergoes both Larmor precession about the magnetic field $\Bv$ and radiative precession about the radiation propagation direction $\kv$. In the ISM, MCs, and star-forming regions, where the radiation field is generally weak and approximately steady, Larmor precession typically dominates. The grain angular momentum therefore aligns with the magnetic field, corresponding to the $\BRAT$ alignment regime \citep{Hoangetal.2022}. In sufficiently strong radiation fields, however, radiative precession can become faster than Larmor precession, causing the alignment axis to switch from $\Bv$ to $\kv$ and giving rise to the $\kRAT$ regime.

\subsection{RAT physics in time-varying radiation fields and expected observational signatures}

When a transient source turns on, the nearby dusty cloud is exposed to a radiation field that evolves over days to months. Throughout this paper, we describe this evolution using the retarded time $t_r$, measured from the arrival of the transient's first light at the cloud and defined quantitatively in Section~\ref{subsec:urad_general}. Following \citet{LazHoang.2007,Hoang.2025}, we classify alignment as \textit{fast} when its characteristic timescale is shorter than the gaseous randomization time and as \textit{slow} when it is longer.

To characterize the transient-alignment regime, we introduce the dimensionless ratio
\bea
\Lambda = \frac{\tau_{\rm source}}{\tau_{\rm gas}},\label{eq:lambda}
\ena
where $\tau_{\rm gas}$ is the gaseous damping time, which also sets the collisional randomization timescale, and $\tau_{\rm source}$ is the characteristic duration of the radiation source. For $\Lambda\gg1$, the radiation field is quasi-steady over a gas damping time and slow alignment can operate. For $\Lambda<1$, the source evolves before slow alignment can be established, leaving fast alignment as the relevant channel.

A transient source modifies every component of the steady-state picture described in the previous subsection. Rapid radiative heating and infrared cooling establish a time dependent equilibrium grain temperature, $T_d(t_r)$, that follows the evolving radiation field. If $T_d(t_r)$ exceeds the sublimation temperature $T_{\rm sub}$, grain sublimation reduces the surviving dust mass. The evolving grain temperature and dust abundance, in turn, produce a time dependent thermal dust emission. Through the Curie law, $\chi(T_d)\propto T_d^{-1}$, transient heating also makes the magnetic susceptibility and therefore the Larmor precession and magnetic relaxation timescales time dependent \citep{Hoang.2026}.

As the radiation field intensifies, faster radiative precession can trigger a transition from $\BRAT$ to $\kRAT$ alignment. Paramagnetic (PM) and superparamagnetic (SPM) grains respond differently to this transition. The susceptibility of PM grains decreases as $T_d$ rises, causing them to lose $\BRAT$ alignment relatively early. By contrast, SPM grains containing iron clusters with populations extending to $N_{\rm cl}\sim10^4$ can maintain sufficiently rapid Larmor precession to retain $\BRAT$ alignment for much longer. This contrast produces a polarization angle reverberation signature that provides a diagnostic of grain magnetism reported in \paperi.

Intense transient radiation also rapidly redistributes grains among their rotational attractor states. Depending on the torque configuration, grains initially aligned at either high-$J$ or low-$J$ attractors can be driven toward new high-$J$ states and potentially disrupted through RAT-D. Consequently, fast alignment and rotational disruption make the critical grain sizes for alignment and disruption functions of time, constituting a key dynamical prediction of the RAT paradigm \citep{LazHoang.2021,Hoang.2025}. At the same time, stronger irradiation can enhance RAT trapping by confining a larger fraction of grains at low-$J$ despite randomizing gas collisions \citep{Hoang.2025}. Because the transition from $\BRAT$ to $\kRAT$ alignment occurs more readily for grains at low-$J$ attractors, RAT trapping can further increase the fraction of $\kRAT$-aligned grains \citep{Hoangetal.2022}.\footnote{Increasing grain rotation rate by rising transient leads to increasingly elongated oblate shapes due to centrifugal force. This effect is not considered in this paper.} 

\subsection{Time-Varying Radiation Field and Dust Properties}
\label{sec:lightcurve}

\subsubsection{Radiation strength for a general light curve \texorpdfstring{$L(t)$}{L(t)}}
\label{subsec:urad_general}

Let $t=0$ denote the emission of the transient's first light. For dust grains at distance $D$ from the radiation source, we define the \emph{retarded time}
\bea
t_{r}=t-D/c = t-\Delta t,\label{eq:tret}
\ena
where the light-travel delay is
\bea
\Delta t=D/c \simeq 3.26 \left(\frac{D}{1\pc}\right)\,{\rm yr}.\label{eq:tdelay}
\ena
Thus, $t_r=0$ marks the arrival of the first light at the cloud.

The transient luminosity as a function of time $t$ is modeled as
\begin{eqnarray}
	L(t) = L_{\rm peak}\, f(t),
	\label{eq:Lt}
\end{eqnarray}
where $f(t)$ is the normalized light curve profile, with $f(t_{\rm peak})=1$ and $f(0)=f(\infty)=0$.

The corresponding local radiation energy density and dimensionless radiation strength are
\begin{eqnarray}
u_{\rm rad}(t_{r},D) = \frac{L(t_{r})}{4\pi D^{2} c},
\quad
U(t_{r},D) \equiv \frac{u_{\rm rad}(t_{r},D)}{u_{\rm ISRF}},
\label{eq:urad}
\end{eqnarray}
where $u_{\rm ISRF}=8.64\times10^{-13}\,\rm erg\,cm^{-3}$ is the reference interstellar radiation energy density of \citet{Mathis.1983}.

Including the ambient ISRF contribution $U_0$, Equation~\eqref{eq:urad} evaluates to
\begin{eqnarray}
	U(t_{r},D) =
	\begin{cases}
		U_0 & t_{r} < 0,\\[4pt]
		\displaystyle U_0 + \frac{L_{\rm peak}\,f(t_{r})}{4\pi D^2\,c u_{\rm ISRF}} & t_{r} \geq 0,
	\end{cases}
	\label{eq:Utotal}
\end{eqnarray}
where $U_0$ is the dimensionless strength of the pre-existing ISRF.

Numerically,
\bea
U(t_{r},D) \simeq U_{0}+1.2\times 10^{4}
\left(\frac{L(t_{r})}{10^{9}L_{\odot}}\right)
\left(\frac{10\,\mathrm{pc}}{D}\right)^{2},
\ena
which implies that, in the absence of attenuation by intervening dust, the peak transient radiation can exceed the standard ISRF ($U_0=1$) out to $D\lesssim1$~kpc.

%\footnote{For SNe~Ia exploding in the diffuse ISM, dust within approximately 1~kpc can therefore be affected by the transient radiation. The gas-randomization time is $\sim10^{6}n_{1}^{-1}$~yr, where $n_{1}=n_{\rm H}/(10\cm^{-3})$, allowing fossil signatures to persist for up to Myr timescales (Section~\ref{sec:discussion_memory}).}

\subsubsection{Specific bolometric light curves used in this work}
\label{subsec:lightcurves_specific}

For numerical results, we adopt the following analytic parametrization of a Type~IIP supernova light curve \citep{Dastidar.2018}
\begin{eqnarray}
{f_{\rm SNII}(t)}\;=\;
\begin{cases}
\dfrac{3.33x^{2}}{[1+x^{\alpha_d}]^{2}}\,
& t \le t_b\quad\text{(rise/peak)}\\[6pt]
0.85\times 10^{-0.002\,t}, & t_b < t \le 60\,\text{d}\\[3pt]
0.63\, & 60\,\text{d}<t \le 120\,\text{d}\\[3pt]
3.98\times 10^{-0.0067\,t}, & t>120\,\text{d}
\end{cases}
\label{eq:LSNII}
\end{eqnarray}
where $x=t/t_b$, $t$ is the time since explosion measured in days, $t_b=10$~d, and the decay index is $\alpha_d=7/2$, and the peak luminosity $L_{\rm peak} \approx 1.74\times 10^{9}\Lsun$ at $t\simeq 0.77\,t_{b}$. The four pieces capture, in order, the shock breakout and rise to peak, the post-peak decline, the photospheric plateau, and the $^{56}$Co radioactive tail.

The effective photospheric temperature of the SNe II is assumed to decrease with time as a power law \citep{Faran.2018}:
\bea
T_{\rm SNII}=T_{0}t^{-\alpha},\label{eq:TSNII}
\ena
with $\alpha=0.38$ and $T_{0}=24000$ K, and the minimum temperature is $T_{\rm SNII}=6000$~K.

\subsubsection{Equilibrium grain temperature}
\label{subsec:Td}

For a given transient luminosity, the equilibrium temperature of a spherical grain of radius $a$ at distance $D$ follows from radiative energy balance,
\begin{eqnarray}
	\pi a^{2} \langle Q_{\rm abs}(a, T_{\rm SN})\rangle_{\rm Planck}\,cu_{\rm rad}(t_{r})
	%\pi a^{2} Q_{\rm abs}(\lambda_{\rm SN},a)\,\frac{L(t_{r})}{4\pi D^{2}}	
	= 4\pi a^{2}\,\langle Q_{\rm abs}(a,\Td)\rangle_{\rm Planck}\,\sigma_{\rm SB}\,\Td(t_{r})^{4},
	\label{eq:Tdbal}
\end{eqnarray}
where $\langle Q_{\rm abs}(a,T)\rangle_{\rm Planck}$ is the absorption efficiency averaged over the Planck function for a blackbody spectrum of temperature $T$, and $T_{\rm SN}$ is the effective blackbody temperature of the transient radiation. Evaporative cooling by sublimation is neglected \citep{2002ApJ...569..780D,Hoangetal.2015}. When the transient luminosity becomes negligible, the local radiation field is just the background ISRF.

To obtain the dust temperature, we solve Equation~\eqref{eq:Tdbal} iteratively using $Q_{\rm abs}$ from Astrodust model \citep{Draine.2021b1e} and the Planck-weighted cooling efficiency at each trial $\Td$. For simple estimates, one can find $T_{d}$ for silicate grains, as follows \citep{Draine.2011book}
\bea
T_{d}(t_{r})\simeq 16.4U(t_{r},D)^{1/6}~\K\simeq 116\,L_{10}^{1/6}D_{1}^{-1/3}~\K,\label{eq:Td_r}
\ena
where $L_{10}=L(t_{r})/10^{10}L_{\odot}$ and $D_{1}=D/10\,\mathrm{pc}$. 

For silicate grains with a sublimation threshold $T_{\rm sub}\approx1500\K$, the sublimation front is defined by $T_d(t_r)=T_{\rm sub}$, yielding
\bea
R_{\rm sub}(t_{r})\simeq 1.4\times 10^{16} L_{10}^{1/2} \left(\frac{T_{\rm sub}}{1500\K}\right)^{-3}~\cm,\label{eq:Rsub}
\ena
which corresponds to $R_{\rm sub}\sim935$~au ($\sim0.005$~pc) for $L_{10}=1$; for the adopted peak luminosity $L_{\rm peak}\approx1.74\times10^{9}\Lsun$, $R_{\rm sub}\approx0.002$~pc. %Molecular clouds surrounding massive star progenitors generally lie beyond $\sim10$~pc, whereas dense clumps produced by stellar outflows can occupy circumstellar regions at $1$--$10$~pc and reach substantially higher temperatures.

\subsubsection{Magnetic susceptibility and Larmor precession}
\label{sec:susceptibility}
The magnetic susceptibility of dust gives rise to a grain magnetic moment, enabling the grain to interact with the ambient magnetic field. The static (zero-frequency) magnetic susceptibility, $\chi_0(\Td)$, determines the Larmor precession frequency and therefore helps set the alignment axis. The frequency-dependent susceptibility, $\chi''(\omega,\Td)$, controls Barnett and paramagnetic relaxation, which govern internal and external alignment.

For paramagnetic (PM) grains with Fe atoms embedded in the silicate matrix, $\chi_{\rm PM}(0)$ follows Curie's law, $\chi\propto 1/\Td$, so the susceptibility \emph{drops} as the transient heats the dust. For superparamagnetic (SPM) grains containing iron clusters of $N_{\rm cl}$ atoms, the susceptibility is enhanced by a factor $\propto N_{\rm cl}$, and the rising $T_{d}(t_{r})$ progressively activates larger clusters, so the effective susceptibility of a cluster size distribution remains elevated \citep{Hoang.2026}. The full formulae---the Curie law for PM grains, the SPM effective susceptibility, and the frequency-dependent susceptibility $\chi_{2}^{cd}(\omega,N_{\rm cl})$ of rotating grains---are given by \citet{Hoang.2026}. %The resulting time evolution of $\chi(0,T_d)$ and $\chi(\omega,T_d)$ for SNe~IIP illumination is shown in Figs.~\ref{fig:chi0_time} and \ref{fig:chi_time}.

A rotating magnetic grain acquires a magnetic moment through the Barnett effect \citep{Barnett.1915}. %Let $a$ be the effective size of the irregular grain, defined as the radius of a sphere of the same volume, $V_{\rm grain}=4\pi a^{3}/3$. 
For a dust grain of volume $V_{\rm grain}=4\pi a^{3}/3$ rotating with angular velocity $\bOmega$, the instantaneous Barnett magnetic moment is
\bea
\bmu_{\rm Bar}=\frac{\chi_{\rm eff}(0) V_{\rm grain}\bOmega}{\gamma_{g}}=-\frac{\chi_{\rm eff}(0) \hbar V_{\rm grain}}{g_{e}\mu_{B}}\bOmega,\label{eq:muBar}
\ena
where $\chi_{\rm eff}(0)$ is the effective magnetic susceptibility of SPM grains with superparamagnetic inclusions (SPIs), $\gamma_{g}=-g_{e}\mu_{B}/\hbar\approx -e/(m_{e}c)$ is the gyromagnetic ratio of an electron, $g_{e}\approx 2$ is the $g-$factor, and $\mu_{B}=e\hbar/2m_{e}c\approx 9.27\times 10^{-21} \erg \G^{-1}$ is the Bohr magneton (see \citealt{Hoang.2026}). 

%For the adopted spherical-equivalent grain model, the moment of inertia about the symmetry axis is $I_{\rm grain}=8\pi\rho a^{5}/15$ with $\rho$ the grain mass density. 

The Larmor precession arising from the magnetic torque between $\bmu_{\rm Bar}$ and an ambient magnetic field $B$ has the characteristic timescale of
\bea
\tau_{\rm Lar}=\frac{2\pi}{|d\phi/dt|}=\frac{2\pi I_{\rm grain}g_{e}\mu_{B}}{\chi_{\rm eff}(0)V_{\rm grain}\hbar B}
\simeq 23.7 a_{-5}^{2}\hat{\rho}\frac{10^{-4}}{\chi_{\rm eff}(0)}\frac{100\,\mu\mathrm{G}}{B}~\mathrm{days},
\label{eq:tauB}
\ena
where $I_{\rm grain}=8\pi\rho a^{5}/15$ with $\rho$ the grain mass density is the grain moment of inertia, $B$ is normalized to the typical MC field strength of $100\,\mu$G, $a_{-5}=a/(10^{-5}\cm)$, and $\hat{\rho}=\rho/(3\g\cm^{-3})$. For SPM grains, the Larmor precession timescale is $\tau_{\rm Lar}\simeq 0.2(N_{\rm cl}/100)^{-1}$~d, much shorter than the characteristic timescales of SNe.

\section{Radiative Torques: Spin-Up, Radiative Precession, Alignment, and Rotational Disruption}
\label{sec:RAT}
We now formulate the RAT processes that govern grain rotational dynamics in a transient radiation field. We assume efficient internal alignment between the grain axis of maximum moment of inertial with $\bJ$ through Barnett, inelastic, and nuclear relaxation \citep{Purcell.1979,LazDraine.1999}, an approximation appropriate for the interstellar grains in MCs considered here \citep{Hoangetal.2022}.

\subsection{Grain rotational damping}

A rotating grain is damped by random collisions with gas atoms/molecules followed by their thermal evaporation. The characteristic gas damping timescale is (see, e.g., \citealt{LazRoberge.1997})
\begin{eqnarray}
	\tau_{\rm gas}=\frac{3}{4\sqrt{\pi}}\frac{I_{\rm grain}}{1.2n_{\rm H}m_{\rm H}
		v_{\rm T}a^{4}}
\simeq 830\hat{\rho}a_{-5}n_{4}^{-1}T_{\gas,1}^{-1/2}~{\rm yr},\label{eq:taudamp_gas}
\end{eqnarray}
where $v_{\rm T}=\left(2k_{\rm B}T_{\rm gas}/m_{\rm H}\right)^{1/2}$ is the thermal velocity of hydrogen atoms of mass $m_{\rm H}$, $n_{4}=n_{\rm H}/(10^{4}\cm^{-3})$, and $T_{\gas,1}=T_{\gas}/(10\K)$.

Gas collisions tend to drive grains to thermal rotation at the angular velocity
\bea
\Omega_{T}=\left(\frac{k_{\rm B}T_{\gas}}{I_{\rm grain}}\right)^{1/2}\simeq 3.2\times 
10^{4}a_{-5}^{-5/2}T_{\gas,1}^{1/2} \s^{-1}.\label{eq:omega_th}
%10^{4}s^{-1/2}a_{-5}^{-5/2}T_{\gas,1}^{1/2} \s^{-1}.\label{eq:omega_th}
\ena

When exposed to a radiation field of energy density $u_{\rm rad}$, grains are heated and rapidly cool through infrared (IR) emission. Emitted photons carry away angular momentum and therefore cause an additional rotational damping process \citep{Draine.1998}. For a grain at equilibrium temperature $\Td$, the IR damping rate is $\tau_{\rm IR}^{-1}=F_{\rm IR}\tau^{-1}_{\gas}$, where the dimensionless IR damping parameter is
\bea
F_{\rm IR}\simeq \left(\frac{3.8\times 10^{-3}}
{a_{-5}}\right)\left(\frac{U^{2/3}}{n_{4}T_{\gas,1}^{1/2}}\right).\label{eq:FIR}
%{s^{1/3}a_{-5}}\right)\left(\frac{U^{2/3}}{n_{4}T_{\gas,1}^{1/2}}\right).\label{eq:FIR}
\ena

The total rotational damping rate is therefore
\bea
\tau_{\rm damp}^{-1}=\tau_{\rm gas}^{-1}(1 + F_{\rm IR}).\label{eq:tau_ran}
\ena

As a transient increases $U$, IR damping strengthens and dominates over gas damping when $F_{\rm IR}\gg1$.

\subsection{Radiative torques of irregular grains}
\label{subsec:rat}

The monochromatic radiative torque exerted on a dust grain of irregular shape by anisotropic radiation is
\begin{eqnarray}
	{\Gamma}_{\lambda}(t_{r})=\pi a_{\rm eff}^{2}
	\gamma_{\rm rad} u_{\lambda}(t_{r}) \left(\frac{\lambda}{2\pi}\right){Q}_{\Gamma},\label{eq:GammaRAT}
\end{eqnarray}
where $a_{\rm eff}$ is the effective grain size, defined as the radius of the equivalent sphere of the same volume, $\gamma_{\rm rad}$ is the anisotropy of the radiation field which is taken to be 1 for unidirectional transient radiation, $u_{\lambda}$ is the spectral energy density of the transient radiation field, and ${Q}_{\Gamma}$ is the RAT efficiency \citep{DraineWein.1997}. For simplicity, we neglect the order-unity factor and set $a_{\rm eff}=a$. The efficiency depends on wavelength, grain shape, and grain size, but only weakly on composition and internal structure \citep{LazHoang.2007,Herranen.2019}. %Appendix~\ref{app:QRAT} gives the adopted power-law form of $Q_{\Gamma}$ and its spectrum-averaged value $\overline{Q}_{\Gamma}$ (Eqs.~\ref{eq:QAMO}--\ref{eq:Qbar_approx}).

Let $\bar{\lambda}=\int \lambda u_{\lambda}d\lambda/u_{\rm rad}$ be the mean wavelength of the radiation spectrum. Using the spectrum-averaged RAT efficiency $\overline{Q}_{\Gamma}$, the averaged RAT can be written as \citep{Hoangetal.2022}
\begin{eqnarray}
	\Gamma_{\rm RAT}(t_{r})=\pi a^{2}
	\gamma_{\rm rad} u_{\rm rad}(t_{r}) \left(\frac{\bar{\lambda}}{2\pi}\right)\overline{Q}_{\Gamma}\simeq 5.6\times 10^{-23}\gamma_{\rm rad} a_{-5}^{4.7}\bar{\lambda}_
	{0.5}^{-1.7}U_{6}(t_{r})~\rm erg
\end{eqnarray}
for $a\lesssim \bar{\lambda}/1.8$, and
\begin{eqnarray}
	\Gamma_{\rm RAT}(t_{r})&\simeq & 8.6\times 10^{-22}\gamma_{\rm rad} a_{-5}^{2}\bar{\lambda}_{0.5}U_{6}(t_{r})~\rm erg
\end{eqnarray}
for $a> \bar{\lambda}/1.8$, where $a_{-5}=a/(10^{-5}\rm cm)$, $\bar{\lambda}_{0.5}=\bar{\lambda}/(0.5~\mu\rm m)$, and $U_6 = U(t_{r},D)/10^6$ is the radiation strength in units of $10^6\,u_{\rm ISRF}$. The grain size at which RAT changes its scaling from $\propto a^{4.7}$ to $a^{2}$ is referred to as the {\it transition size}.

The energy density and the mean wavelength of the radiation, $\bar{\lambda}$, are determined using the bolometric luminosity and effective photospheric temperatures of transients, which are shown in Eqs. (\ref{eq:LSNII}) and (\ref{eq:TSNII}) for the case of SNeII considered in this paper. 

\subsection{Radiative precession}
RATs induce grain precession about the radiation direction $\bk$. For a grain with angular momentum $J$, the radiative precession time is \citep{Hoangetal.2014}
\bea
\tau_{k}&=&\frac{2\pi}{|d\phi/dt|}=\frac{2\pi J}{\pi a^{2}\gamma_{\rm rad} u_{\rad}(t_{r})(\bar{\lambda}/2\pi)Q_{e3}},\nonumber\\
&\simeq& 359.2
\hat{\rho}^{1/2}T_{2}^{1/2}\hat{s}^{-1/6}a_{-5}^{1/2}\St\left(\frac{1.2\mum}{\gamma_{\rm rad} \bar{\lambda}\hat{Q}_{e3}U(t_{r},D)}\right)\,
\yr,\label{eq:tauk_Qe3}
\ena 
where $d\phi/dt$ is the precession rate with $\phi$ the azimuthal angle, $T_{2}=T_{\gas}/(100\K)$, $\hat{Q}_{e3}=Q_{e3}/10^{-2}$ with $Q_{e3}$ the third component of RAT efficiency $Q_{\Gamma}$ that induces the grain precession around $\bk$ \citep{LazHoang.2007}, and $\St=\Omega(t_{r})/\Omega_{T}$ is the suprathermal parameter.

The instantaneous radiative precession time depends on the grain rotation rate. At low-$J$ attractors, one has $\St\sim 1$, and the precession time decreases as $\tau_{k}^{\rm low-J}\propto 1/U(t_r,D)$ increases. At high-$J$ attractors, we use the time dependent angular velocity $\Omega(a,t_r)$ obtained from the spin-up equation below, giving $\tau_{k}^{\rm high-J}\propto\Omega(a,t_r)/U(t_r,D)$.

\subsection{Spin-up and grain angular velocity}
\label{subsec:eom}

Before the transient, grains with $a>a_{\rm align}^{\rm ISRF}$ are partitioned between low-$J$ and high-$J$ attractors by the steady radiation field (Section~\ref{sec:initial}). The onset of the transient increases $\Gamma_{\rm RAT}$ and drives the subsequent angular-momentum evolution. We hold the local gas properties fixed, an approximation appropriate for grains sufficiently far from the source to remain unaffected by the shock.

The equation of motion for irregular grains subject to RATs and rotational damping is described by
\begin{eqnarray}
	\frac{I_{\rm grain}d\Omega}{dt_{r}} = \Gamma_{\rm RAT}(t_{r})-\frac{I_{\rm grain}\Omega}{\tau_{\rm damp}},\label{eq:domega}\label{eq:omegaEoM}
\end{eqnarray}
where $\Gamma_{\rm RAT}(t_r)$ is the time dependent RAT and $\tau_{\rm damp}$ is the combined rotational damping time due to gas collisions and IR emission.

We solve Equation~\eqref{eq:omegaEoM} using the exponential integrator scheme derived in Appendix~\ref{app:spinup} (Eq.~\ref{eq:omegaUpdate}). A calculation with gas damping alone provides an upper limit to the RAT driven angular momentum, and IR damping supplies an additional sink and lowers the rotation rate.

\subsection{Rotational disruption (RAT-D) and the \texorpdfstring{disr-$J$}{disr-J} structural level}
\label{subsec:ratd}
\label{sec:RAT-D}
A grain is disrupted into fragments when the centrifugal stress exceeds the
tensile strength of the grain material $\Smax$:
\begin{eqnarray}
\Omdisr(a,\Smax) = \frac{2}{a}\sqrt{\frac{\Smax}{\rho}}.
\label{eq:omegadisr}
\end{eqnarray}
The exact value of the tensile strength for cosmic dust is uncertain, but for numerical results, we assume $\Smax=10^{7}\erg\cm^{-3}$ for large grains of $a>0.1\mum$ and $\Smax=10^{9}\erg\cm^{-3}$ for smaller compact grains, approximately reflecting the dependence of the tensile strength on the grain size and internal structures \citep{Tatsuuma.2019}.

At each epoch, we evaluate $\Omega(a,t_r)$ on a logarithmic grain size grid and identify the disruption band $[\adisr^{\min}(t_r),\adisr^{\max}(t_r)]$, within which $\Omega\ge\Omdisr$. Disruption occurs only over a \emph{bounded} size range. Although $\Omdisr\propto a^{-1}$, the RAT-driven rotation rate first increases with size and then declines for grains much larger than the dominant radiation wavelength, where the torque efficiency decreases. Grains above $\adisr^{\max}$ therefore rotate slower than the disruption limit and remain intact \citep{HoangTram.2020}. Because grain regrowth to micron sizes requires Myr timescales, disruption is effectively irreversible over the $\lesssim1$~yr intervals considered here. We represent this irreversibility by accumulating the instantaneous disruption bands, using the running minimum and maximum of their lower and upper edges,
\begin{eqnarray}
\adisr^{\min,\rm eff}(t_{r}) = \min_{t'\le t_{r}} \adisr^{\min}(t'),\qquad
\adisr^{\max,\rm eff}(t_{r}) = \max_{t'\le t_{r}} \adisr^{\max}(t'),
\label{eq:adisrirrev}
\end{eqnarray}
so once disrupted, grains remain small.

%To describe the RAT-D, we introduce a third grain J-level, the \textbf{disr-$J$ level}, defined as the set of grains inside the disruption band $a\in[\adisr^{\min}(t_{r}),\adisr^{\max}(t_{r})]$.  

\section{The TransRAT Framework: Time-Dependent Grain Populations and Rotational States}
\label{sec:TransRATmodel}

This section combines the physics of Sections~\ref{sec:lightcurve} and \ref{sec:RAT-D} into the {\it TransRAT} population model. The critical sizes for fast alignment and disruption ($\afast$, $\adisr^{\min}$, and $\adisr^{\max}$) evolve with the transient light curve and partition the GSD into components with time dependent boundaries. Rotational $J$ states provide a second dimension to this partition: they determine the alignment efficiency and disruption fate of each component, while $\BRAT$/$\kRAT$ switching determines its alignment axis. We proceed from the pre-transient ISRF initial condition (Section~\ref{sec:initial}) to the partition by critical size and $J$ state (Section~\ref{sec:components}), the fast alignment size (Section~\ref{sec:fastalign}), the component mass weights (Section~\ref{sec:mass_weights}), and the $\BRAT$/$\kRAT$ transition (Section~\ref{sec:brat_krat}). The resulting population weights, alignment efficiencies, and alignment axes provide the inputs to the observable calculations in Section~\ref{sec:observables}.

Table~\ref{tab:transrat_params} summarizes the principal model parameters. The fiducial values of $\fHiISRF$ and $\fhif$ are motivated by numerical calculations \citep{Hoang.2025}.

\begin{table*}
	\centering
	\caption{Key parameters of the TransRAT framework.}
	\label{tab:transrat_params}
	\setlength{\tabcolsep}{4pt}
	\begin{tabular}{llll}
		\toprule
		Symbol & Meaning & Default & Reference \\
		\midrule
		\multicolumn{4}{l}{\textit{Critical sizes}}\\
		$\aali^{\rm ISRF}$ & Minimum alignment size under the pre-transient ISRF (static) & --- & \S\ref{sec:initial} \\
		$\afast(t_{r})$ & Fast-alignment size: smallest grain spun up to suprathermal rotation by $t_{r}$ & --- & Eq.~\eqref{eq:afastmin} \\
		$\adisr^{\min}(t_{r}),\,\adisr^{\max}(t_{r})$ & Edges of the RAT-D disruption band (irreversibly accumulated) & --- & Eq.~\eqref{eq:adisrirrev} \\
		$a_{\rm trans}(t_{r})$ & $\BRAT$/$\kRAT$ transition size, where $\tkRAT<\tLar$ first holds & --- & \S\ref{sec:brat_krat} \\
		\midrule
		\multicolumn{4}{l}{\textit{Rotational $J$-level fractions}}\\
		$\fHiISRF$ & High-$J$ fraction under the steady ISRF ($t_{r}=0$); $\fLoISRF=1-\fHiISRF$ & 0.5 & \S\ref{sec:initial} \\
		$\fhif$ & High-$J$ fraction of fast-aligned grains ($t_{r}>0$) & 0.25 & \S\ref{sec:fastalign} \\
		$\fDisrJ$ & Disrupted (high-$J$) mass fraction inside the band & --- & Eq.~\eqref{eq:fdisr_eff_parallel} \\
		\midrule
		\multicolumn{4}{l}{\textit{Alignment efficiencies and weights}}\\
		$\Rhigh$, $\Rlow$ & Rayleigh reduction factors at high-$J$ and low-$J$ attractors & 1.0, 0.1 & \S\ref{sec:initial} \\
		$f_{\rm align}$ & Bare per-grain alignment efficiency: $\Rhigh$, $\Rlow$, or 0 (unaligned/disrupted) & --- & Table~\ref{tab:populations} \\
		$f_{\rm align}^{\rm eff}$ & Population-weighted effective alignment (fraction $\times$ Rayleigh factor) & --- & Eqs.~\eqref{eq:falign_pre}, \eqref{eq:falign_fast} \\
		$w_{X}$ & Dust-mass weight of component $X$ (carries the population fraction) & --- & Eq.~\eqref{eq:weights} \\
		\midrule
		\multicolumn{4}{l}{\textit{Alignment axis (B/$\kRAT$ switching)}}\\
		$\tLar$ & Larmor precession timescale about $\Bv$ & --- & Eq.~\eqref{eq:tauB} \\
		$\tkRAT$ & Radiative precession timescale about $\kv$ & --- & Eq.~\eqref{eq:tkRAT} \\
		$N_{B\to k}$, $N_{k\to B}$ & Precession cycles required before an axis switch (delay clocks) & 5, 5 & Eqs.~\eqref{eq:delay_bk}, \eqref{eq:delay_kb} \\
		$\psi$ & Angle between radiation direction $\kv$ and magnetic field $\Bv$ & $45^{\circ}$ & \S\ref{sec:brat_krat} \\
		\bottomrule
	\end{tabular}
	\tablecomments{Quantities marked ``---'' are computed per epoch from the light curve $L(t_{r})$ and local gas/dust conditions rather than set as inputs. $\psi=0^{\circ}$ is used for the validation case (Appendix~\ref{app:kpB_validation}).}
\end{table*}

\subsection{Initial condition (pre-transient ISRF)}
\label{sec:initial}
The pre-transient state at $t_r=0$ provides the simplest {\it TransRAT} partition. The steady ISRF, characterized by radiation strength $U_{0}$ and mean wavelength $\bar\lambda_{0}$, defines the static minimum alignment size $\aali^{\rm ISRF}$ through the standard RAT condition (e.g., \citealt{Hoangetal.2022}). The two rotational attractors then divide the grains into three components (Figure~\ref{fig:BRAT_kRAT_transient}, left): grains below $\aali^{\rm ISRF}$ rotate thermally and have no net alignment ($f_{\rm align}=0$; component~A; \citealt{HoangLaz.2016,Hoang.2025}), whereas grains above $\aali^{\rm ISRF}$ occupy the high-$J$ and low-$J$ attractors with fractions $\fHiISRF$ and $\fLoISRF=1-\fHiISRF$ (components~B and C, respectively).

We quantify the alignment efficiency of grains of a given size through the Rayleigh reduction factor $R$ \citep{Greenberg.1998,LazRoberge.1997}. Grains at high-$J$ and low-$J$ attractors have the alignment efficiency $\Rhigh$ and $\Rlow$, respectively, with typical ISRF values $\Rhigh\simeq1$ and $\Rlow\sim0.1$--$0.2$ \citep{HoangLaz.2008,HoangLaz.2016,Hoang.2025}. Throughout this paper, $f_{\rm align}$ denotes the per-grain alignment efficiency, whereas the superscript ``eff'' denotes the population weighted value (population fraction multiplied by the Rayleigh factor). The high-$J$ and low-$J$ contributions are therefore
\bea
\fHiISRFeff &&= \fHiISRF\Rhigh,\\
\fLoISRFeff &&= (1-\fHiISRF)\Rlow.
\label{eq:fhiloJ_ISRF}
\ena

The population weighted size-dependent alignment function at $t_{r}=0$ is given by
\begin{eqnarray}
	f_{\rm align}^{\rm eff}(a, t_{r}=0) =
	\begin{cases}
		0, & a<\aali^{\rm ISRF}, \ \text{(A)}\\[3pt]
		\fHiISRFeff, & a\ge \aali^{\rm ISRF}, \ \text{high-$J$ (B)}\\[3pt]
		\fLoISRFeff, & a\ge \aali^{\rm ISRF}, \ \text{low-$J$ (C)}.
	\end{cases}
	\label{eq:falign_pre}
\end{eqnarray}

If all grains have the same alignment axis--the magnetic field, components B and C can be combined:
\bea
f_{\rm align}^{\rm eff,B+C}(a\ge \aali^{\rm ISRF}, t_{r}=0)=\fHiISRFeff+\fLoISRFeff= \fHiISRF\Rhigh + (1-\fHiISRF)\Rlow.\label{eq:fali_ISRF}
\ena
This expression recovers the step-like alignment function of \citet{Hoang.2014b} and underlies RAT implementations in \textsc{DustPOL-py} \citep{Tram.2023} and \textsc{POLARIS} \citep{Reissl.2016,Giangetal.2024}. The combined treatment is adequate when both populations share an alignment axis. In a transient radiation field, however, low-$J$ and high-$J$ grains can have distinct alignment axes, so components B and C must be modeled separately.

\subsection{Modeling time dependent RAT alignment and disruption}
\subsubsection{Grain-population partition and alignment efficiency}
\label{sec:components}

The time-varying radiation field from transients generalizes the pre-transient partition of Section~\ref{sec:initial} in both grain size and rotational state. Two time dependent size boundaries emerge: the fast alignment size $\afast(t_r)$, which extends alignment below $\aali^{\rm ISRF}$, and the RAT-D band $[\adisr^{\min}(t_r),\adisr^{\max}(t_r)]$. Their evolution with the light curve drives the changing extinction, emission, and polarization. To distinguish the different alignment axes available to low-$J$ and high-$J$ grains \citep{Hoangetal.2022} and to represent disruption, we augment the classical two-attractor RAT picture \citep{HoangLaz.2008,HoangLaz.2016,Hoang.2025} with a third, disrupted structural state.

For each grain size, the resulting three-state hierarchy is
\begin{eqnarray}
\underbrace{\text{low-}J}_{\text{poorly aligned}}
\;\longrightarrow\;
\underbrace{\text{high-}J}_{\text{well aligned,}\;a<\adisr^{\min}\;\text{or}\;a>\adisr^{\max}}
\;\longrightarrow\;
\underbrace{\text{disr-}J}_{\text{disrupted,}\;\adisr^{\min}\le a\le\adisr^{\max}}.
\label{eq:three_levels}
\end{eqnarray}

\begin{figure*}
	\includegraphics*[width=0.49\textwidth]{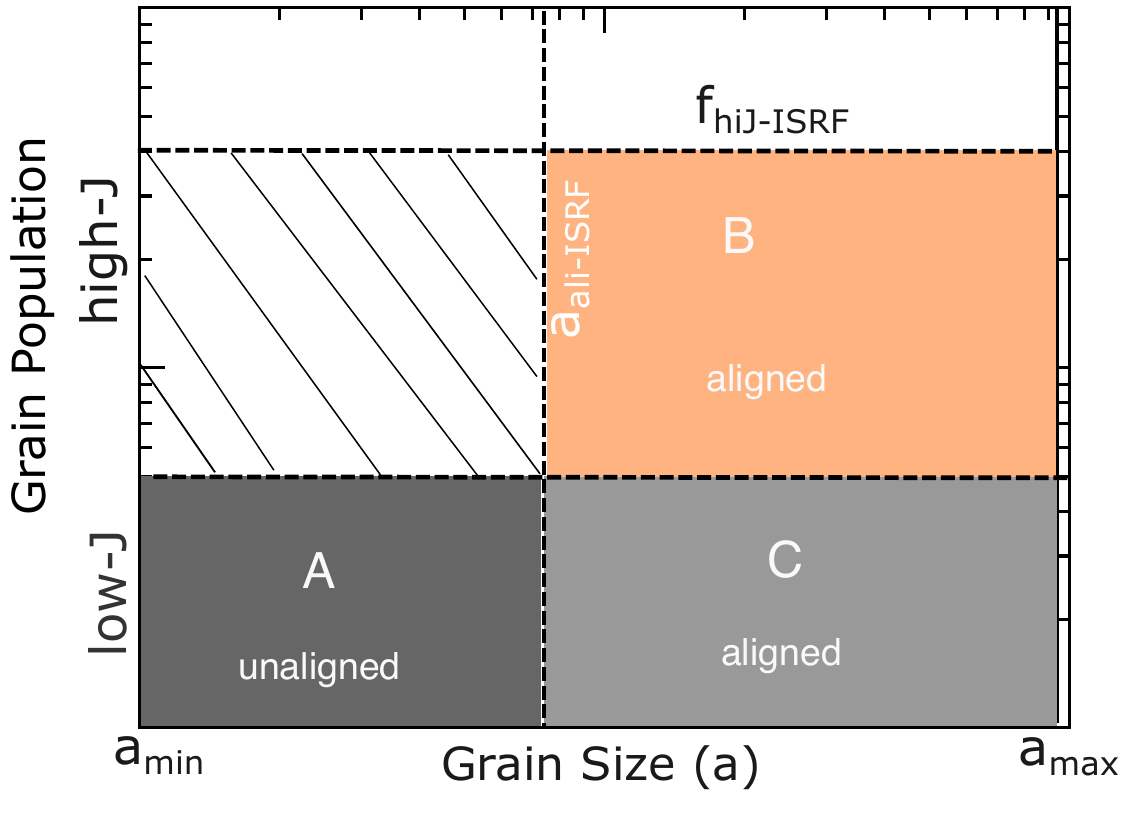}
	\includegraphics*[width=0.49\textwidth]{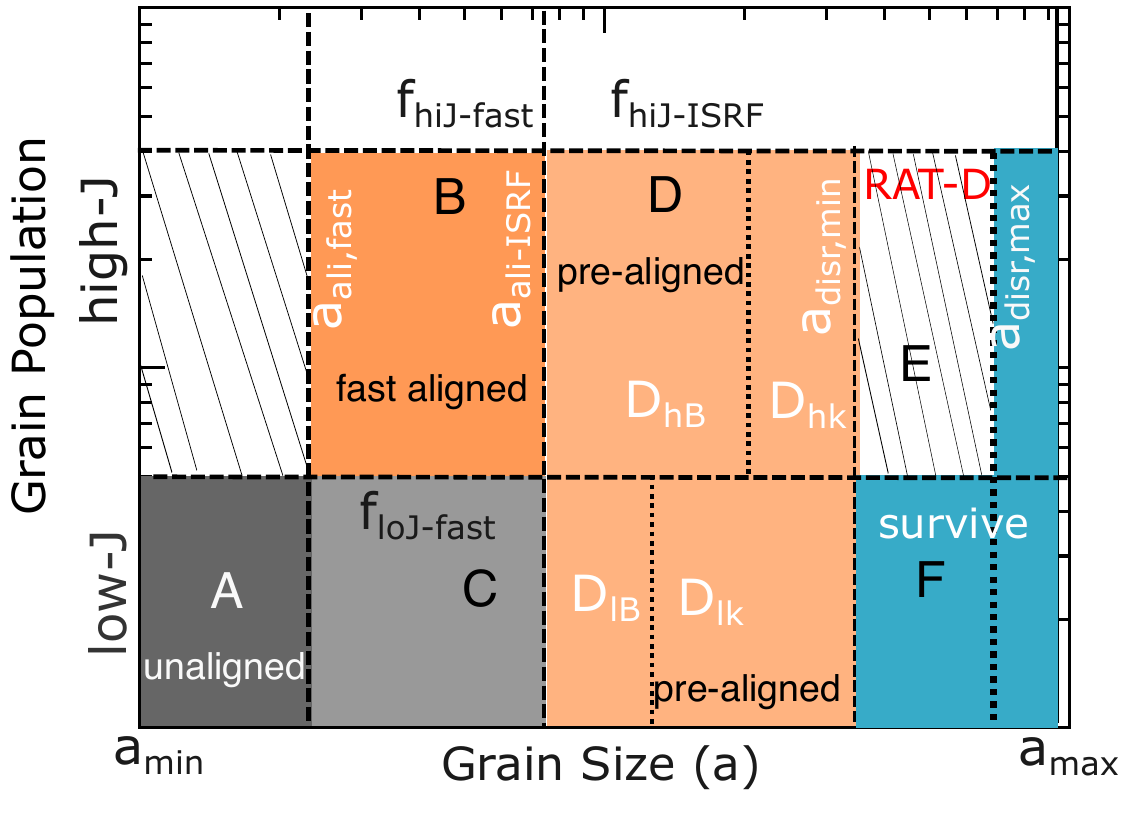}
	\caption{Grain population partition in the {\it TransRAT} size--$J$ plane. \textit{Left:} Before the transient, the steady ISRF cannot align grains with $a<\aali^{\rm ISRF}$ (unaligned; A) and drives larger grains to aligned high-$J$ (B) and low-$J$ (C) attractors. \textit{Right:} During the transient, $\afast$ extends alignment to smaller grains (B and C), while the band $[\adisr^{\min},\adisr^{\max}]$ contains disrupted high-$J$ grains (E) and surviving low-$J$ grains (F). Pre-aligned grains outside the disruption band form the high-$J$ and low-$J$ D components and may be further divided into $\BRAT$ and $\kRAT$ subpopulations. Vertical boundaries mark the critical sizes; Table~\ref{tab:populations} gives the corresponding size ranges, fractions, and alignment efficiencies.}
	\label{fig:BRAT_kRAT_transient}
\end{figure*}

Figure~\ref{fig:BRAT_kRAT_transient} (right) displays the partition in the size--$J$ plane, where the intersections of the size intervals with these $J$ states define components A, B, C, $D_h$, $D_l$, E, and F. Table~\ref{tab:populations} lists the corresponding size ranges, population fractions, and alignment efficiencies. Relative to the pre-transient state, components D, E, and F describe the consequences of RAT-D, whereas B and C describe fast alignment induced by the transient.

Two independent fractions govern the populations at different epochs. The \emph{ISRF equilibrium fraction} $\fHiISRF$ gives the fraction driven to high-$J$ attractors by the steady ISRF and sets the initial condition at $t_r=0$. The \emph{fast alignment fraction} $\fhif$ gives the fraction that reaches high-$J$ under the rapid action of the transient \citep{Hoang.2025} and controls the newly aligned components at $t_r>0$. These parameters are physically distinct; setting $\fHiISRF=\fhif$ recovers the adopted steady-state partition.

Combining the fast alignment fractions with the Rayleigh factors gives the population weighted effective alignment degrees of fast-aligned grains at high-$J$ and low-$J$ attractors, analogous to Equation~(\ref{eq:fhiloJ_ISRF}):
\begin{eqnarray}
\fHiFasteff &&= \fhif\Rhigh,\\
\fLoFasteff &&= (1-\fhif)\Rlow,
\label{eq:feff_definitions}
\end{eqnarray}
with $f_{\rm align}=0$ for the disr-$J$ level (structural label, no observables; grains are disrupted and carry no coherent alignment).

The pre-transient fractions ($\fHiISRFeff$, $\fLoISRFeff$) are used for the ISRF baseline ($t_{r}=0$) and for the $\BRAT$ sub-range of D-range grains ($a\ge\aali^{\rm ISRF}$) that retain their ISRF alignment properties without undergoing a $\BRAT$ $\to$ $\kRAT$ axis switch. The {\it fast} combinations ($\fHiFasteff$, $\fLoFasteff$) give the population weighted alignment \emph{contributions} of the fast-aligned high-$J$ and low-$J$ grains and characterizes the effective efficiency of the combined B$+$C interval (Eq.~\ref{eq:falign_fast}). In the per-component decomposition of Table~\ref{tab:populations}, each component instead carries its bare alignment efficiency $f_{\rm align}$ ($\Rhigh$ or $\Rlow$). For numerical results, we adopt typical values of $\fHiISRF=0.5$, $\fhif=0.25$, $\Rlow=0.1$, $\Rhigh=1.0$ (see e.g.,\citealt{Hoang.2025}).

\begin{table*}
	\centering
	\caption{Grain populations defined by rotational level and size range.}
	\label{tab:populations}
	\footnotesize
	\setlength{\tabcolsep}{4pt}
	\begin{tabular}{llllll}
		\toprule
		$J-$Level & Component & Size range & Fraction & Efficiency $f_{\rm align}$ & Notes \\
		\midrule
		low-$J$ & $\mathbf{A}$ & $a_{\min}\le a < \afast(t_{r})$ & -- & $0$ & Thermally rotating \\
		low-$J$ & $\mathbf{C}$ & $\afast(t_{r})\le a < a_{\rm align}^{\rm fast,top}$ & $1-\fhif$ & $\Rlow$ & Fast-aligned $\to$ low-$J$ \\
		low-$J$ & $\mathbf{D_{l}}$ & $[\aali^{\rm ISRF},\adisr^{\min})\cup[\adisr^{\max},a_{\max}]$ & $1-\fHiISRF$ & $\Rlow$ & ISRF-aligned $\to$ low-$J$ (outside band) \\
		low-$J$ & $\mathbf{F}$ & $\adisr^{\min}(t_{r})\le a \le \adisr^{\max}(t_{r})$ & $1-\fDisrJ$ & $\Rlow$ & Low-$J$ survivors inside the band \\
		\midrule
		high-$J$ & $\mathbf{B}$ & $\afast(t_{r})\le a < a_{\rm align}^{\rm fast,top}$ & $\fhif$ & $\Rhigh$ & Fast-aligned $\to$ high-$J$ \\
		high-$J$ & $\mathbf{D_{h}}$ & $[\aali^{\rm ISRF},\adisr^{\min})\cup[\adisr^{\max},a_{\max}]$ & $\fHiISRF$ & $\Rhigh$ & ISRF-aligned $\to$ high-$J$ (outside band); see \S\ref{sec:dsplit} \\
		\midrule
		disr-$J$ & $\mathbf{E}$ & $\adisr^{\min}(t_{r})\le a \le \adisr^{\max}(t_{r})$ & $\fDisrJ$ & $0$ & Structural label only; no observables\\
		\bottomrule
	\end{tabular}
	\tablecomments{$a_{\rm align}^{\rm fast,top} = \min(\aali^{\rm ISRF},\,\adisr^{\min}(t_{r}))$; $\afast(t_{r})$ is the fast alignment size defined in \S\ref{sec:fastalign}. RAT-D disrupts only grains inside the band $[\adisr^{\min},\adisr^{\max}]$; grains above $\adisr^{\max}$ survive and extend the ISRF-aligned components $D_h,D_l$. The tabulated $f_{\rm align}$ is the \emph{bare} Rayleigh factor of each component.}
\end{table*}

Component~A grains lie below the fast RAT alignment threshold and remain thermally rotating with \emph{zero} net alignment ($f_{\rm align}=0$). Component~F grains (inside the band, $\adisr^{\min}\le a\le\adisr^{\max}$) survive RAT-D because they were at low-$J$ (only the high-$J$ fraction $\fDisrJ$ is disrupted; see Eq.~\ref{eq:fdisr_eff_parallel}). Since \emph{all} F survivors are low-$J$ grains, they share the bare low-$J$ Rayleigh reduction factor $\Rlow$, independent of whether they were ISRF- or fast-aligned before disruption.  Like all other low-$J$ components, $F$ is split between $\BRAT$ and $\kRAT$ sub-ranges at the low-$J$ precession-timescale transition size $a_{\rm tr,F}^{\rm lo}$ (see \S\ref{sec:brat_krat}); this split changes only the alignment \emph{axis} ($\Bv$ vs.\ $\kv$), not the efficiency.

For components B and C, the combined, population weighted effective alignment of the fast-aligned interval is
\begin{eqnarray}
	f_{\rm align}^{\rm eff,fast} =
	\fHiFasteff + \fLoFasteff = \fhif\,\Rhigh + (1-\fhif)\,\Rlow.
	\label{eq:falign_fast}
\end{eqnarray}

\subsubsection{Alignment axis switching between $\mathbf{B}$ and $\mathbf{k}$}
In the special case where the radiation direction is parallel to the magnetic field ($\bk\|\bB$), the alignment axis never changes, and the transient radiation only redistributes grains among the three $J$ levels. Appendix~\ref{app:kpB_validation} shows the results for this testing case. 

For the general case in which $\bk$ is not aligned with $\Bv$, grains that retain their $\BRAT$ alignment axis behave, at any angle $\psi$, identically to the $\kv\parallel\Bv$ case. Otherwise, some grains can switch between $\BRAT$ and $\kRAT$, altering both the net polarization degree and position angle. Following \citet{LazHoang.2021,Hoang.2025}, the possible transitions include:
\begin{itemize}
\item \emph{High-$J$ grains: axis-dependent alignment.} A fraction of D-range grains ($a\ge\aali^{\rm ISRF}$; component $D_h$) pre-aligned at high-$J$ by the ISRF may undergo a $\BRAT$ $\to$ $\kRAT$ axis switch, acquiring fast alignment fraction $\fhif$. The axis choice affects only the alignment geometry entering the polarization: because the fast transient re-partitions the $J$-levels of all band grains regardless of their initial alignment state, the disruption efficiency throughout the band is $\fhif$ uniformly (\S\ref{sec:mass_weights}). This contrasts with the $\kv\parallel\Bv$ case, where no axis switch occurs and D-range band grains retain the ISRF high-$J$ fraction $\fHiISRF$ (Eq.~\ref{eq:fdisr_eff_parallel}).

\item \emph{Low-$J$ grains: fast-align only if the axis switches.} D-range grains ($a\ge\aali^{\rm ISRF}$; component $D_l$) at low-$J$ attractors that undergo a $\BRAT$ $\to$ $\kRAT$ switch fast-align along $\kv$ with fraction $\fhif$. Grains that retain their $\BRAT$ axis remain at the ISRF-level efficiency $\fLoISRFeff$ due to RAT trapping, with no fast alignment event in this sub-range.

\item \emph{Unaligned grains: fast-align via $\kRAT$.} Grains below the ISRF threshold ($a<\aali^{\rm ISRF}$) fast-align, predominantly via $\kRAT$ (\S\ref{sec:brat_krat}), with fraction $\fhif$; the disruption efficiency in this B/C-range is also $\fhif$.
\end{itemize}

Therefore, our overall modeling strategy is described as follows. We follow the grain populations across the three $J$-levels; on each level, grains can align via $\BRAT$ or $\kRAT$ depending on their precession rate, each with an effective efficiency $f_{\rm align}^{\rm eff}$. Rotational disruption depletes the high-$J$ population and leaves survivors at low-$J$, whereas fast alignment can transport grains from low-$J$ to high-$J$ attractors. The resulting time dependent level populations determine the dust extinction, emission, and polarization. The remainder of this section describes this strategy quantitatively: the fast alignment size $\afast$ (\S\ref{sec:fastalign}), the component mass weights (\S\ref{sec:mass_weights}), and the $\BRAT$/$\kRAT$ switching that sets the alignment axis (\S\ref{sec:brat_krat}).

\subsection{Fast alignment grain size under transient RATs}
\label{sec:fastalign}

For grains initially at low-$J$ ($a<\aali^{\rm ISRF}$), the transient RAT spin-up is governed by Equation~\eqref{eq:omegaEoM}. The fast alignment timescale
$\tau_{\rm fast}(a)$ is the time at which $\Omega(t_{r})$ first crosses $\zeta_{\rm align}\Omth$, with suprathermal threshold $\zeta_{\rm align}=3$:

\begin{eqnarray}
\Omega(\tau_{\rm fast}(a)) = \zeta_{\rm align}\,\Omth(a).
\label{eq:tau_fast}
\end{eqnarray}

This crossing is permanent (the angular velocity does not subsequently fall below threshold within the damping time $\tdamp$, which far exceeds the timespan considered, for the size range
considered). We define $\afast(t_{r})$ as the smallest grain that has crossed threshold by time $t_{r}$:
\begin{eqnarray}
\afast(t_{r}) = \min\bigl\{\,a < \aali^{\rm ISRF}\,:\,\tau_{\rm fast}(a) \le t_{r}\,\bigr\}.
\label{eq:afastmin}
\end{eqnarray}

Once fast-aligned, a fraction $\fhif$ of grains reaches the high-$J$ attractor while the remaining $1-\fhif$ stays at low-$J$ (\S\ref{sec:components}).

\subsection{Mass weights: dust mass fraction in each component}
\label{sec:mass_weights}
\label{sec:gsd}
We assume that the pre-transient GSD follows the \citet{Mathis.1977} (hereafter MRN) power law, $dn/da=Za^{q}$ with $q=-3.5$ (Appendix~\ref{app:gsd}, \ref{eq:gsd_mrn}), with $Z$ constrained by the gas-to-dust mass ratio fixed to $100:1$. In addition, a nanodust population of size $a\le a_{\rm junc}=5$~nm is considered and assumed to follows a log-normal size distribution. Because the nanodust population is assumed to be unaligned and does not experience RAT-D, the analysis below is performed on the MRN part; the nanodust is tracked separately through its amplitude, which RAT-D rescales (Eq.~\ref{eq:ratd_amp}). 
We now quantify the dust-mass fraction in each component: regions A--F of Figure~\ref{fig:BRAT_kRAT_transient}, with the ISRF-aligned range~D resolved into its high-$J$ part $D_h$ and low-$J$ part $D_l$.

The disrupted grains are not vanished in mass but fragment into smaller pieces and are immediately redistributed through the new size distribution according to the models of Appendix~\ref{app:gsd}. The quantity $M_{\rm frag}$ (Eq.~\ref{eq:Mfrag}) records the mass transferred out of the disruption band, which is not a separate mass added to the active grain components. \emph{RAT-D acts only on grains that are rotating suprathermally, i.e., on the high-$J$ attractor population.} The low-$J$ population is trapped by RATs and retains the original MRN distribution up to $a_{\max}$. The resulting effective size distribution and its mass-conserving normalization are derived in Appendix~\ref{app:gsd}.

\paragraph{Sublimation convention.}
For the sake of simplicity, we neglect the grain size dependence of sublimation and adopt a sharp destruction approximation: once the grain temperature exceeds sublimation limit $T_{\rm sub}=1500$K for silicates, the entire dust population is removed. The mass partition and conservation relations below therefore apply only while the dust survives; after sublimation, all dust observables are set to zero.

For the initial MRN distribution, the dust mass in the size range $[a_{1},a_{2}]$ is proportional to the mass measure
\begin{eqnarray}
\mathcal{M}(a_1,a_2)\equiv\int_{a_1}^{a_2} a^{q+3}\,da = 2\bigl(a_{2}^{1/2}-a_{1}^{1/2}\bigr),
\quad q=-3.5,
\label{eq:Mdef}
\end{eqnarray}  
where the common factor containing the MRN amplitude, grain density, and geometrical normalization is omitted because it cancels from every mass fraction. RAT-D changes the shape of the size distribution but not its total mass. We therefore define the corresponding post-redistribution mass measure
\begin{eqnarray}
\widetilde{\mathcal{M}}(a_1,a_2;t_r)
\equiv \frac{1}{Z}\int_{a_1}^{a_2}a^3
\left(\frac{dn}{da}\right)_{\!\rm eff}(a,t_r)\,da,
\label{eq:Mtdef}
\end{eqnarray}
where $(dn/da)_{\rm eff}$ includes both the undisrupted grains and the fragments redistributed according to Appendix~\ref{app:gsd}. In the pre-transient limit, $\widetilde{\mathcal{M}}(a_1,a_2;0)=\mathcal{M}(a_1,a_2)$. 

The conserved total dust mass is
\begin{eqnarray}
M_{\rm tot}\equiv M_0
\equiv\widetilde{\mathcal{M}}(a_{\min},a_{\max};t_r)
=\mathcal{M}(a_{\min},a_{\max}),
\label{eq:Mtot}
\end{eqnarray}
at every epoch. Let $\mathcal{M}_E = \mathcal{M}(\adisr^{\min},\adisr^{\max})$ denote the original mass measure inside the disruption band. The disrupted part of this original mass is transferred to smaller sizes and is already contained in $\widetilde{\mathcal{M}}$; it is not a separate additive mass reservoir.

\paragraph{Effective disruption fraction $\fDisrJ$.}
The disruption efficiency depends on whether the alignment axis changes during the transient, i.e., on the relative orientation of $\kv$ and $\Bv$:
\begin{itemize}
\item \emph{$\kv\parallel\Bv$ ($\BRAT$ maintained):} The alignment axis does not change, so grains in the band that were already at high-$J$ by the ISRF ($a\geq\aali^{\rm ISRF}$) are further spun up by the transient and disrupted with efficiency $\fHiISRF$.  Band grains below the ISRF threshold ($\adisr^{\min}\le a < \aali^{\rm ISRF}$) are newly spun up by the transient and disrupted with efficiency $\fhif$. The effective disruption fraction is therefore the mass weighted average over the band
\begin{eqnarray}
\fDisrJ = \frac{\fhif\,\mathcal{M}(\adisr^{\min},\,\tilde{a}) + \fHiISRF\,\mathcal{M}(\tilde{a},\,\adisr^{\max})}
              {\mathcal{M}(\adisr^{\min},\,\adisr^{\max})},
\quad \tilde{a}\equiv\mathrm{clip}\!\bigl(\aali^{\rm ISRF},\,\adisr^{\min},\,\adisr^{\max}\bigr),
\label{eq:fdisr_eff_parallel}
\end{eqnarray}
where $\mathrm{clip}(x,x_{1},x_{2})=\min[\max(x,x_{1}),x_{2}]$ restricts $\aali^{\rm ISRF}$ to the band; this reduces to $\fHiISRF$ when $\adisr^{\min}\ge\aali^{\rm ISRF}$ (all band grains are D-range).
\item \emph{$\kv\nparallel\Bv$ ($\BRAT$ $\to$ $\kRAT$ axis transition possible):} All disrupted grains are driven to high-$J$ by the fast transient regardless of their initial state, so $\fDisrJ = \fhif$ uniformly.
\end{itemize}
The surviving low-$J$ fraction is $(1-\fDisrJ)$, which constitutes component~F.

Because RAT-D acts only inside the band, the grains above $\adisr^{\max}$ survive intact and extend the ISRF-aligned population. Including redistributed fragments wherever they fall in size, the total mass assigned to the ISRF-aligned range is
\begin{eqnarray}
\widetilde{\mathcal{M}}_{D}(t_r) \equiv
\widetilde{\mathcal{M}}(\aali^{\rm ISRF},\adisr^{\min};t_r)
+ \widetilde{\mathcal{M}}(\adisr^{\max},a_{\max};t_r).
\label{eq:MtD}
\end{eqnarray}

The mass fractions of the six active components (A, B, C, $D_h$, $D_l$, and F) are
\begin{eqnarray}
\begin{aligned}
w_{A} &= \frac{\widetilde{\mathcal{M}}(a_{\min},\afast;t_r)}{M_{\rm tot}},&
w_{B} &= \fhif\,\frac{\widetilde{\mathcal{M}}(\afast,a_{\rm align}^{\rm fast,top};t_r)}{M_{\rm tot}},\\[4pt]
w_{C} &= (1-\fhif)\,\frac{\widetilde{\mathcal{M}}(\afast,a_{\rm align}^{\rm fast,top};t_r)}{M_{\rm tot}},&
w_{D_{h}} &= \fHiISRF\,\frac{\widetilde{\mathcal{M}}_{D}}{M_{\rm tot}},\\[4pt]
w_{D_{l}} &= (1-\fHiISRF)\,\frac{\widetilde{\mathcal{M}}_{D}}{M_{\rm tot}},&
w_{F} &= \frac{(1-\fDisrJ)\,\mathcal{M}_E}{M_{\rm tot}},
\end{aligned}
\label{eq:weights}
\end{eqnarray}
satisfying $w_A+w_B+w_C+w_{D_{h}}+w_{D_{l}}+w_F = 1$. Here the $D_h$ and $D_l$ weights integrate over \emph{both} ISRF-aligned intervals (below and above the disruption band) through $\widetilde{\mathcal{M}}_D$ (Eq.~\ref{eq:MtD}). The high-$J$/low-$J$ split places fraction $\fHiISRF$ of this mass in $D_h$ (efficiency $\Rhigh$) and $1-\fHiISRF$ in $D_l$ (efficiency $\Rlow$), so that $w_{D_h}\Rhigh+w_{D_l}\Rlow = (\fHiISRFeff+\fLoISRFeff)\,\widetilde{\mathcal{M}}_{D}/M_{\rm tot}$ reproduces the combined ISRF-aligned contribution exactly.

Each weight $w_{X}$ is the fraction of the conserved total dust mass assigned to component $X$, after fragment redistribution, so that $M_X=w_XM_{\rm tot}$ and $\sum_Xw_X=1$. The same $w_X$ also serves as the normalization correction factor that makes the per-component \texttt{DustPOL\_py} subrange calculations additive (Appendix~\ref{app:bookkeeping}).

The disr-$J$ diagnostic weight
\begin{eqnarray}
w_{E}(t_{r}) = \frac{\fDisrJ\,\mathcal{M}_E}{M_{\rm tot}}
\label{eq:wE_def}
\end{eqnarray}
tracks the dust mass fraction disrupted by RAT-D.

\paragraph{Global mass conservation and fragment redistribution.}
For diagnostic purposes, the mass transferred out of the disruption band is
\begin{eqnarray}
M_{\rm frag}(t_{r}) \equiv \fDisrJ(t_{r})\,\mathcal{M}_E(t_{r}) = w_E(t_{r})\,M_{\rm tot},
\label{eq:Mfrag}
\end{eqnarray}
but this mass has already been redistributed among the active components through $(dn/da)_{\rm eff}$ and $\widetilde{\mathcal{M}}$. It must therefore not be added again to their masses. At every epoch,
\begin{eqnarray}
M_A+M_B+M_C+M_{D_h}+M_{D_l}+M_F
= M_{\rm tot}\equiv M_{0}.
\label{eq:massbudget}
\end{eqnarray}

Thus $M_{\rm frag}$ measures how much mass has changed size, rather than an additional reservoir outside A--F. Equivalently, there is no declining parent-mass normalization: the component mass represented by the post-redistribution distribution is always $M_{\rm tot}$. Equation~\eqref{eq:massbudget} is verified numerically at every epoch by integrating the complete post-RAT-D distribution (Appendix~\ref{app:massbudget}, Figure~\ref{fig:massbudget_fractions}).

\paragraph{Alignment efficiency of component~F.}
As described in \S\ref{sec:components}, all F survivors carry the bare low-$J$ factor $\Rlow$ (the surviving fraction $1-\fDisrJ$ is already in $w_F$, Equation~\ref{eq:weights}), and the $\BRAT$/$\kRAT$ split at $a_{\rm tr,F}^{\rm lo}$ changes only the alignment axis: grains with $a < a_{\rm tr,F}^{\rm lo}$ remain aligned along $\Bv$ and grains with $a \ge a_{\rm tr,F}^{\rm lo}$ along $\kv$, both with efficiency $\Rlow$.

The time dependent quantities that drive these mass weights for the case of SNe-MC are summarized in Figure~\ref{fig:diagnostics}, including the radiation strength $U(t_{r},D)$, the grain temperature $\Td(t_{r})$, and the critical sizes for fast alignment ($\afast$) and rotational disruption ($\adisr$), together with the SN bolometric light curve.

\subsection{$\BRAT$ versus $\kRAT$ alignment-axis switching}
\label{sec:brat_krat}

Whether a grain aligns along $\Bv$ ($\BRAT$) or along the radiation direction $\kv$ ($\kRAT$) is determined by along which axis the grain precession dominates. The following subsections give the instantaneous criterion, the alignment axis-switching delay model, the per-grain size transition boundary, and the D-component realignment logic.

\subsubsection{Instantaneous precession timescale criterion}
\label{subsec:brat_crit}
At low-$J$ the relevant rotation rate is approximately $\Omth$; at high-$J$ it is the time dependent angular velocity $\Omega(a,t_{r})$ obtained from the RAT
spin-up ordinary differential equation (ODE; Section~\ref{sec:RAT-D}). The radiative precession timescale about $\kv$ can be evaluated by
\begin{eqnarray}
\tkRAT^{\rm low\text{-}J}(a,t_{r}) = \frac{2\pi I_{\rm grain}\,\Omth(a)}{\Gamrat(a,t_{r})},
\qquad
\tkRAT^{\rm high\text{-}J}(a,t_{r}) = \frac{2\pi I_{\rm grain}\,\Omega(a,t_{r})}{\Gamrat(a,t_{r})},
\label{eq:tkRAT}
\end{eqnarray}
where $\Omega(a,t_{r})$ carries a thermal floor $\max[\Omega(a,t_{r}),\Omth]$ so
that high-$J$ timescales are always $\ge$ low-$J$ ones \citep{Hoangetal.2022}. Here we have approximated the precession-inducing torque by the full RAT magnitude, i.e., $Q_{e3}\approx Q_{\Gamma}$, by disregarding the order-unity coefficient, so the above equation is equivalent to Equation~\ref{eq:tauk_Qe3}.

The faster precession determines the alignment axis:
\begin{eqnarray}
\tkRAT < \tLar &\;\Longrightarrow\;& \text{$\kRAT$, with axis }\kv,\nonumber\\
\tkRAT > \tLar &\;\Longrightarrow\;& \text{$\BRAT$, with axis }\Bv.
\label{eq:brat_krat}
\end{eqnarray}

Because $\tkRAT$ is independent of $\chi_0$ while $\tLar\propto 1/\chi_0$, the criterion is most stringent for PM grains and most easily satisfied
(toward $\kRAT$) for grains with weak magnetic susceptibility. We consider five magnetic cases spanning three model classes (PM, single-size SPM of $\Ncl=10^{2}-10^{4}$) and an SPM power-law distribution of cluster sizes, $dn_{\rm cl}/dN_{\rm cl}\propto N_{\rm cl}^{-\eta}$, with the typical slope $\eta=11/6$ \citep{Hoang.2026}.

The per-grain size transition boundary $a_{\rm trans}(t_{r})$ within each component range is found by scanning the grain size grid and locating the
smallest $a$ where $\tkRAT(a,t_{r})<\tLar(a,t_{r})$ first holds. Grains in $[a_{\rm lo}, a_{\rm trans})$ contribute as $\BRAT$; grains in $[a_{\rm trans}, a_{\rm hi})$ contribute as $\kRAT$ (see Section~\ref{subsec:stokes}).

\subsubsection{Alignment axis-switching delays}
\label{subsec:switch_delay}

The instantaneous criterion Equation~\eqref{eq:brat_krat} gives the equilibrium alignment axis, but a grain cannot switch axes instantaneously because it must
complete several precession cycles before settling into the new attractor. The code implements two independent delay clocks, one per direction:

\paragraph{$\BRAT$ $\to$ $\kRAT$ (luminosity rise).}
We assume that a pre-aligned dust grain via $\BRAT$ exposed to a brightening SN needs $N_{B\to k}$ full radiative precession periods about $\kv$ before it stably becomes aligned via $\kRAT$. The switch to $\kRAT$ is therefore delayed by
\begin{eqnarray}
\Delta t_{B\to k} = N_{B\to k}\,\tkRAT,
\label{eq:delay_bk}
\end{eqnarray}
with the default value chosen as $N_{B\to k}=5$. During this window, the grain is still treated as $\BRAT$, and the $a_{\rm trans}$ value from the last confirmed $\BRAT$ time step is held fixed.

\paragraph{$\kRAT$ $\to$ $\BRAT$ (luminosity decay).}
A dust grain aligned via $\kRAT$ whose $\tkRAT$ has grown back above $\tLar$ (SN fading) must complete $N_{k\to B}$ Larmor precession periods about $\Bv$ before $\BRAT$ is restored:
\begin{eqnarray}
\Delta t_{k\to B} = N_{k\to B}\,\tLar,
\label{eq:delay_kb}
\end{eqnarray}
with the default $N_{k\to B}=5$. During this window, the grain retains $\kRAT$, and the $a_{\rm trans}$ value from the last confirmed $\kRAT$ time step is held fixed.

The two clocks are mutually exclusive at every time step. When the instantaneous state is $\kRAT$, the $B\to k$ clock is reset, and when it is $\BRAT$, the $k\to B$ clock is reset. A quasi-static resolution guard on the time step is described in Appendix~\ref{app:bookkeeping}.

\subsubsection{Component D --- \texorpdfstring{$k$}{k}-RAT realignment}
\label{sec:dsplit}

The fast-aligned components B and C are likewise split at their B/$\kRAT$ axis boundaries by the per-size precession criterion of \S\ref{sec:brat_krat}; the additional realignment machinery below is specific to the ISRF-aligned range~D (components $D_h$ and $D_l$), whose grains carry a pre-existing $\Bv$ alignment from the ISRF and must \emph{actively} transition to $\kv$ rather than aligning from the beginning.

Pre-SN ISRF-aligned grains now occupy the two disjoint intervals $[\aali^{\rm ISRF},\adisr^{\min})$ (below the band) and $[\adisr^{\max},a_{\max}]$ (above the band), with combined post-redistribution mass measure $\widetilde{\mathcal{M}}_D$ (Eq.~\ref{eq:MtD}). They comprise the high-$J$ component $D_h$ and the low-$J$ component $D_l$ (Table~\ref{tab:populations}); both initially have their alignment axis along $\Bv$, with combined per-mass alignment quality $\fHiISRFeff+\fLoISRFeff$. The realignment construction below is applied \emph{identically} to each interval, each with its own per-size transition sizes $a_{\rm tr,D}^{\rm hi}$, $a_{\rm tr,D}^{\rm lo}$, and the resulting sub-range weights are summed; we write it out for the below-band interval $[\aali^{\rm ISRF},\adisr^{\min})$ (replace $\adisr^{\min}\to a_{\max}$ and $\aali^{\rm ISRF}\to\adisr^{\max}$ for the above-band interval). When the SN turns on they can realign from $\Bv$ to $\kv$ if and only if (i)~$\kv \nparallel \Bv$,
(ii)~$\tkRAT^{\rm high\text{-}J}(D)<\tLar$, and (iii)~$t_{\rm elapsed} \ge N_{B\to k}\,\tkRAT^{\rm high\text{-}J}(D)$ (the same B$\to$k delay of Eq.~\ref{eq:delay_bk} applies).

When all three conditions hold, grains within $D_h$ and $D_l$ are split at the per-size high-$J$ $\kRAT$ transition $a_{\rm tr,D}^{\rm hi}$: grains below it retain their $\BRAT$ axis and ISRF alignment quality, while grains above it re-align along $\kv$ with the fast fraction --- $\fhif$ to high-$J$ ($\Rhigh$) and $1-\fhif$ to low-$J$ ($\Rlow$, split further at $a_{\rm tr,D}^{\rm lo}$). The use of $\fhif$ (not $\fHiISRF$) reflects that the B$\to$k axis transition is itself a fast alignment process: above the $\kRAT$ transition the ISRF high-$J$/low-$J$ partition is re-rolled to the fast fraction. The resulting sub-range mass weights (Eq.~\ref{eq:Dsplit}), the $d_{\rm split}=\mathrm{False}$ fallback, and the four sub-range calls of component~F are collected in Appendix~\ref{app:bookkeeping}.

With the mass weights and effective alignment efficiencies of all grain populations now specified, we can proceed to model the dust observables, including extinction, emission, and polarization, as described in the next section.

\section{Time-Dependent Dust Extinction, Emission, and Polarization}
\label{sec:observables}

We now describe the general formulae used to predict the observational signatures of grain alignment and disruption in transient radiation fields. We first specify the alignment axis of each grain component and its contribution to the plane-of-sky (POS) Stokes parameters (Section~\ref{subsec:stokes}), then formulate extinction and extinction polarization (Section~\ref{subsec:ext_pol}), and thermal dust emission and polarization (Section~\ref{subsec:em_pol}). Section~\ref{subsec:nummethods} summarizes the numerical implementation.

%The new module exercises the latest RAT physics developed for the radiation-dominated regime in \citep{Hoang.2025} and the effective magnetic susceptibility for SPIs with size distribution from \citep{Hoang.2026}.  For each grain population with size range $[a_1,a_2]$, effective alignment fraction $f_{\rm align}^{\rm eff}(a)$, axis-to-LOS angle $\psi$, and radiation parameter $U$, \texttt{DustPOL\_py} returns the four quantities below; the multi-population results are combined as weighted sums. 

\subsection{Magnetic field and radiation direction configuration}
\label{subsec:stokes}
The magnetic field $\Bv$ and radiation direction $\kv$ are specified by its inclination $\gamma$ from the POS and its projected position angle $\psi$: $(\gamma_B,\psi_B)$ for the magnetic field and $(\gamma_k,\psi_k)$ for the radiation direction. Unless stated otherwise, we assume both axes in the POS ($\gamma_B=\gamma_k=0^{\circ}$), so that $\psi$ is their projected position angle difference. We determine the alignment axis $\Bv$ vs. $\kv$ of each grain population by comparing its characteristic precession timescales (see Appendix~\ref{app:bookkeeping}). 

If the cloud instead lies directly along the sightline to the transient, $\kv$ is parallel to the line of sight ($\gamma_k=90^{\circ}$), and $\kRAT$-aligned grains produce no dichroic polarization. Components that retain $\BRAT$ alignment still polarize the transient light, with the amplitude reduced by the corresponding projection factor. The $\cos^{2}\gamma$ geometric factor (maximum at $\gamma=0^{\circ}$, axis fully in the POS; zero at $\gamma=90^{\circ}$, axis along the LOS) enters Equations~\eqref{eq:QUext} and \eqref{eq:QU} explicitly; $\hat{p}_{\rm ext}^X/N_{\rm H}$ and $\hat{I}_{\rm pol}^X$ are the \texttt{DustPOL\_py} quantities evaluated at $\gamma=0^{\circ}$.  Position angles follow Equation~\ref{eq:PAext_X}: emission PA $=$ axis PA $+90^{\circ}$; extinction PA $=$ axis PA. Stokes $(Q,U)$ are assembled via Equations~\eqref{eq:QU} and \eqref{eq:QUext}.  The condition $\kv\parallel\Bv$ is tested via $|\Bv\times\kv|<10^{-4}$ (3-D unit-vector cross product); when satisfied the D split is suppressed and only RAT spin-up modifies the alignment degree. Mixed $\BRAT$/$\kRAT$ populations with different $\psi$ partially depolarise via $Q/U$ cancellation; the rotation of the position angle $\theta$ occurs when the dominant axis switches from $\Bv$ to $\kv$.

\subsection{Extinction and dichroic polarization}
\label{subsec:ext_pol}
Consider a background star viewed through the MC, including the special case in which the dust cloud lies between the transient and the observer. Full-range size integrals in this section use the post-redistribution distribution $(dn/da)_{\rm eff}$ of Equation~\eqref{eq:Mtdef}, whereas per-component integrals use the reference distribution $(dn/da)_{X}^{\rm norm}$, renormalized to the total dust mass on the component interval (Appendix~\ref{app:gsd}); $1/N_{\rm H}$ converts either to the distribution per H atom.

The wavelength-dependent extinction in magnitudes per H column is
\begin{eqnarray}
\frac{A(\lambda)}{N_{\rm H}}
= \frac{1.086}{N_{\rm H}}
  \int_{a_{\min}}^{a_{\max}} \pi a^{2}\,Q_{\rm ext}(\lambda,a)\,
  \left(\frac{dn}{da}\right)_{\!\rm eff}da,
\label{eq:Alam}
\end{eqnarray}
where $Q_{\rm ext}(\lambda,a) = Q_{\rm abs}(\lambda,a)+Q_{\rm sca}(\lambda,a)$
is the total extinction efficiency from \cite{Draine.2021b1e}, and $a_{\min},a_{\max}$ are the lower and upper cutoffs of the entire GSD. %The factor $1.086 = 2.5\log_{10}e$ converts optical depth to magnitudes.

Aligned non-spherical grains produce wavelength-dependent linear polarization of background starlight due to dichroic extinction. For oblate grains with a symmetry axis $\ba_{1}$, the polarization efficiency due to dichroic extinction is 
\begin{eqnarray}
 Q_{\rm pol}^{\rm ext}(\lambda,a)
  \equiv \frac{1}{2}\left[ Q_{\rm ext}^{\perp}(\lambda,a) - Q_{\rm ext}^{\parallel}(\lambda,a)\right],
\label{eq:Qpol_ext}
\end{eqnarray}
where $\perp$ and $\parallel$ refer to the $E$-field orientation relative to the grain symmetry axis $\ba_{1}$. For numerical results, we adopt the fixed axial ratio of $s=1/1.4$ and use cross-sections from \cite{Draine.2021b1e}.

For a single grain population $X$ with size range $[a_1,a_2]$ and alignment efficiency $f_{\rm align}^{X}(a)$ (the bare Rayleigh factor of Table~\ref{tab:populations}), the extinction polarization efficiency evaluated at $\gamma=0^{\circ}$ (axis fully in the POS) is
\begin{eqnarray}
\frac{\hat{p}_{\rm ext}^{X}(\lambda)}{N_{\rm H}}
= \frac{1}{N_{\rm H}}
  \int_{a_1}^{a_2}
    \pi a^{2}\,f_{\rm align}^{X}(a)\,
    Q_{\rm pol}^{\rm ext}(\lambda,a)\,
    \left(\frac{dn}{da}\right)_{\!X}^{\rm norm}da,
%\quad [\%\,\mathrm{cm}^{2}].
\label{eq:pextX}
\end{eqnarray}
where we have assumed that the polarized optical depth $\tau_{\rm pol}\ll 1$ (see e.g., \citealt{BaoHoang.2025}).

The actual extinction polarization efficiency for a component $X$ whose alignment axis has inclination $\gamma_X$ above the POS is
$p_{\rm ext}^X(\gamma_X)/N_{\rm H} = \cos^2\!\gamma_X\,\hat{p}_{\rm ext}^X/N_{\rm H}$, with $\gamma_X \in \{\gamma_B, \gamma_k\}$
depending on whether the component is $\BRAT$ or $\kRAT$ aligned.

For extinction, the $E$-vector of polarized light is \emph{parallel} to the projected alignment axis ($\Bv$ or $\kv$) on the sky (the axis preferentially absorbs the perpendicular component), so the extinction polarization position angle of component $X$ is
\begin{eqnarray}
\psi_{X}^{\rm ext} = \psi^{\rm axis}.
\label{eq:PAext_X}
\end{eqnarray}

Because the alignment axis depends on grain size, magnetic properties, and rotational state, the six active grain populations (A, B, C, $D_h$, $D_l$, and F) can contribute at different polarization angles. We therefore calculate and combine their Stokes parameters separately. The mass-weighted Stokes parameters for extinction polarization are built by summing over all active populations $X$:
\begin{eqnarray}
q_{\rm ext}(\lambda)
  = \sum_{X} w_{X}\,\frac{\hat{p}_{\rm ext}^{X}(\lambda)}{N_{\rm H}}\,\cos^{2}\!\gamma_{X}\,
    \cos 2\psi_{X}^{\rm ext},
\qquad
u_{\rm ext}(\lambda)
  = \sum_{X} w_{X}\,\frac{\hat{p}_{\rm ext}^{X}(\lambda)}{N_{\rm H}}\,
    \cos^{2}\!\gamma_{X}\,\sin 2\psi_{X}^{\rm ext},
\label{eq:QUext}
\end{eqnarray}
where $X$ runs over all active $\BRAT$ and $\kRAT$ subranges (up to 18 in the general $\bk\nparallel\bB$ case; the full enumeration is given in Appendix~\ref{app:bookkeeping}). Component~A ($f_{\rm align}=0$) contributes zero to polarized intensities but is retained for bookkeeping. The weight $w_{X}$ appears because $\hat{p}_{\rm ext}^{X}$ is the \texttt{DustPOL\_py} output for the reference distribution $(dn/da)_{X}^{\rm norm}$, internally renormalized to the total dust-to-gas mass ratio on $[a_1,a_2]$ (Eq.~\ref{eq:Chi_norm}); multiplication by $w_{X}$ restores the physical component mass (Eq.~\ref{eq:gsd_eff}).

The total extinction polarization efficiency is
\begin{eqnarray}
\frac{p_{\rm ext}(\lambda)}{N_{\rm H}}
  = 100\times \sqrt{q_{\rm ext}^{2}(\lambda) + u_{\rm ext}^{2}(\lambda)}
\quad [\%\,\mathrm{cm}^{2}].
\label{eq:Pext_NH}
\end{eqnarray}

\begin{comment}
For finite optical depth the observed fractional polarization is obtained
from the exact dichroic-absorption formula
\begin{eqnarray}
p_{\rm ext}(\lambda,t_{r})
  = \tanh\!\!\left(\frac{P_{\rm ext}(\lambda)}{N_{\rm H}}\cdot\frac{N_{\rm H}}{100}\right)
    \times 100\%,
\label{eq:pext_obs}
\end{eqnarray}
which reduces to $p_{\rm ext} \approx P_{\rm ext}/N_{\rm H} \cdot N_{\rm H}$ in the
optically thin limit $P_{\rm ext} N_{\rm H} \ll 100\%$ and saturates at
$100\%$ for large columns.  
\end{comment}

The polarization position angle is evaluated with the quadrant-aware two-argument arctangent,
\begin{eqnarray}
\theta_{\rm ext}(\lambda,t_{r})
  = \tfrac{1}{2}\operatorname{atan2}\!\left[u_{\rm ext}(\lambda)/q_{\rm ext}(\lambda)\right].
\label{eq:theta_ext}
\end{eqnarray}

Equations~\eqref{eq:QUext}--\eqref{eq:theta_ext} are evaluated exactly at
every time step of the numerical pipeline (\S\ref{subsec:nummethods}).

\subsection{Thermal dust emission intensity and polarization}
\label{subsec:em_pol}
Grains heated by the transient reemit radiation in infrared. Their specific emission intensity is
\begin{eqnarray}
I_{\rm em}(\lambda,t_{r}) =
  N_{\rm H} \int_{a_1}^{a_2}
    \pi a^{2}\,Q_{\rm abs}(\lambda,a)\,
    B_\lambda\!\bigl(\Td(a,t_{r})\bigr)\,
    \frac{1}{N_{\rm H}}\left(\frac{dn}{da}\right)_{\!\rm eff}da,
\label{eq:Iem}
\end{eqnarray}
where $B_{\lambda}$ is the Planck function and $\Td(a,t_r)$ is the grain equilibrium temperature (Eq.~\ref{eq:Tdbal}). Here, we ignore the temperature distribution of grains for simplicity, which is valid for large grains with thermal equilibrium. The limits $(a_1,a_2)$ span the size range of the component considered; for the total emission entering Equation~\eqref{eq:pem}, $(a_1,a_2)=(a_{\min},a_{\max})$ of the effective post disruption distribution. %Unaligned PAHs contribute to the total intensity but not to the polarized intensity.

The specific intensity of linearly polarized thermal emission, evaluated at the maximum polarization geometry ($\gamma=0^{\circ}$, alignment axis fully in the
plane of sky), is
\begin{eqnarray}
\hat{I}_{\rm pol}^{X}(\lambda,t_{r})
= N_{\rm H} \int_{a_1}^{a_2}
    \pi a^{2}\,f_{\rm align}^{X}(a)\,
   Q_{\rm pol}^{\rm em}(\lambda,a)\,
    B_\lambda\!\bigl(\Td(a,t_r)\bigr)\,
    \frac{1}{N_{\rm H}}\left(\frac{dn}{da}\right)_{\!X}^{\rm norm}da,
\label{eq:Ipol_X}
\end{eqnarray}
where 
\bea
Q_{\rm pol}^{\rm em}(\lambda,a) = \frac{1}{2}\left[ Q_{\rm abs}^{\perp}(\lambda,a) - Q_{\rm abs}^{\parallel}(\lambda,a)\right],\label{eq:Qpol_em}
\ena
is the emission polarization efficiency, and the optically thin regime is assumed \citep{HoangBao.2024}.

The actual polarized intensity for a component with alignment-axis inclination
$\gamma_X$ above the POS is $I_{\rm pol}^{X}(\lambda,\gamma_X) = \cos^{2}\!\gamma_X\,\hat{I}_{\rm pol}^{X}(\lambda)$, with $\gamma_X = \gamma_B$ for $\BRAT$ components or $\gamma_X = \gamma_k$ for $\kRAT$ components.

The emission position angle of the $E$-vector is perpendicular to the
projected alignment axis,
\bea
\psi^{\rm em} = \psi^{\rm axis}+90^{\circ}.
\ena
%with $\hat{I}_{\rm pol}^{X}$ the DustPOL polarized intensity evaluated at $\gamma_X=0^{\circ}$.

The mass-weighted emission Stokes parameters normalized to the total emission intensity are
\begin{eqnarray}
q_{\rm em}(\lambda)
  = \sum_{X} w_{X}\,\frac{\hat{I}_{\rm pol}^{X}(\lambda)}{I_{\rm em}}\,\cos^{2}\!\gamma_{X}\,\cos 2\psi_{X}^{\rm em},
\qquad
u_{\rm em}(\lambda)
  = \sum_{X} w_{X}\,\frac{\hat{I}_{\rm pol}^{X}(\lambda)}{I_{\rm em}}\,\cos^{2}\!\gamma_{X}\,\sin 2\psi_{X}^{\rm em},
\label{eq:QU}
\end{eqnarray}
where $X$ runs over the same active sub-ranges as in Equation~\eqref{eq:QUext}, with mass weights from Equation~\eqref{eq:weights}.%, and $I_{\rm em}(\lambda)$ is the total thermal emission evaluated from the post-RAT-D distribution $(dn/da)_{\rm eff}$ and therefore already includes the redistributed fragments.

The fraction and position angle of thermal dust polarization are
\begin{eqnarray}
p_{\rm em}(\lambda,t_{r})
  = 100\times \sqrt{q_{\rm em}^{2}(\lambda)+u_{\rm em}^{2}(\lambda)}[\%],
\qquad
\theta_{\rm em}(\lambda,t_{r})
  = \tfrac{1}{2}\operatorname{atan2}\!\left[u_{\rm em}(\lambda)/q_{\rm em}(\lambda)\right].
\label{eq:pem}
\end{eqnarray}
%Their polarized or unpolarized contributions are assigned through the appropriate A--F component according to grain size and alignment state.% No additional $I_{\rm frag}$ term is added, because doing so would count the redistributed mass twice.

\begin{comment}
For finite optical depth the code applies an exact uniform-slab correction to $p_{\rm em}$:
\begin{eqnarray}
p_{\rm em}^{\rm thick}(\lambda,t_{r})
  = \frac{e^{-\tau}\sinh(p_{\rm thin}\tau)}{1 - e^{-\tau}\cosh(p_{\rm thin}\tau)},
\label{eq:pem_thick}
\end{eqnarray}
where $p_{\rm thin} = P_{\rm em}/I_{\rm em}$ is the optically thin fraction and $\tau(\lambda)$ is the dust optical depth at emission wavelength $\lambda$.  The total intensity is simultaneously corrected by $h(\tau)=(1-e^{-\tau})/\tau \to 1$ as $\tau\to 0$.  Both corrections matter only for very dense, nearby clouds.
\end{comment}

In the general case where $\Bv\neq\kv$, each component in the sums of Equations~\eqref{eq:QU} and \eqref{eq:QUext} is further split at its per-grain size B/$\kRAT$ transition boundary $a_{\rm trans}$ (Section~\ref{sec:brat_krat}): grains in $[a_{\rm lo}, a_{\rm trans})$ contribute with $\BRAT$ position angle $\psi_{B}$, and grains in $[a_{\rm trans}, a_{\rm hi})$ with $\kRAT$ position angle $\psi_{k}$. The sub-range decomposition of components D and F is detailed in Appendix~\ref{app:bookkeeping}.

\subsection{Numerical Implementation}
\label{subsec:nummethods}

To obtain the time dependent alignment and disruption by RATs and resulting observables, we solve the coupled system of equations in Sections~\ref{sec:lightcurve}--\ref{sec:observables} with the modular pipeline summarized in Figure~\ref{fig:pipeline} (Appendix~\ref{app:bookkeeping}). The calculation couples three elements: (i) a transient source described by a generalized bolometric light curve $L(t_r)$, (ii) time dependent grain physics, including sublimation, magnetic susceptibility, RAT spin-up, fast alignment and rotational disruption, and precession processes; and (iii) polarized radiative transfer, evaluated with an extended version of \texttt{DustPOL\_py} \citep{Tram.2023} that returns extinction, extinction polarization, thermal emission intensity, and thermal emission polarization for each component. We evolve the models to $T_{r}=3650$ days (10 years) after the arrival of the SN radiation at the cloud surface, thereby capturing both the full transient light curve and the subsequent long-term evolution of the dust. The per-epoch sequence of operations---from the update of $U$ and $\Td$ through the spin-up solution, critical sizes, size--$J$ repartition, and assembly of the Stokes parameters via Equations~\eqref{eq:QUext} and \eqref{eq:QU}---is shown in Figure~\ref{fig:pipeline}. Appendix~\ref{app:bookkeeping} describes the implementation details, including the adaptive refinement of the time grid near the onset of RAT-D and the $\BRAT$$\to$$\kRAT$ transition. %Because the modules exchange only $U(t_r,D)$, $\Td(t_r)$, and the critical grain sizes, a different transient can be modeled by supplying its light curve, spectrum, and cloud parameters. 

\begin{table}[ht]
\centering
\caption{Fiducial model parameters.}
\label{tab:fid}
\begin{tabular}{lll}
\toprule
Parameters & Symbols & Values\\
\midrule
Dust model & --- & Astrodust$+$PAH\\
Minimum grain size & $a_{\min}$ & $3.5$~\AA\\
Maximum grain size & $a_{\max}$ & $0.5\mum$\\
Gas number density & $\nH$ & $10^{4}\,\rm cm^{-3}$\\
Gas temperature & $\Tgas$ & $20\,$K\\
Magnetic field strength & $B$ & $100\,\mu$G\\
Tensile strength & $\Smax$ & $10^{7}{\rm erg\,cm^{-3}}$ for $a>0.1\mu$m and $10^{9} {\rm erg\,cm^{-3}}$ for $a<0.1\mu$m\\
Grain density & $\rho$ & $3\,\rm g\,cm^{-3}$\\
Pre-SN ISRF mean wavelength & $\bar\lambda_{0}$ & $1.2\,\mu$m\\
Pre-SN ISRF strength & $U_{0}$ & $0.1$\\
%SN radiation mean wavelength & $\bar\lambda_{\rm SN}$ & $0.3\,\mu$m\\
Suprathermal threshold & $\zeta_{\rm align}$ & $3$\\
ISRF high-$J$ fraction & $\fHiISRF$ & $0.5$\\
SN fast alignment fraction & $\fhif$ & $0.25$\\
Rayleigh factors & $(\Rlow,\Rhigh)$ & $(0.1, 1.0)$\\
Angle between $\kv$ and $\Bv$ & $\psi$ & $45^{\circ}$ ($0^{\circ}$ for validation)\\
%Distance grid & $D$ & $\{0.1,1,5\}\,$pc\\
Time grid & $t_r$ & 0.5--3650\,d, $N_T=70$ log-spaced $+$ adaptive refinement\\
Grain size grid & $a$ & $a_{\min}$--$a_{\max}$, 80 log-spaced\\
\bottomrule
\end{tabular}
\end{table}

\section{Numerical Results for a Type~IIP Supernova-MC}
\label{sec:results}
\subsection{Model setup}
We now apply {\it TransRAT} to an idealized scenario in which a dense MC is subject to time dependent radiation field from a Type IIP SN having the piecewise bolometric light curve in Equation~\eqref{eq:LSNII}. To isolate the time dependent dust physics, we represent the illuminated cloud by a homogeneous, single-phase, one-zone model instead of treating three-dimensional cloud structure. We further assume that the zone is geometrically thin with depth $L_{\rm zone}\ll D$, optically thin to the incident radiation so that the variations in the incident radiation field across the zone are disregarded. We adopt $L_{\rm zone}=0.01$~pc, corresponding to $N_{\rm H}\simeq3\times10^{20}\,\mathrm{cm^{-2}}$ for $n_{\rm H}=10^{4}\,\mathrm{cm^{-3}}$. The gas density determines the grain-damping and relaxation timescales, whereas the zone depth sets only the column normalization. The fiducial parameters are listed in Table~\ref{tab:fid}.

The simple one-zone treatment allows us to follow the coupled evolution of grain rotation, alignment, disruption, and the GSD without introducing radiative transfer effects. Because the principal results are expressed as $A_\lambda/N_{\rm H}$, $P_{\rm ext}/N_{\rm H}$, $R_V$, and $p_{\rm em}$, they are independent of the adopted zone depth in the optically thin limit. The predictions should therefore be interpreted as the local dust response at a given source--cloud distance $D$. Spatially integrated predictions for extended clouds require integration over radiation attenuation, light-travel-time effects, and internal variations in the gas, dust, and magnetic-field properties.

\subsection{Grain alignment function and grain size distribution}
\label{sec:diag_falign}
%We first show the numerical results for time dependent grain alignment and size distribution functions.

\begin{figure}
	\centering
	\includegraphics[width=0.7\textwidth]{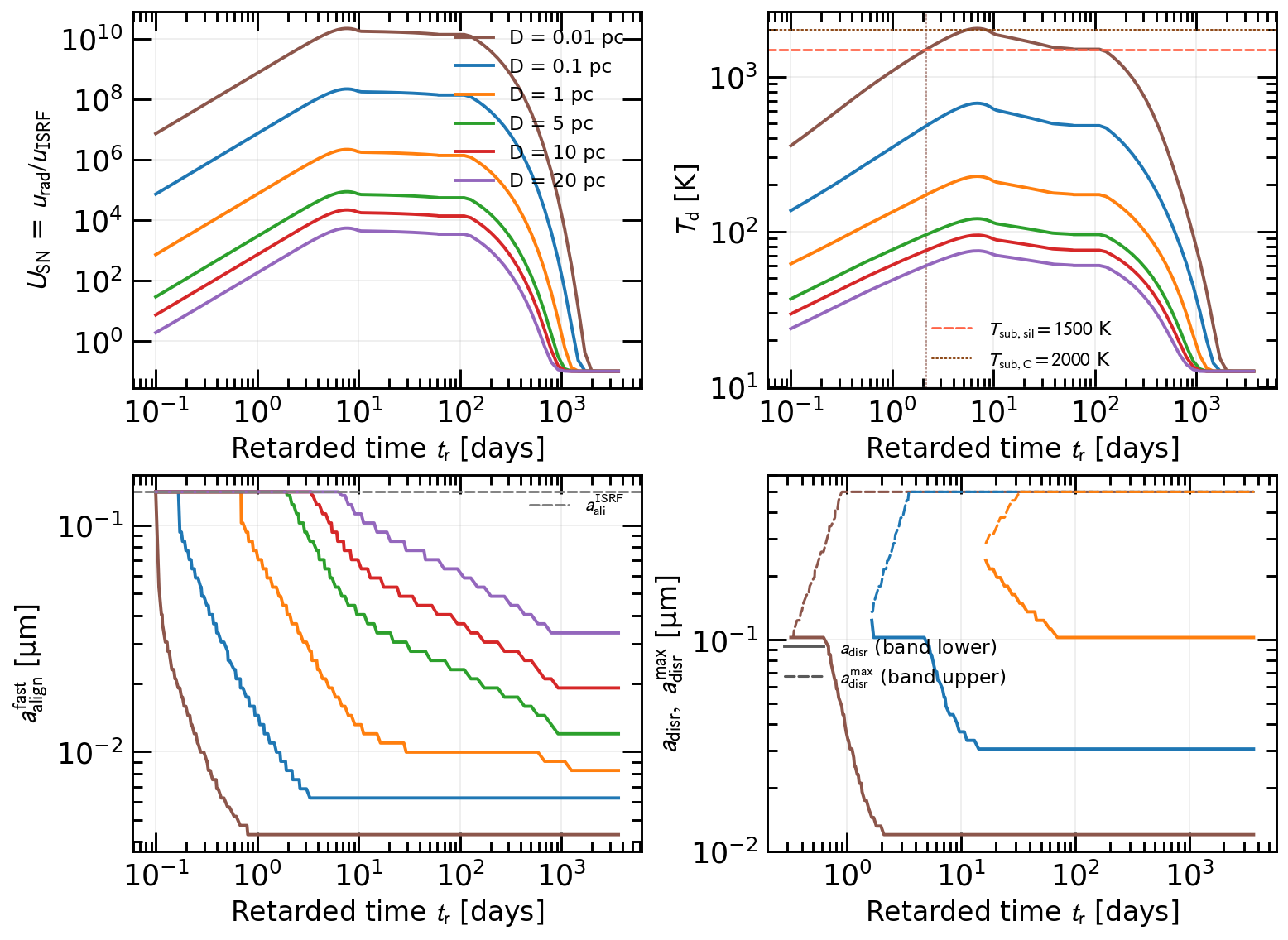}
	\caption{Time-dependent key quantities for grain rotational dynamics for the six source--cloud distances indicated by color. From upper left to lower right: radiation strength $U(t_r,D)$; equilibrium dust temperature $\Td(t_r,D)$, with silicate and carbon sublimation thresholds; the fast alignment size $\afast(t_r,D)$, compared with $\aali^{\rm ISRF}$; the lower (solid) and upper (dashed) edges of the RAT-D band. Stronger irradiation at small $D$ heats the dust and rapidly shifts the alignment and disruption boundaries to smaller sizes, whereas dust at $D\gtrsim5$~pc remain close to their pre-SN state. After SN radiation fades away, the alignment and disruption sizes remain constant.}
	\label{fig:diagnostics}
\end{figure}

Figure~\ref{fig:diagnostics} presents the evolution of the key quantities governing grain rotational dynamics for ten years, including the radiation strength $U(t_r,D)$, dust temperature $\Td(a=0.1,\mu{\rm m},t_r,D)$, and the critical sizes of grain alignment and disruption, $\afast(t_r,D)$ and $\adisr(t_r,D)$. The radiation strength and dust temperature broadly track the SN luminosity light curve (see Eq.~\ref{eq:LSNII}). The radiation strength rises rapidly to its peak, remains nearly constant during the plateau phase, and then declines sharply, becoming negligible after $\sim1000$ days, at which point it is set to the pre-SN value of $U_{0}=0.1$. Only for the closest dust zones, at $D=0.01$pc, does the dust temperature exceed $1500$~K and sublimation can happen.%\footnote{The $D=0.01$ pc case is included only to illustrate sublimation and is not a quantitative one-zone MC model.} 
In contrast, the evolution of grain alignment and disruption sizes exhibits a delay because RATs require finite time to spin up and align or disrupt grains. The fast alignment size $\afast$ initially decreases rapidly and then approaches an approximately constant value once the SN radiation becomes negligible (panel (d)). Remarkably, it does not return to the pre-SN alignment size, $a_{\rm align}^{\rm ISRF}$; instead, the difference between the pre- and post-SN alignment sizes becomes larger for dust closer to the SN. The disruption size $\adisr$ for nearby zones ($D=0.01,0.1, 1$~pc) shows a similar evolution, decreasing during strong irradiation and then remaining nearly unchanged after the SN radiation fades, and disruption does not occur for distant zones ($D=5,10,20$~pc). For $D=1$pc, small grains below $a=0.1\mu$m cannot be disrupted due to their higher tensile strength than larger grains (i.e., $S_{\max}=10^{9}\erg\cm^{-3}$ vs $10^{7}\erg\cm^{-3}$, see Table \ref{tab:fid}). Thus, both the enhanced alignment and disruption induced by the SN become effectively frozen into the post-SN dust population. We refer to these long-lived changes as \emph{physical memory effects}.

\begin{figure}
	\begin{overpic}[width=\textwidth]{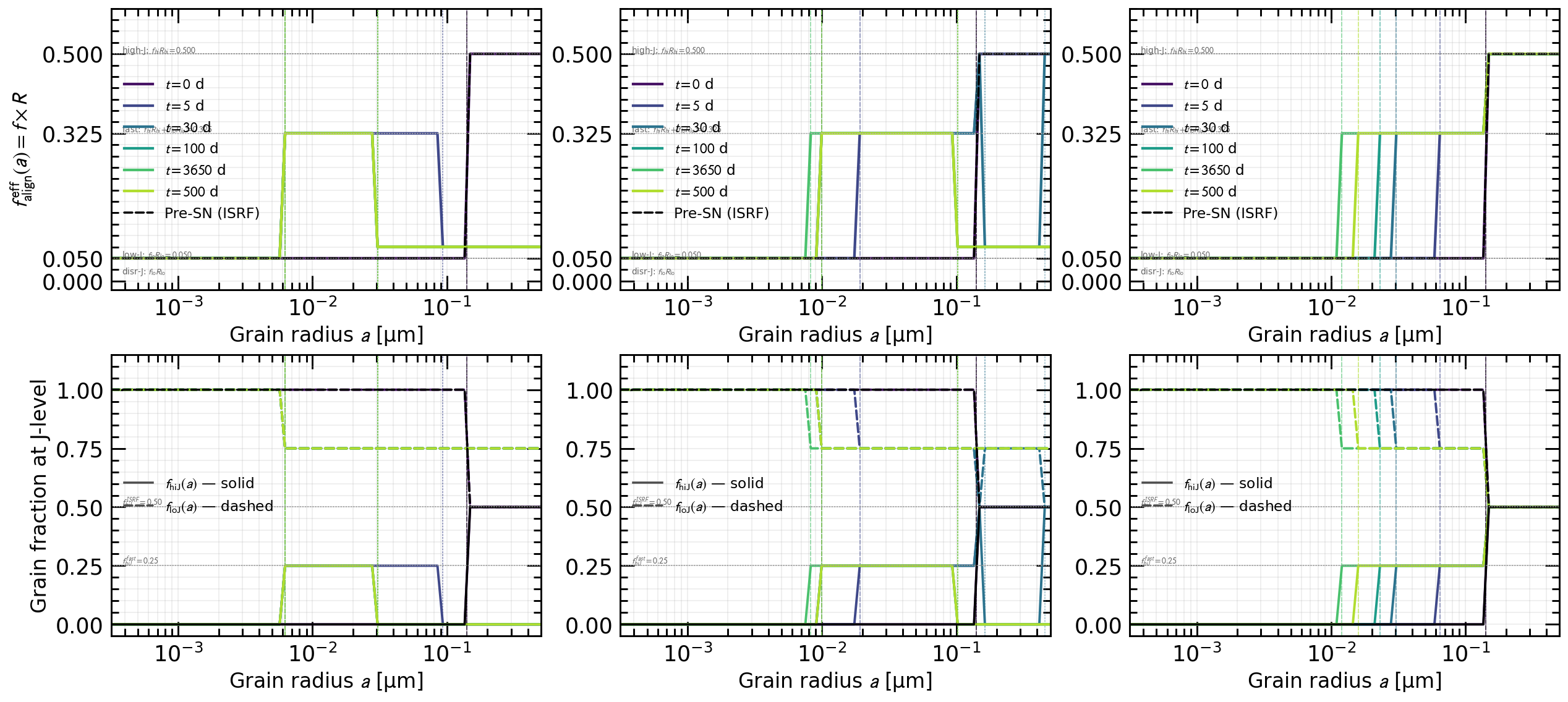}
	\put(15,42){\textbf{D=0.1 pc}}\put(45,42){\textbf{D=1 pc}}\put(75,42){\textbf{D=5 pc}}
	\end{overpic}
	
	\caption{Evolution of grain alignment function across three representative distances of $D=0.1$, 1~pc and 5~pc. For each distance, the upper panel shows the effective alignment efficiency $f_{\rm align}^{\rm eff}(a)$ and the lower panel shows the high-$J$ (solid) and low-$J$ (dashed) population fractions at different time epochs; the black dashed curve is the pre-SN ISRF state. Vertical markers locate $\afast$, $\aali^{\rm ISRF}$, and $\adisr^{\min,\max}$. Fast alignment extends to smaller grains as $\afast$ decreases, while RAT-D removes the high-$J$ population within the disruption band.}
	\label{fig:falign}
\end{figure}

\begin{figure}
	\begin{overpic}[width=\textwidth]{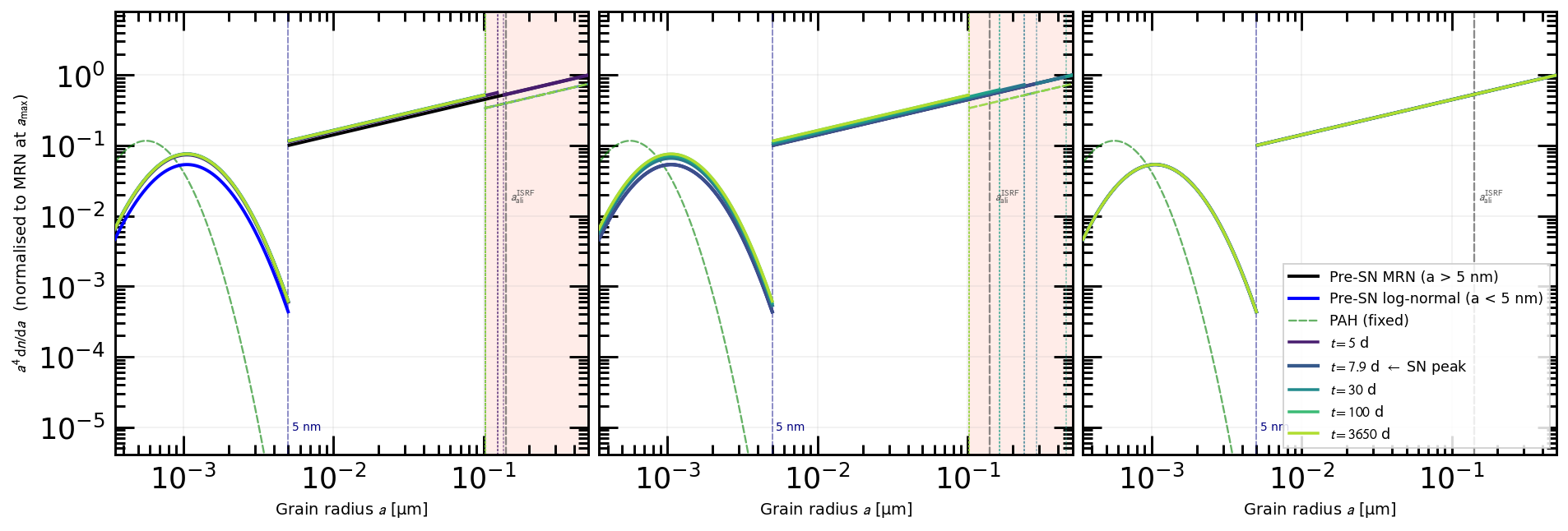}
 \put(10,30){\textbf{(a) D=0.1 pc}}
 \put(40,30){\textbf{(b) D=1 pc}}
 \put(72,30){\textbf{(c) D=5 pc}}
	\end{overpic}
	\caption{Mass-weighted GSD, $a^4dn/da$, for the six modeled zone distances, arranged as in Figure~\ref{fig:falign}, at representative epochs. The black and blue curves show the pre-SN MRN and Astrodust log-normal components, respectively; the green dashed curve is the fixed PAH component, and colored curves show the post-SN distributions. RAT-D depletes grains within $[\adisr^{\min},\adisr^{\max}]$ and transfers their mass to the nanodust and smaller-grain reservoirs, most strongly at small $D$. Grains above $\adisr^{\max}$ remain intact because their RAT-driven rotation is insufficient for disruption.}
	\label{fig:gsd_time}
\end{figure}

Figure~\ref{fig:falign} shows the time dependent effective alignment efficiency (upper panels) and the fractions of grains at the high-$J$ and low-$J$ levels (lower panels) for each source-cloud distance. The step-like profile $f_{\rm align}^{\rm eff}(a)$ is shown at different epochs together with the pre-transient baseline, for which only grains above $\aali^{\rm ISRF}\approx0.15\,\mu$m are aligned (Section~\ref{sec:fastalign}). The SN radiation drives $\afast$ toward progressively smaller sizes, extending fast alignment down to $a\approx0.01\,\mu$m within days at $D\leq1$~pc. Acting in the opposite direction, the disruption band $[\adisr^{\min},\adisr^{\max}]$ removes grains from the high-$J$ population, so the effective alignment efficiency within the band collapses to the weakly aligned low-$J$ floor; grains above $\adisr^{\max}$ rotate too slowly to disrupt and remain intact. At $D\geq5$~pc, the band is narrow or absent, and the evolution is dominated by the advance of $\afast$ alone. In particular, the post-SN alignment function at $t_{r}=3650$d is very similar to that at $t_{r}=100$d, and both are significantly different from pre-SN function. This originates from the {\it memory effect of alignment.}

Figure~\ref{fig:gsd_time} shows the corresponding mass-weighted size distribution $a^{4}\,dn/da$, separating the fixed polycyclic aromatic hydrocarbon (PAH) component, the Astrodust log-normal component ($a<5$~nm), and the MRN power law ($a>5$~nm). As RAT-D proceeds, the disruption band widens and its lower edge moves toward smaller sizes; the mass removed from the band is transferred to the nanodust reservoir, raising the log-normal component above its pre-SN level, and to the smaller grains of the MRN range below $a_{\rm disr}^{\min}$. This evolution is fastest at small $D$, whereas at $D\geq5$~pc, the post-SN distributions are nearly indistinguishable from the pre-SN state.  In particular, the post-SN grain distribution at $t_{r}=3650$d is very similar to that at $t_{r}=100$d, and both are significantly different from pre-SN function. This originates from the {\it memory effect of disruption.}

\subsection{Extinction curve}
\label{sec:diag_RV}

\begin{figure*}
	\begin{overpic}[width=0.33\textwidth]{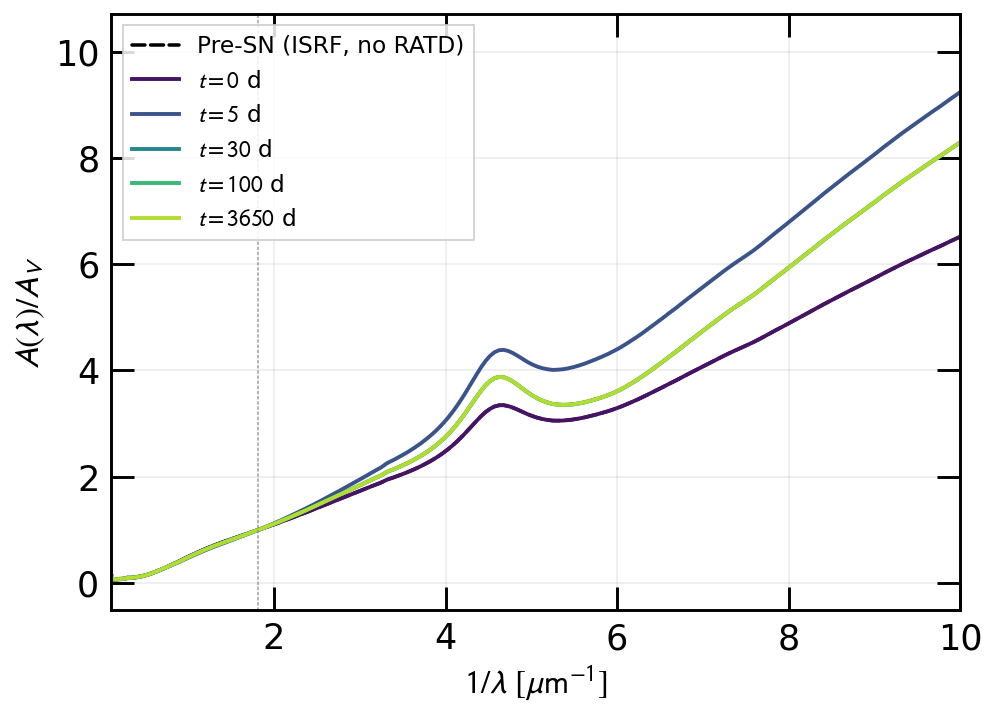}
		\put(60,15){\textbf{(a) D=0.1 pc}}
	\end{overpic}
	\begin{overpic}[width=0.33\textwidth]{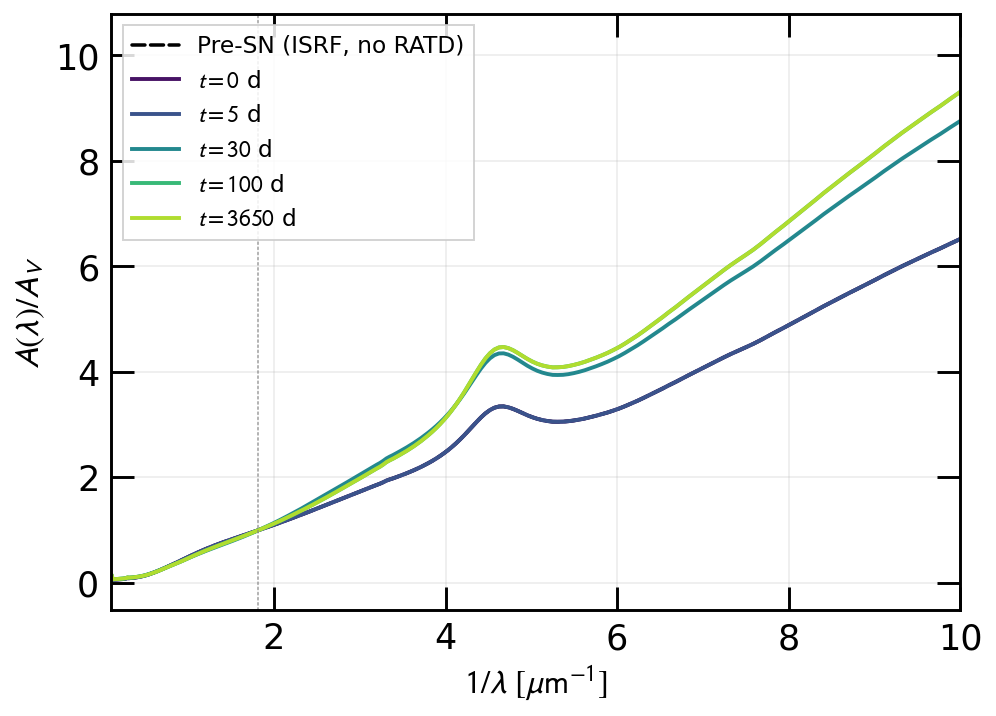}
		\put(62,15){\textbf{(b) D=1 pc}}
	\end{overpic}
	\begin{overpic}[width=0.33\textwidth]{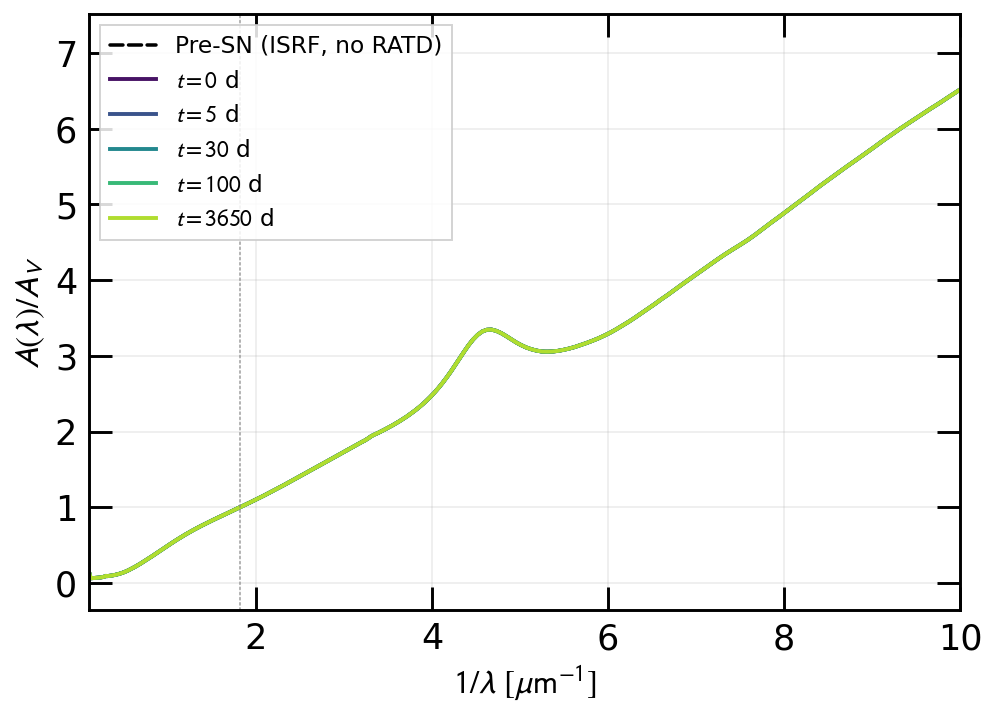}
		\put(62,15){\textbf{(c) D=5 pc}}
	\end{overpic}
	\caption{Normalized extinction curves $A(\lambda)/A_V$ versus inverse wavelength for dust clouds at $D=0.1$, 1, and 5 pc, at different time epochs. The black dashed curve is the pre-SN distribution without RAT-D, and the vertical dotted line marks the $V$-band normalization. The expanding disruption band steepens the UV extinction most strongly at small $D$, while the 2175~\AA\ feature persists because PAHs are not disrupted in the adopted model. Extinction curve is constant over time for $D=5$~pc.}
	\label{fig:Aext}
\end{figure*}

\begin{figure}
	\centering
	\includegraphics[width=\textwidth]{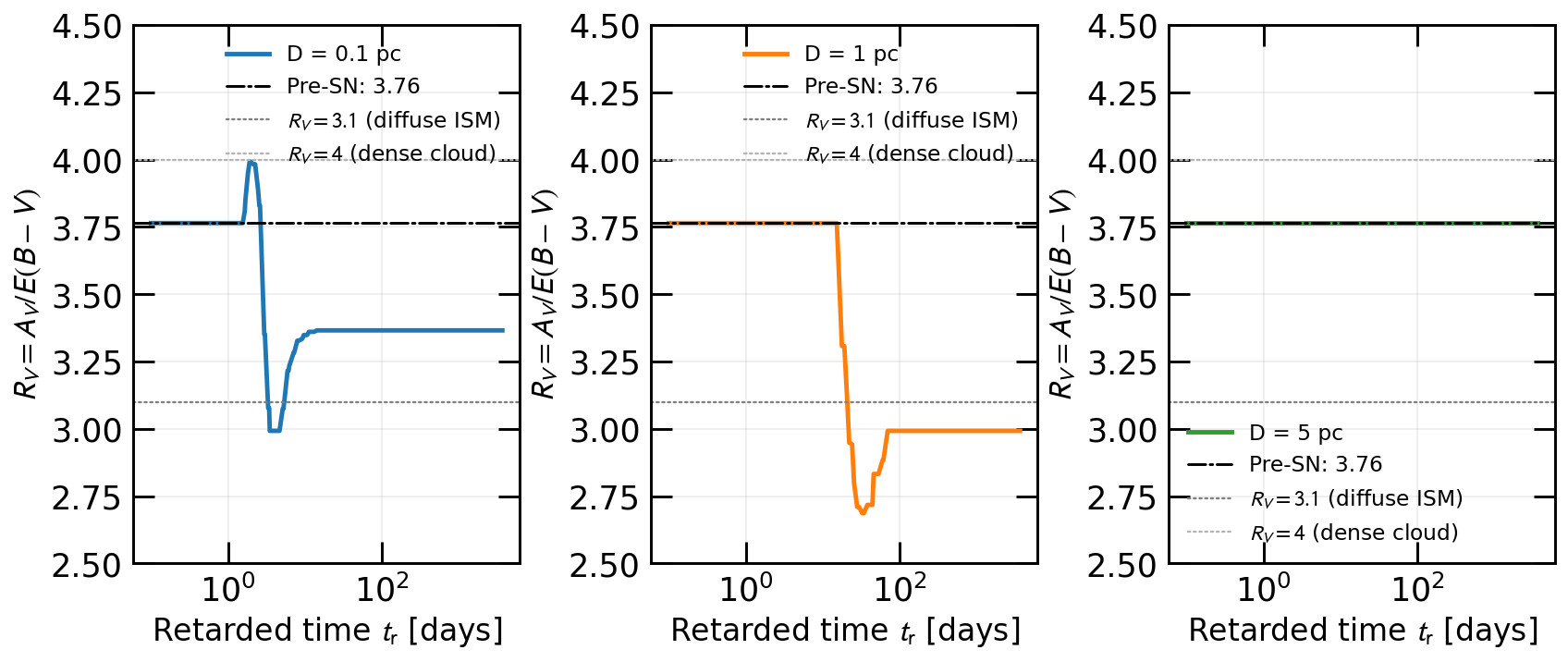}
	\caption{Time-evolution of $R_V=A_V/E(B-V)$ for the three modeled distances. The black dash--dotted line marks the pre-SN value $R_V=3.76$, and gray dotted lines show the diffuse-ISM and dense-cloud reference values. At small $D=0.1$~pc, RAT-D first removes grains near the optical-efficiency peak, raising $R_V$, and then broadens to smaller and larger sizes, driving $R_V$ downward. The variation is negligible for $D=5$~pc.}
	\label{fig:RV_time}
\end{figure}

The time dependent alignment function (Figure~\ref{fig:falign}) and grain size distribution (Figure~\ref{fig:gsd_time}) are then used to model dust extinction, thermal dust emission, and polarization.

Figure~\ref{fig:Aext} shows $A(\lambda)/\Av$ at selected epochs for dust at $D\in\{0.1,1,5\}$~pc, normalized at $V=0.55\,\mu$m and plotted against inverse wavelength $\invlam$. In the adopted dust model, the 2175~\AA\ feature at $\invlam\approx4.6~\mu\mathrm{m}^{-1}$ persists because PAHs are not subject to RAT-D, whereas the dust continuum evolves as the disruption band expands. At $D=0.1$~pc, the far-UV extinction at $\invlam=10\,\mu{\rm m}^{-1}$ steepens by $\approx$20\% by $t_r=100$~d, and the 2175~\AA\ bump strengthens relative to the continuum as RAT-D reduces $A_V$; at $D=1$~pc, the steepening reaches only a few percent by the same epoch, and at $D\geq5$~pc, the curve is unchanged, consistent with the static size distributions of Figure~\ref{fig:gsd_time}.

Figure~\ref{fig:RV_time} shows $\RVsn(t_r)=A_V/E(B-V)$, where $E(B-V)=A_B-A_V$, for each distance, together with the reference values $\RVsn=3.1$ for the diffuse ISM and $\RVsn=4$ for dense clouds. At the onset of RAT-D, the disruption band opens near the transition size $a \sim \bar{\lambda}/2\sim 0.1$--$0.2\,\mu$m; the removal of these grains initially raises $R_V$ from the pre-SN value of 3.76 to $\approx$4.0. As the band extends to both smaller and larger sizes---with $\adisr^{\min}$ falling below $\sim0.02\,\mu$m and $\adisr^{\max}$ approaching $a_{\max}$ under the most intense irradiation (Figure~\ref{fig:diagnostics})---$R_V$ subsequently drops to a minimum of $\approx$3.0, at the diffuse-ISM value, and then settles at $\approx$3.4. The phase reached in this rise-and-decline sequence is set by the received fluence: at $D=0.1$~pc, the full sequence completes within a few days, whereas at $D=1$~pc, $R_V$ stays at its pre-SN value until $t_r\approx20$~d and then falls monotonically---the transient rise is absent---reaching a minimum of $\approx$2.7 near 50~d before settling at $\approx$3.0. Dust zones at $D\geq5$~pc show no measurable variation ($\Delta R_V<0.1$) and provide a quasi-static reference. This nonmonotonic evolution is a distinctive RAT-D prediction; with an appropriate intrinsic-color and extinction calibration, time-resolved $B$- and $V$-band photometry can track it.
	
\subsection{Extinction polarization}
\label{sec:diag_pext}
%\paperi~ presents the polarization light curves, alignment-axis reverberation, and observing diagnostics in detail. Here we focus on three complementary outputs of the full Stokes calculation: the wavelength-dependent polarization spectra, the decade-long evolution of the $V$-band polarization degree and angle, and the polarization peak wavelength $\lambda_{\max}$, which together show how the observables follow from the evolving grain populations of the preceding subsections.

\begin{figure*}
	\begin{overpic}[width=0.99\textwidth]{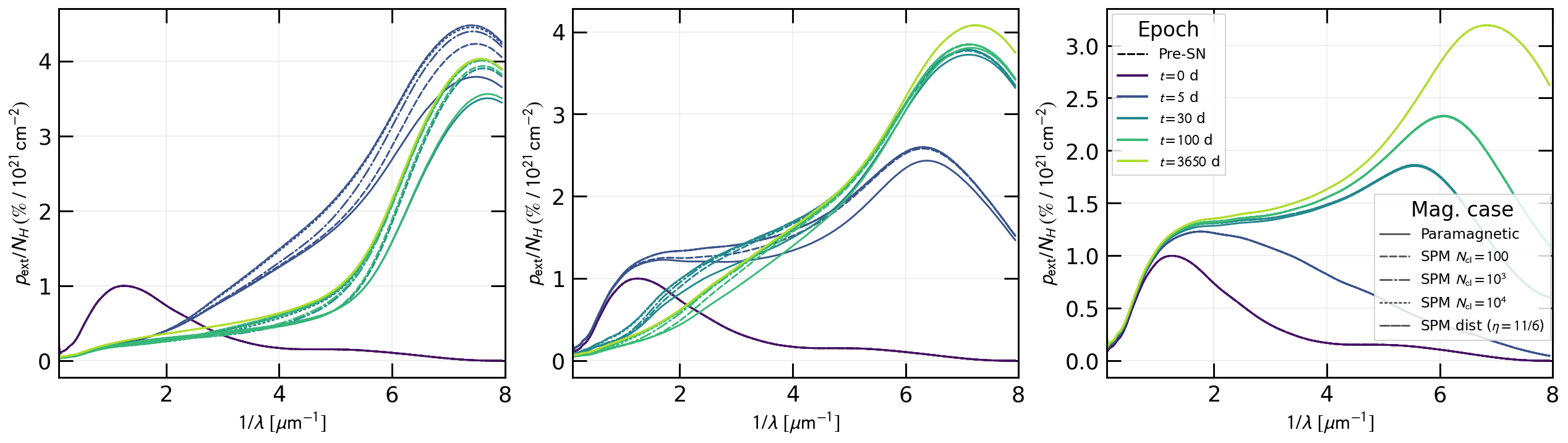}
		\put(7,22){\textbf{(a) D=0.1 pc}}\put(38,22){\textbf{(b) D=1 pc}}\put(68.5,22){\textbf{(c) D=5 pc}}
	\end{overpic}
	\caption{Extinction polarization spectra $p_{\rm ext}(\lambda)/N_{\rm H}$ for three different distances in the fiducial geometry $\psi=45^\circ$. Colors denote epoch, line styles distinguish the five magnetic-susceptibility models, and the black dashed curve is the pre-SN spectrum. Fast alignment enhances the polarization and shifts its peak to shorter wavelengths, while subsequent RAT-D suppresses the signal at the smallest distances.}
	\label{fig:pext_spectra_Bk45}
\end{figure*}

Figure~\ref{fig:pext_spectra_Bk45} shows the extinction polarization spectra for the fiducial $\psi=45^\circ$ geometry. The pre-SN spectrum peaks near $\invlam\approx1.3\,\mu{\rm m}^{-1}$ ($\lambda_{\max}\approx0.8\,\mu$m), reflecting ISRF alignment of only the largest grains. Due to SN radiation, fast alignment of progressively smaller grains both strengthens the polarization and shifts its peak into the UV. At $D=0.1$~pc, the peak amplitude grows by a factor of $\sim$4 within days and moves to $\invlam\approx7\,\mu{\rm m}^{-1}$, and the later epochs then fall to $\sim$3.9 times the pre-SN peak as RAT-D removes large aligned grains. At $D=1$~pc, the response is slower: the peak reaches only $\sim$2.5 times the pre-SN value at $t_r=5$~d before growing to a factor of $\sim$4 by $t_r=100$~d. At $D\geq5$~pc, the same spectral evolution proceeds more gradually still, reaching a factor of $\sim$2 by $t_r=100$~d and $\sim$3 by 10~yr. The five magnetic-susceptibility models nearly coincide in these spectra: the spectrum primarily constrains which grain sizes are aligned, while the susceptibility dependence appears in the timing of the angle response discussed below. For some cases, the polarization curve exhibits two peaks, one in optical/NIR and the new one in far-UV with $\lambda^{-1}>6$ induced by fast alignment of small grains. Between these peaks, the polarization fraction changes from the standard falling to rising trend with the inverse wavelength. 

In Appendix~\ref{app:kpB_validation}, we show the results for the case $\bk\parallel\bB$ 
 in which no axis switching can occur. The calculation shows that the position angle remain constant as expected from the alignment axis, while the polarization degree follows the expected limiting behavior: a rise followed by a rapid RAT-D-driven decline in nearby dust and a rise to saturation in distant ones, with the amplitude of the initial rise set by the fast alignment fraction $\fhif$ (shown for thermal emission in Figures~\ref{fig:pem_mag_Bk0}).

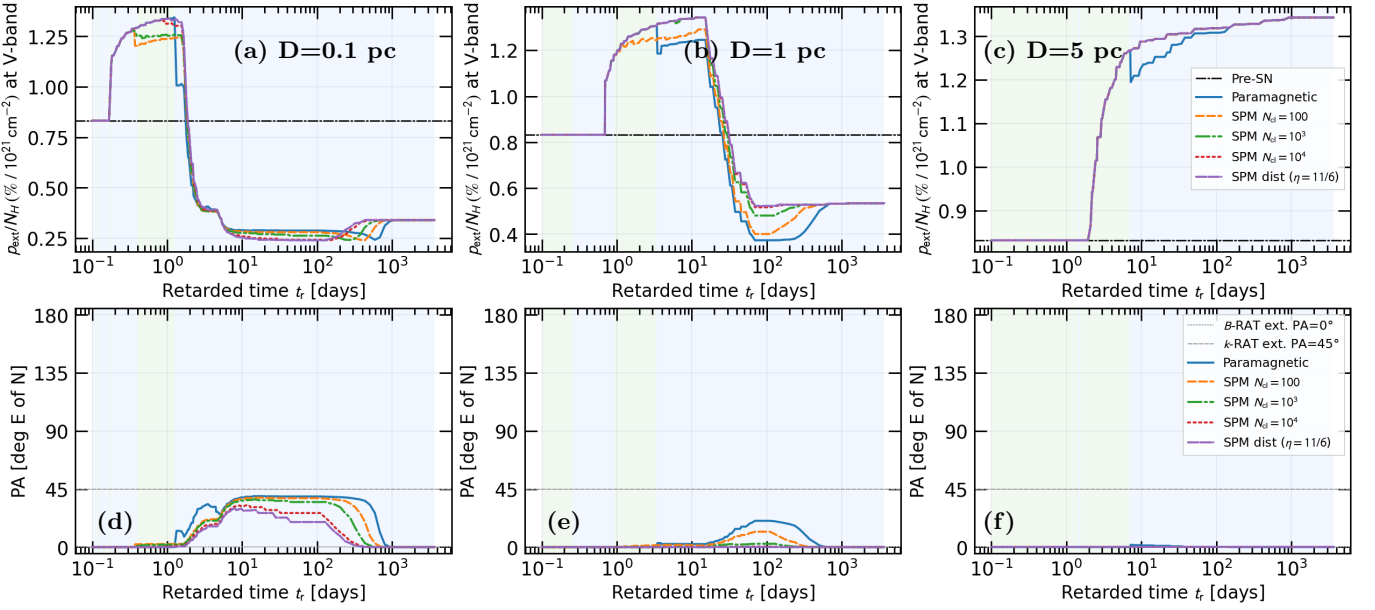
\begin{figure*}
	\centering
	\begin{overpic}[width=\textwidth]{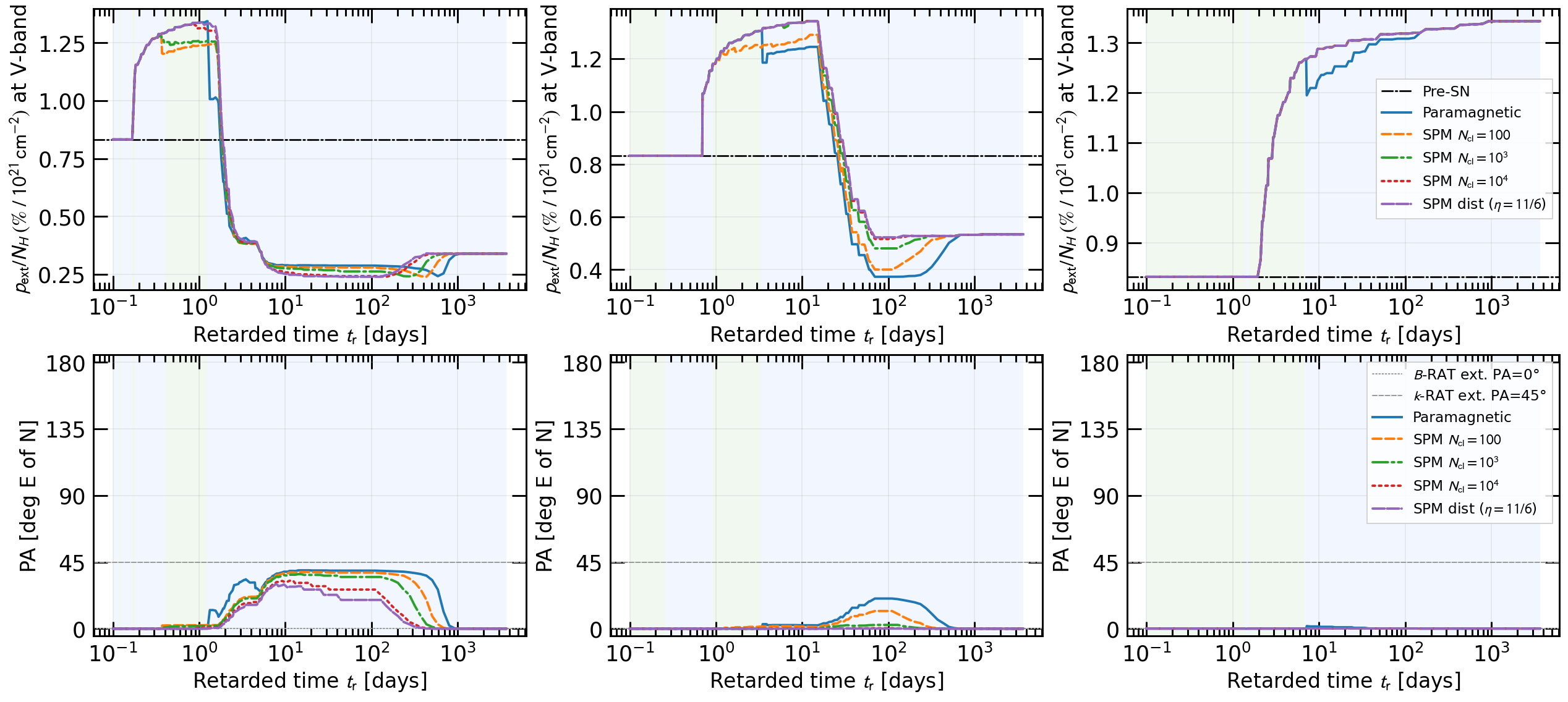}
		\put(17.5,40.5){\textbf{(a) D=0.1 pc}}\put(50.5,40.5){\textbf{(b) D=1 pc}}\put(72.5,40.5){\textbf{(c) D=5 pc}}
		\put(7.5,6){{\textbf{(d)}}}\put(40.5,6){{\textbf{(e)}}}\put(72.5,6){{\textbf{(f)}}}
	\end{overpic}
	\caption{Time evolution of the $V$-band extinction polarization (upper panels) and its position angle (lower panels) for clouds at $D=0.1$, 1, and 5~pc (left to right). Colors distinguish the five magnetic-susceptibility models; the black dash--dotted line marks the pre-SN fraction, and horizontal gray lines mark the $\BRAT$ and $\kRAT$ extinction position angles. The nearby clouds exhibit a flare, a RAT-D-driven dip, and a temporary $\BRAT$$\to$$\kRAT$$\to$$\BRAT$ angle excursion, whereas the more distant cloud retains a long-lived polarization enhancement with a smaller angle response.}
	\label{fig:pext_time_fossil}
\end{figure*}

Figure~\ref{fig:pext_time_fossil} follows the $V$-band polarization degree and position angle for three similar distances. At $D=0.1$~pc, the polarization fraction increases by $\sim1.6$ times its pre-SN value within the first day and then decreases below the pre-SN level once RAT-D destroys the large aligned grains (upper panels). Concurrently, the polarization position angle rotates from the $\BRAT$ orientation ($0^\circ$) toward the $\kRAT$ orientation ($45^\circ$) and returns as the radiation weakens, with a timing and duration that depend on the magnetic susceptibility model (lower panels). The polarization degree remains a factor of $\sim$2--3 below its pre-SN value long after the angle has recovered. At $D=1$~pc, the polarization enhancement persists for $\sim$20~d before a milder RAT-D-driven decline to $\approx$0.65 of the pre-SN level, accompanied by only a brief, small angle excursion (reaching $\approx$20$^\circ$) in the weakest susceptibility cases, whereas at $D=5$~pc the enhancement is long-lived and the angle barely responds. The polarization degree and angle thus separate the irreversible modification of the grain population from the temporary alignment axis competition; the detailed flare--dip sequence and its susceptibility dependence are analyzed in \paperi.

\begin{figure}
	\begin{overpic}[width=\textwidth]{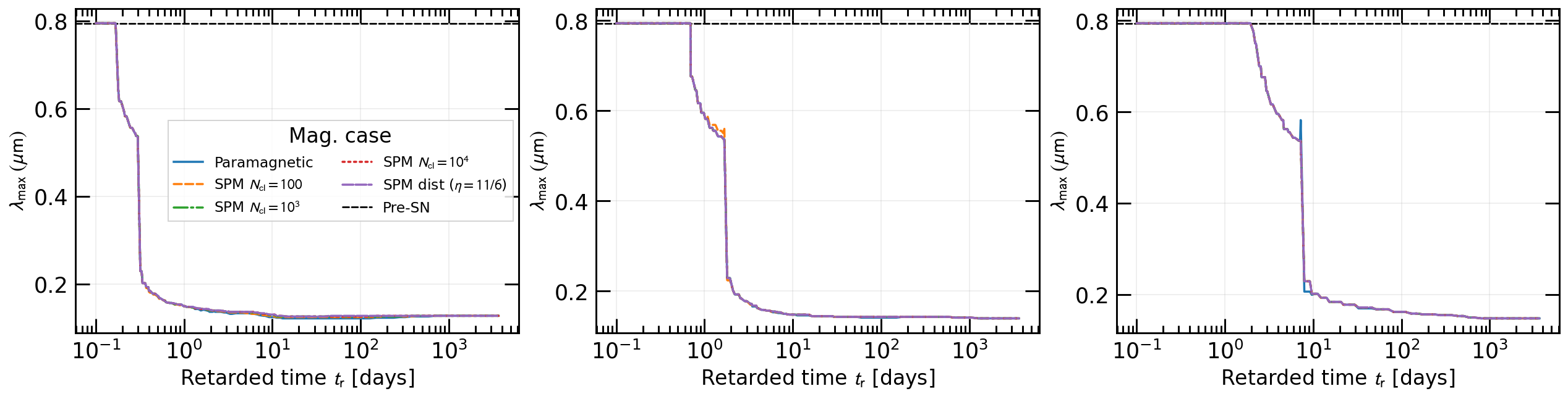}
				\put(15,22){\textbf{(a) D=0.1 pc}}\put(45,22){\textbf{(b) D=1 pc}}\put(80,22){\textbf{(c) D=5 pc}}
		\end{overpic}
	\caption{Peak wavelength $\lambda_{\max}$ of the extinction polarization spectrum over 10~yr for three representative source-cloud distances in the fiducial $\psi=45^\circ$ geometry. Colored curves show the five magnetic susceptibility models, and the black dashed line marks the pre-SN peak. Fast alignment and RAT-D shift $\lambda_{\max}$ from its pre-SN value near $0.8\,\mu$m to the UV; the persistent post-SN blueshift records the fossil memory of the transient.}
	\label{fig:peakwavelen_time_fossil}
\end{figure}

Figure~\ref{fig:peakwavelen_time_fossil} shows the extinction polarization peak wavelength, $\lambda_{\max}$, determined by the global maximum of the polarization curves. The peak wavelength first gradually decrease from the pre-SN value of $\lambda_{\max}\approx0.8\,\mu$m to $\lambda_{\rm max}\sim 0.55\mu$m, followed by an abrupt drop to $\lambda_{\max}\approx$0.12--0.15$\,\mu$m due to the fast alignment timescale---within $\sim$0.3~d at $D=0.1$~pc, $\sim$2~d at 1~pc, and $\sim$7~d at 5~pc---and is essentially independent of the magnetic model. Crucially, $\lambda_{\max}$ never recovers over the 10~yr shown: the newly aligned small grains retain their alignment against slow gas randomization, while the large grains that produce the pre-SN peak have been partially removed by RAT-D. The persistent blueshift of $\lambda_{\max}$ is therefore the most robust fossil imprint of the transient, complementing the amplitude deficit and the temporary angle excursion of Figure~\ref{fig:pext_time_fossil}.

\subsection{Thermal dust polarization}
\label{sec:diag_SED}

%Paper~I analyzes the polarization light curves and polarization angle reverberation evaluated at $\lambda=850\mum$. Here we use two complementary outputs of the full Stokes calculation---the wavelength-dependent polarization spectra and the ten-year $850\mum$ polarization histories---to show how those signatures arise from the evolving grain populations and how long they persist after the SN fades. 
	
\begin{figure*}
	\centering
	\begin{overpic}[width=0.99\textwidth]{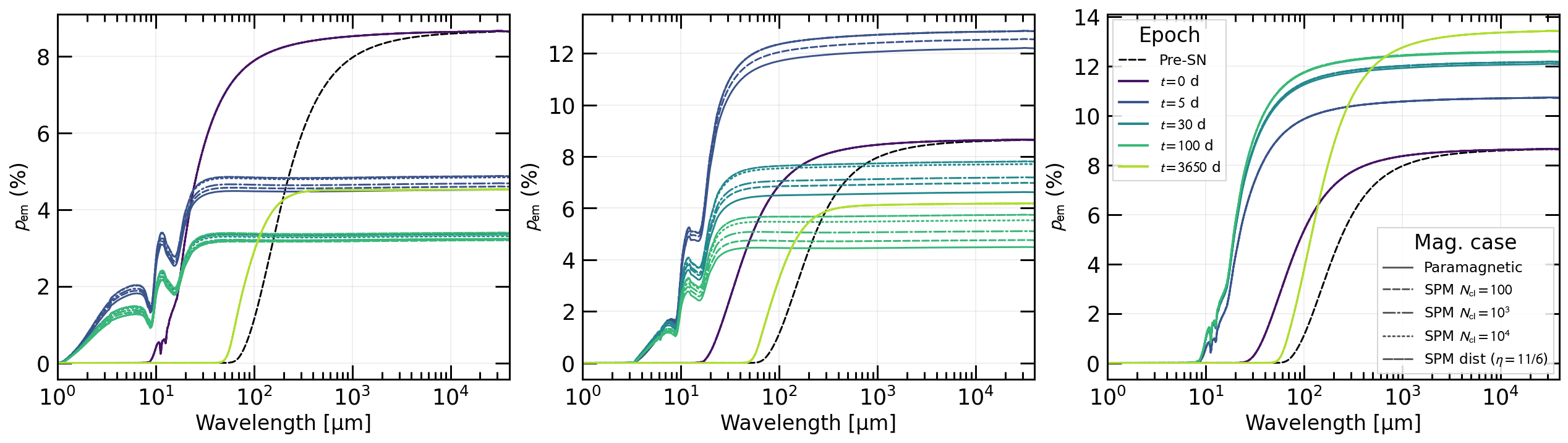}
		\put(16,7){{\textbf{(a) D=0.1 pc}}}\put(46,7){{\textbf{(b) D=1 pc}}}\put(76,7){{\textbf{(c) D=5 pc}}}
	\end{overpic}
	\caption{Thermal emission polarization spectra $p_{\rm em}(\lambda)$ for the three modeled distances in the fiducial $\psi=45^\circ$ geometry. Colors denote epoch, line styles distinguish the five magnetic-susceptibility models, and the black dashed curve gives the pre-SN spectrum. Emission from weakly aligned nanoparticles suppresses the polarization at short wavelengths. Fast alignment shifts the polarized rise toward shorter wavelengths and raises the far-IR/submillimeter plateau; RAT-D subsequently reduces this plateau in nearby clouds.}
	\label{fig:pem_spectra_Bk45}
\end{figure*}

Figure~\ref{fig:pem_spectra_Bk45} presents the wavelength-dependent polarization of thermal dust emission. Below $\sim10\mum$, emission by PAHs and Astrodust nanoparticles, which are unaligned, strongly dilutes the polarization and imprints structure near the aromatic bands. At longer wavelengths, aligned grains dominates and $p_{\rm em}$ rises to a plateau. In the pre-SN state, the polarization rise occurs near $\lambda\sim100\mum$ and the plateau reaches $p_{\rm em}\approx8.6$\%. Due to increasing transient radiation, fast alignment of progressively smaller grains shifts the polarization rise to $\lambda\sim10$--20$\mum$ and lifts the plateau to $\approx$11--14\% at $D=1$--5~pc. In nearby clouds, this enhancement is transient and RAT-D sets in before the plateau can be raised: at $D=0.1$~pc, RAT-D reduces the large aligned grains that dominate the polarized intensity and depresses the plateau to $\approx$3.3\% by $t_r=30$--100~d. By contrast, at $D=5$~pc, weak disruption allows the alignment-driven enhancement to grow and persist over the epochs shown. Magnetic susceptibility sets the partition between the $\BRAT$ and $\kRAT$ populations and therefore produces the largest spectral separation near an alignment-axis transition for $D\lesssim 1$pc.% Thus, the position angle remains the more direct diagnostic of that transition, as developed in Paper~I. 
In the validation geometry $\bk\parallel\bB$, the projected axes coincide and the calculated angle remains fixed, as required (Figures~\ref{fig:pem_mag_Bk0}, Appendix~\ref{app:kpB_validation}).

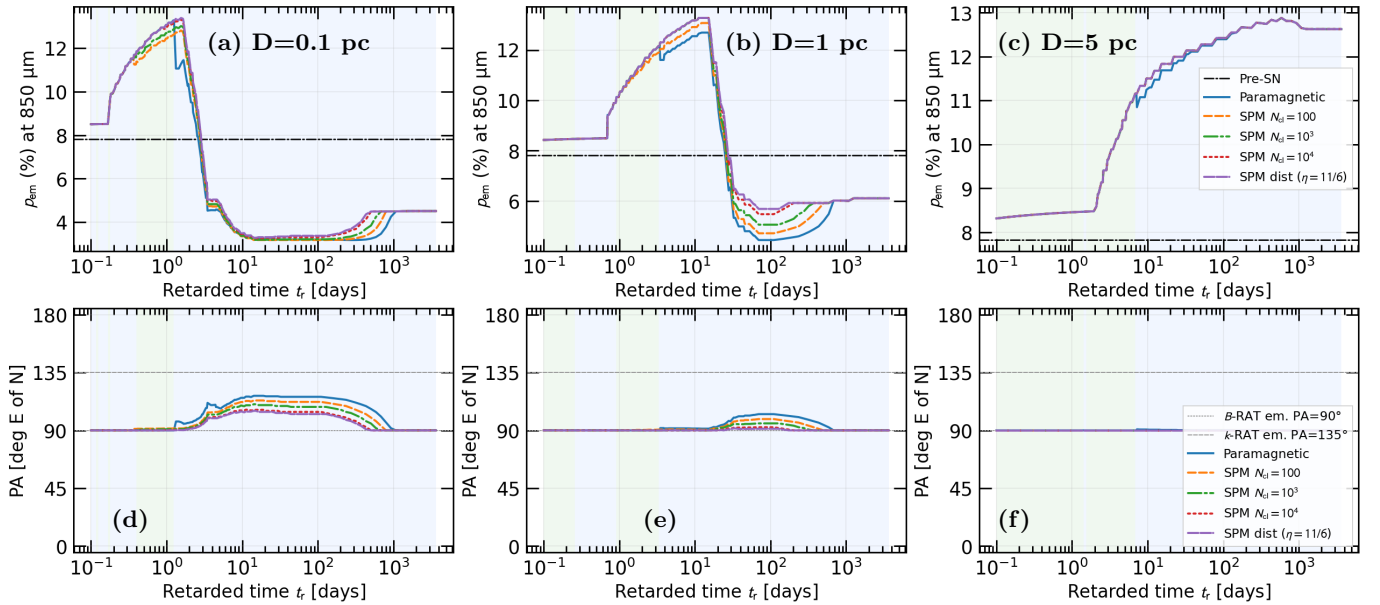
\begin{figure*}
	\centering
	\begin{overpic}[width=\textwidth]{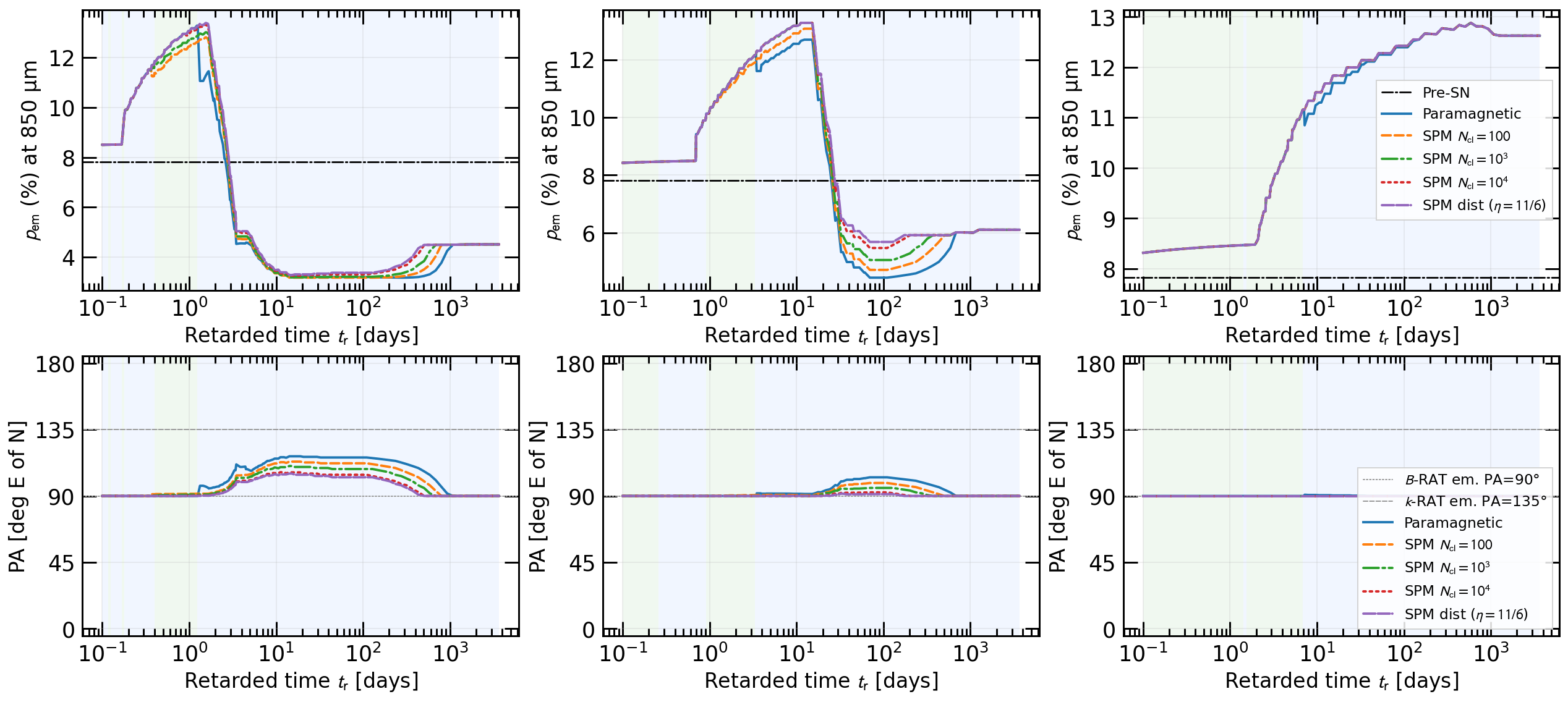}
	\put(15,41){\textbf{(a) D=0.1 pc}}\put(53,41){\textbf{(b) D=1 pc}}\put(73,41){\textbf{(c) D=5 pc}}
	\put(8,6){{\textbf{(d)}}}\put(47,6){{\textbf{(e)}}}\put(73,6){{\textbf{(f)}}}	
	\end{overpic}
	\caption{Time evolution of the $850\,\mu$m thermal emission polarization fraction (upper panels) and position angle (lower panels) for clouds at $D=0.1$, 1, and 5~pc (left to right). Colors distinguish the five magnetic susceptibility models; the black dash--dotted line marks the pre-SN fraction, and horizontal gray lines mark the $\BRAT$ and $\kRAT$ emission angles. At $D=0.1$~pc, the alignment flare is followed by a deep RAT-D-driven dip and a temporary $\BRAT$$\to$$\kRAT$$\to$$\BRAT$ angle excursion. At $D=1$~pc the flare is longer-lived but disruption still drives the late-time fraction below the pre-SN level, with a smaller angle response; at $D=5$~pc disruption is negligible and the enhancement persists.}
	\label{fig:pem_time_fossil}
\end{figure*}

Figure~\ref{fig:pem_time_fossil} follows the $850\mum$ polarization over time. At $D=0.1$~pc, the polarization fraction flares from its pre-SN value of $\approx$7.9\% to $\approx$13.3\% within a day and then collapses to $\approx$3.5\% as RAT-D destroys the large aligned grains (upper panels). During this dip, the polarization position angle rotates from the $\BRAT$ orientation ($90^\circ$) toward the $\kRAT$ orientation ($135^\circ$), reaching a maximum excursion of $\approx$120$^\circ$, and returns as the radiation field fades, by $\sim$600--1000~d depending on the magnetic susceptibility (lower panel). A partial recovery of the polarization fraction to $\approx$4.5\% accompanies the angle return as the surviving grains realign with $\bB$. The angle evolution is therefore a reverberation rather than a permanent fossil, whereas the polarization fraction remains a factor of $\sim$1.8 below its pre-SN value after the angle has recovered because RAT-D irreversibly depleted the large aligned grains. At $D=1$~pc, the polarization climbs to $\approx$13.4\% by $\sim$15~d and then declines to a minimum of $\approx$5\% near 100~d, settling at $\approx$6\%---still below the pre-SN level---with a modest angle excursion reaching $\approx$105$^\circ$ near 100~d, and at $D=5$~pc, weak disruption and slow gas randomization leave a monotonic, long-lived enhancement to $\approx$12.6\%. The magnetic cases differ mainly in the timing and extent of the temporary angle response; the persistent amplitude offset records the surviving alignment state and the post-RAT-D size distribution. Together with Figures~\ref{fig:pext_time_fossil} and \ref{fig:peakwavelen_time_fossil}, these histories show that the polarization angle diagnoses the temporary $\BRAT$/$\kRAT$ competition and the grain magnetic response, whereas the spectrum and the long-time polarization amplitude provide the fossil imprint of alignment and disruption.

\section{Discussion}
\label{sec:discussion}

\subsection{{\it TransRAT} Framework: Linking time dependent dust physics to time-domain dust observables}
\label{subsec:method_summary}
%We first recap the key components of TransRAT, discuss the real-time dust physics and dust observables.

We have introduced the {\it TransRAT} framework to model dust rotational dynamics under transient irradiation and predict time dependent extinction, emission, and polarization. For an arbitrary light curve, $L(t_r)$, we solve the equation of motion for grains subject to RATs caused by transient radiation, and gas and infrared damping. The time dependent grain rotation velocity is used to determine critical sizes of fast alignment and rotational disruption. These critical sizes are used to partition the entire dust population over the grain size distribution, which is further split based on the J-levels. Grain alignment is modeled following the RAT theory, including low-$J$ and high-$J$, with respective alignment efficiency depending on the radiation field and local environments. The alignment axis ($\Bv$ or $\kv$) of low-$J$ and high-$J$ aligned grains is treated separately, and the $\Bv$-$\kv$ switching is modeled using their Larmor and radiative precession times.

We have applied the {\it TransRAT} framework to a special case of a CCSN illuminated a dense molecular cloud--approximated as a single-phase one-zone model. We derived the time dependent critical sizes of fast RAT alignment and disruption (Fig. \ref{fig:diagnostics}) and alignment function (Fig. \ref{fig:falign}). For the nearby clouds of distance $D\lesssim 1$pc, we found that RAT-D can disrupt large grains aligned via $\BRAT$ at high-J attractors into smaller grains, resulting in the time-evolving grain size distribution. Due to RAT-D, only large grains at low-J attractors which are aligned by $\kRAT$ can survive and produce dust polarization. The time-evolution of grain alignment (critical size and alignment axis) and size distribution induce time-varying polarization signatures (see Sec \ref{sec:results}), including the polarization flare--dip cycle, the blueward shift of $\lambda_{\max}$, the $\BRAT\rightarrow\kRAT$ polarization-angle rotation, and the subsequent $\kRAT\rightarrow\BRAT$ reverberation, as well as systematic changes in $\RVsn$ (e.g., steeper UV extinction curves).

\subsection{Physical memory effects and fossil imprints in dust observables}
\label{sec:discussion_memory}
%We now move on to memory effects and fossil imprints in dust observables. 
%A central prediction of the time dependent framework is that irradiated dust does not immediately relax to its pre-transient state after the luminosity fades but retains memory of transient-induced physical effects.

Transient irradiation of dust lasts from days to years, yet the physical changes it induces can persist for much longer. As shown in Figure~\ref{fig:diagnostics}, the characteristic sizes for grain alignment and disruption remain nearly unchanged even after the SN radiation has completely faded, up to ten years. In \paperi, we identified four such \emph{physical memory effects}: \emph{alignment axis}, \emph{fast alignment}, \emph{disruption}, and \emph{shape deformation} memory. These effects are characterized by different relaxation, or memory, timescales, $\tau_{\rm relax}$. Alignment-axis memory is the shortest-lived, persisting only until Larmor precession again dominates over radiative precession. Fast-alignment memory persists until gas collisions fully randomize grains that were rapidly aligned by the transient radiation, whereas disruption memory survives until grain growth or coagulation replenishes the large grains destroyed by RAT-D. Shape-deformation memory persists for as long as grains retain the deformed shapes acquired under strong centrifugal stress (see \paperi).%\footnote{Fast-alignment memory is analogous to a spinning top that continues to rotate for some time after the external torque is removed.}

These memory effects produce \emph{fossil imprints} in dust observables, including extinction, emission, and polarization, as demonstrated in Section~\ref{sec:results}. Unlike thermal emission intensity, which responds nearly instantaneously to radiative heating, these fossil imprints retain information about the past radiation field and can therefore reveal the effects of a transient long after its direct radiation has disappeared. As discussed in \paperi, fossil imprints form a hierarchy governed by their respective relaxation timescales. Polarization angles rotated due to changes in the alignment axis relax on the Larmor precession timescale. The enhanced polarization resulting from fast alignment decays over $\tau_{\rm rel}\sim10\,\tau_{\rm gas}$, whereas the post-RAT-D GSD can persist until grain growth or coagulation restores the disrupted grains, requiring $\gtrsim100$~kyr in dense gas. Consequently, the longest-lived fossil imprints are encoded in persistent changes to the extinction curve, $\RVsn$, polarization degree, and polarization peak wavelength.

\subsection{Polarized infrared echoes and spatially resolved fossil imprints}
\label{subsec:ir_echo}
A transient's impact on dust manifests observationally in two ways: a \emph{real-time} signature---the polarized infrared echo---that traces the dust currently being irradiated, and the \emph{fossil} signatures of Section~\ref{sec:discussion_memory}, which persist after the radiation front has passed. We discuss both for circumstellar and local interstellar dust. Circumstellar dust at $D\sim10^{-3}$--$0.3$~pc experiences strong radiation of $U\gtrsim10^{8}$, responds within hours to days, and may sublimate inside $R_{\rm sub}$ (Eq.~\ref{eq:Rsub}), whereas dust in molecular clumps at $D\sim1$--$100$~pc responds over weeks to years and survives thermal sublimation because the radiation field there is only moderately strong. Echo delay, response time, and, for Galactic sources, spatial separation can distinguish these reservoirs; Table~\ref{tab:transients} lists them separately for CCSNe.

An infrared echo identifies the dust currently heated by the transient. The absorbed and re-emitted power rises strongly with $U(t_r,D)$, while heating shifts the emission peak from the far-infrared toward shorter wavelengths (Eq.~\ref{eq:Td_r}). A polarized infrared echo reveals the real-time alignment of grains heated by the transient and can trace magnetic fields in the environment prior to shock arrival. The IR echo can therefore be used to identify illuminated dust and to devise follow-up polarization observations.

%Circumstellar echoes may also be polarized if their grains are aligned, and the $D=0.1$~pc model illustrates a possible $\BRAT$ $\to$ $\kRAT$ transition (Figure~\ref{fig:pem_spectra_Bk45}). This remains a qualitative prediction because dense CSM conditions ($\nH\sim10^{6}$--$10^{10}\cm^{-3}$; Table~\ref{tab:transients}) require recalculation of damping, alignment, and disruption.

%For a pulse of duration $\Delta t_{\rm SN}\sim 1$~yr, however, the illuminated region is only a shell of thickness $c\Delta t_{\rm SN}\simeq0.3$~pc. The observed delay follows the isodelay surface $t_{\rm delay}=r(1-\cos\theta)/c$ \citep{Dwek.1983}; consequently, the one-zone slab model is quantitatively applicable only to coherently illuminated clumps of comparable or smaller size. Predictions for larger clouds require integration over the three-dimensional geometry, attenuation, dust mass, and observing band. 

The more distinctive signatures of transient radiation are the \emph{spatially extended fossil imprints} left behind the IR-bright shell. Because the long-lived imprints of Section~\ref{sec:discussion_memory} are carried by the alignment state and the grain size distribution rather than by grain temperature, a cloud that has already cooled back to its ISRF-heated value---and is therefore no longer conspicuous in the infrared---should still show those imprints varying systematically with distance from the transient source.

\begin{figure*}
	\includegraphics[width=\textwidth]{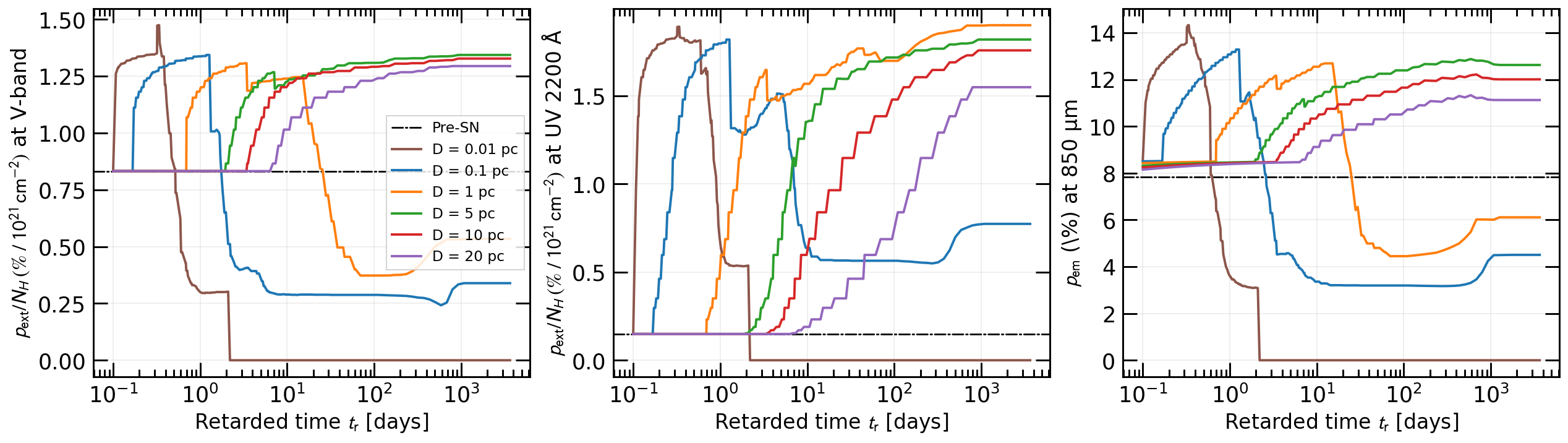}
	\caption{Polarization fraction vs. retarded time $t_r$ (Eq.~\ref{eq:tret}) for cloud zones at different distances $D$ from the transient: (a) $p_{\rm ext}/N_{\rm H}$ in the $V$ band, (b) the same at the $2200$~\AA\ bump, and (c) $p_{\rm em}$ at $850\mum$. Colored curves show $D=0.01$, $0.1$, $1$, $5$, $10$, and $20\pc$; black dash-dotted lines mark the pre-SN steady state. Every zone flares above its pre-SN level, but only those at $D\gtrsim5\pc$ retain the enhancement; closer zones end in a permanent deficit set by RAT-D and, at $0.01\pc$, by sublimation.}
	\label{fig:compareR}
\end{figure*}

Figure~\ref{fig:compareR} quantifies this gradient, following three polarimetric observables for zones at $D=0.01$--$20$~pc. Because each zone corresponds to a different $D$, a cut at a fixed late time directly yields the predicted spatial profile. However, since the curves are plotted as functions of retarded time, comparison with observations requires resampling them along the iso-delay paraboloid,
$t_{\rm delay}=r(1-\cos\theta)/c$ with $r$ the distance from the SN and $\theta$ is the angle between SN-to-dust and dust-to-observer directions \citep{Dwek.1983}. This resampling combines zones with different $D$ values evaluated at different $t_r$.

The flare onset scales with distance, from $\sim0.1$~d at $D=0.01\pc$ to $\sim5$~d at $D=20\pc$, because the response time lengthens as $U(t_r,D)$ weakens. All zones first rise above the pre-SN baselines, as stronger RATs drive more grains to the high-$J$ attractor and push $\aali$ to smaller sizes, but the outcome then splits by whether RAT-D and sublimation act. At $D=0.01\pc$ the polarization vanishes within days once $\Td$ exceeds the sublimation temperature. At $D=0.1\pc$ RAT-D promptly removes the large grains carrying the signal, and all three observables settle well below their pre-SN values, a permanent \emph{deficit}. At $D=1\pc$ disruption is delayed to $\sim20$--$100$~d and leaves a shallower but still permanent deficit. At $D\ge5\pc$ RATs never disrupt grains and the enhanced alignment saturates into a long-lived \emph{surplus}.

The late-time radial profile is therefore non-monotonic, peaking near $D\approx5\pc$: the innermost zones are polarization-poor not because they were weakly irradiated but because they were irradiated too strongly. The deficit-to-surplus crossover lies between $1$ and $5\pc$ both in the $V$ band and at $850\mum$, but its \emph{depth} is wavelength dependent: at $D=1\pc$ the late-time level is $\approx0.6$ of the pre-SN value in the $V$ band and $\approx0.8$ at $850\mum$, whereas at $2200$~\AA\ every surviving zone ends above its pre-SN value---by an order of magnitude at $D\gtrsim1\pc$. This ordering follows from the blueward shift of $\lambda_{\max}$ from $\simeq0.8\mum$ to $0.12$--$0.15\mum$ once grains above $\adisr$ are removed (Figure~\ref{fig:peakwavelen_time_fossil}), which leaves the $2200$~\AA\ band near the new polarization peak and the $V$ band far on its red side. A joint UV--optical--submillimeter map thus tests the picture in a way no single band can: the same zone should appear depolarized in the optical and submillimeter yet strongly polarized in the UV.

Resolving this profile sets a practical constraint on target selection. The structure of Figure~\ref{fig:compareR} develops over $D\lesssim10\pc$, which subtends $\sim34'$ at $1$~kpc but only $\sim40''$ in the Magellanic Clouds and $\sim3''$ at the distance of M31. Spatially resolved tests are therefore restricted to Galactic and Magellanic SNRs; for more distant hosts, only the disk-averaged signal is accessible, in which the near-side deficit and far-side surplus partially cancel and the net imprint is diluted.

\subsection{Present polarization as a probe of past irradiation}
\label{subsec:fossil_probe}

Dust grains in the ISM and molecular clouds are ubiquitously observed to be aligned. In light of the memory effects discussed above, a new question naturally arises: \emph{when did these grains become aligned?} Is their alignment driven primarily by the present-day ISRF, or does their current polarization retain a fossil imprint of stronger irradiation from past transient events, such as nearby SNe? Physical modeling of dust extinction and polarization within the RAT framework can constrain the radiation field strength required to produce the observed grain alignment, $U_{\rm align}$ \citep{Tram.2026}. Independently, the current radiation field strength, $U_{\rm current}$, can be inferred from the dust temperature obtained by SED fitting. Any significant discrepancy between $U_{\rm align}$ and $U_{\rm current}$, particularly $U_{\rm align} \gg U_{\rm current}$, could therefore provide evidence that the present grain alignment was established during an earlier, more intense irradiation episode and has subsequently persisted through alignment memory. Such a comparison offers a new way to reconstruct the irradiation history of interstellar dust, opening the possibility of using present-day dust polarization as a fossil record of past radiation fields and energetic explosions.

Moreover, polarization modeling of observational data can provide constraints on the maximum grain size, $a_{\rm max}$ \citep{Tram.2025}. If the inferred $a_{\rm max}$ is significantly smaller than expected from grain growth physics in dense environments (e.g., \citealt{Lebreuilly.2022}), this discrepancy may indicate that the GSD has been modified by past energetic irradiation, for example through RAT-D. Very recently, detailed physical modeling of ALMA dust polarization toward NGC~6334I by \cite{Rawat.2026} showed that a grain population dominated by submicron-sized grains is required to reproduce the low polarization observed in its dense protostellar clusters where micron-sized grains are expected due to grain growth. Their calculations further demonstrated that such a population of small grains can naturally arise from RAT-D driven by past accretion outbursts in NGC~6334I. Further studies along this direction are needed to establish present-day dust polarization as a probe of past irradiation and episodic outburst activity in YSOs.

\subsection{Observing prospects for testing TransRAT}
\label{subsec:multiwav}
Fast alignment and disruption, alignment-axis switching, and memory effects are distinctive consequences of the time dependent interaction between dust grains and transient radiation. Observational tests of these phenomena would extend our understanding of dust physics into a dynamical regime inaccessible under steady irradiation. \paperi~outlined prospects for both real-time monitoring and archaeological searches for fossil imprints. Here, we focus on three promising observational targets.

\subsubsection{MCs near young SNRs}
Newly discovered SNe (e.g., by the Vera C. Rubin Observatory Legacy Survey of Space and Time, LSST) can be cross-matched with molecular clouds within $\sim100$~pc of the explosion site to identify targets for prospective monitoring. A cloud at three-dimensional distance $D$ is illuminated a time $\Delta t=D/c$ after the explosion (Eq.~\ref{eq:tdelay}), from years at parsec-scale separations to centuries at $\sim100$~pc. What an observer measures, however, is the \emph{echo} delay $D(1-\cos\theta)/c$ between the observed SN peak and the observed cloud response, where $\theta$ is the angle between the cloud and the line of sight; this vanishes for clouds on the near side and reaches $2D/c$ for those directly behind. Useful advance notice is therefore available only for clouds beyond the plane of the sky, and converting a projected separation into $D$ requires an independently constrained geometry---for which the echo itself, mapped over several epochs, is the natural tool. Once illuminated, the radiation front crosses a $1$~pc clump in $\sim3.3$~yr, allowing the radiative response to be monitored before the much slower SN blast wave reaches the same material. During this pre-shock interval, the $\BRAT$-aligned component traces the ambient, unshocked magnetic field. %The detection of hot cores within $\sim3$--$10$~pc of an SNR \citep{Shimonishi.2026} demonstrates that suitably dense targets can exist sufficiently close to an SN for such observations.

The most direct test combines multi-epoch extinction polarization, measuring the polarization degree, $\lambda_{\max}$, and position angle, with thermal emission polarization of the same material. With sufficient cadence, these observations could capture within a single event the predicted flare--dip evolution of the polarization, the blueward migration of $\lambda_{\max}$, rotation of the polarization angle, and the subsequent reverberating response. 

When real-time monitoring is unavailable, a complementary retrospective test can compare the polarization properties of clouds illuminated by recent SNe with those of matched control clouds illuminated only by the ISRF. Such differential measurements can search for persistent imprints, including enhanced polarization, smaller $\lambda_{\max}$, and systematic changes in $\RVsn$. The distinctive prediction is not merely that these quantities differ from control values, but that they vary with distance from the explosion site in the non-monotonic, wavelength-dependent way of Figure~\ref{fig:compareR}: a depolarized inner region, a maximum at intermediate $D$, and a crossover radius that shifts between the UV, optical, and submillimeter. ISRF-irradiated control clouds, illuminated by a weakly anisotropic field, should instead show no such organized radial structure.

Calibrated BVRI photometry can provide an independent measurement of $\RVsn(t_r)$, while infrared echoes detected with JWST or SPHEREx can localize the illuminated dust and trigger targeted follow-up. Joint modeling of optical extinction, infrared emission, and submillimeter polarization can constrain the evolving GSD, the absorbed and reprocessed luminosity, and the aligned large-grain population, respectively. The combined data can also help distinguish changes in grain alignment from changes in the underlying dust GSD. PRIMA and AtLAST would further extend this wavelength leverage into the far-infrared and submillimeter regimes \citep{Clements.2025,Orlowski-Scherer.2024,Fischer.2024}.

\subsubsection{Superbubble shells as fossil imprints}
Superbubble shells are produced by interstellar matter swept up by successive supernova explosions; well-known examples include the Local Bubble, the Loop~I bubble, and the Orion--Eridanus supershell. Dust grains in these shells were irradiated by SN radiation and are therefore natural targets for the $U_{\rm align}$ versus $U_{\rm current}$ test of Section~\ref{subsec:fossil_probe}. UV/optical polarization observations toward background stars behind superbubble shells, compared with those of stars elsewhere, can establish whether the grains are aligned by the present ISRF or by past irradiation.

Existing observations may already exhibit such a fossil signature. Numerical modeling of the super-Serkowski polarization by \cite{Tram.2026} requires $U_{\rm align}\sim 10^{2}$--$10^{4}$, which is 2--4 orders of magnitude higher than both the typical strength of the current ISRF and the value implied by the current dust temperatures of the local superbubble shell. For a superbubble shell radius of $R\sim 100$~pc, the peak SN radiation strength is $U\simeq 10^{2}L_{9}(R/100\pc)^{-2}$, where $L_{9}=L/10^{9}L_{\odot}$ (see Eq.~\ref{eq:Utotal}), consistent with the required $U_{\rm align}$. The timescales are also consistent: the most recent SNe in the Local Bubble occurred $\sim2$--$3$~Myr ago, as indicated by the $^{60}$Fe deposition record \citep{Knie.2004,Breitschwerdt.2016}, comparable to the alignment memory time $\tau_{\rm rel}\sim10\,\tau_{\rm gas}\sim1$--$10$~Myr for swept-up shell densities $\nH\sim10$--$100\cm^{-3}$. This suggests that the grain alignment responsible for the observed super-Serkowski curves may be a fossil of past supernova explosions rather than a product of the current ISRF. The alternative explanation discussed in \cite{Tram.2026} is RAT alignment by extreme UV radiation from B stars in the Per~OB3 association. However, for a Per~OB3 association of 30 B stars \citep{Zeeuw.1998} with an average luminosity of $L_{B}\lesssim 10^{4}L_{\odot}$, the total radiation strength at the bubble shell of radius $R=30$ pc is $U_{\rm tot}\simeq 0.33\,(L_{B}/10^{4}L_{\odot})(R/30\pc)^{-2}$, far below the required value. Note that the longer mean wavelength of stellar radiation only slightly modifies the RAT efficiency, which scales as $\bar{\lambda}^{-1.7}$ (see Eq.~\ref{eq:GammaRAT}).

\subsubsection{The diffuse ISM}
SN radiation is expected to enhance the local radiation field to $U>100$ out to a radius of $D<100$ pc (see Eq.~\ref{eq:Utotal}), disregarding attenuation. The alignment memory there is correspondingly long: $\tau_{\rm gas}\sim10(\nH/1\cm^{-3})^{-1}$~Myr (Eq.~\ref{eq:taudamp_gas}), so $\tau_{\rm rel}\sim10\,\tau_{\rm gas}$ exceeds the local SN recurrence time, and a given parcel of diffuse dust may carry the superposed imprint of several explosions. The resulting dust polarization would then depend not only on the present radiation field and magnetic geometry, but also on the transient irradiation history of the dust. Such history dependence could provide an additional, spatially structured contribution to polarized Galactic foregrounds relevant to CMB observations---one that is uncorrelated with present-day dust temperature and would therefore not be captured by foreground models keyed to the current radiation field.

\subsection{Applications of TransRAT framework to cosmic transients}
Our {\it TransRAT} framework accepts an arbitrary source light curve $L(t_r)$ (Sec.~\ref{sec:lightcurve}). Thus, it is straightforward to extend beyond CCSNe to SNe~Ia, novae, GRB afterglows, TDEs, FBOTs, AGN flares, and YSO outbursts. For each transient class, the source spectrum and temporal evolution, together with the local gas and dust conditions, can be adjusted accordingly. The framework can also be applied to dense circumstellar environments around CCSNe by adopting the appropriate progenitor density profiles \citep{Dessart.2026}.

Two particularly promising extensions probe complementary environments. GRBs occur in dusty star forming regions where rapid RAT-D may alter the GSD and thereby modify host-cloud extinction. TDEs, by contrast, illuminate dense circumnuclear tori. Their large and rapidly variable optical polarization \citep{Floris.2025,Koljonen.2025} may contain a contribution from dust scattering \citep{Uno.2025}, while the intense radiation field and elevated temperatures of the innermost surviving grains may favor $\kRAT$ alignment. Quantitative predictions for both classes will require calculations tailored to their distinct geometries, density structures, radiation spectra, and dust environments. Table~\ref{tab:transients} summarizes the characteristic physical regimes of these and other transient classes.
Limitations
\subsection{Limitations}
\label{subsec:limitations}

The numerical results presented in this paper are based on a homogeneous, one-zone model of a dense cloud illuminated by a Type II supernova (SN). This simplified model intended to focus on dust physics neglects radiation attenuation, internal radiation sources, and spatial variations in the cloud properties. Therefore, our quantitative predictions apply most directly to coherently illuminated clumps whose sizes are comparable to or smaller than the light-crossing length of the transient pulse. Modeling larger, optically thick clouds requires integration over the three-dimensional cloud geometry, accounting for radiation attenuation and the spatial distribution of dust mass. Such 3D radiative transfer calculations coupled with time dependent grain physics are beyond the scope of this paper and will be presented in future work. In addition, we adopt fixed values for several key parameters governing grain alignment, as listed in Table~\ref{tab:fid}. Varying these parameters may change the predicted polarization levels, but is not expected to alter the general features of the results, particularly the physical memory effects and their fossil imprints.

% ================================================================

\begin{table*}
	\centering
	\caption{Potential applications of the TransRAT framework to cosmic transients and their dusty environments.}
	\label{tab:transients}
	\footnotesize
	\setlength{\tabcolsep}{5pt}
	\begin{tabular}{llcllcc}
		\toprule
		Transient & $L_{\rm peak}$ ($L_\odot$) & Duration & Environment & $D$ (pc) & $\nH$ (cm$^{-3}$) & $\Lambda$ \\
		\midrule
		CCSNe (this work) & $10^{9}$--$10^{10}$ & $\sim$100~d & molecular clouds & $0.01$--$100$ & $10^{3}$--$10^{6}$ & $10^{-4}$--$10^{-1}$ \\
		CCSNe (CSM) & $10^{9}$--$10^{10}$ & $\sim$100~d & RSG wind/CSM & $10^{-3}$--$0.3$ & $10^{6}$--$10^{10}$ & $10^{-1}$--$10^{3}$ \\
		SNe~Ia & $10^{9}$--$10^{10}$ & $\sim$40~d & foreground ISM & $0.1$--$50$ & $10$--$10^{4}$ & $10^{-6}$--$10^{-3}$ \\
		Novae & $10^{4}$--$10^{5}$ & weeks--months & nova shells, CSM & $10^{-4}$--$10^{-2}$ & $10^{6}$--$10^{10}$ & $10^{-2}$--$10^{3}$ \\
		GRB afterglows & $\lesssim 10^{14}$$^{\rm a}$ & min--days & host MCs & $0.01$--$50$ & $10^{2}$--$10^{6}$ & $10^{-10}$--$10^{-3}$ \\
		FBOTs & $10^{10}$--$10^{11}$ & days--weeks & dense CSM & $10^{-3}$--$0.1$ & $10^{6}$--$10^{10}$ & $10^{-3}$--$10^{2}$ \\
		TDEs & $10^{10}$--$10^{11}$ & months--years & nuclear torus & $0.01$--$1$ & $10^{5}$--$10^{8}$ & $10^{-2}$--$10^{2}$ \\
		AGN flares & $10^{11}$--$10^{13}$ & months--years & torus, NLR clouds & $0.1$--$10$ & $10^{3}$--$10^{7}$ & $10^{-4}$--$10$ \\
		YSO outbursts & $10^{2}$--$10^{3}$ & yr--decades & disk-envelope & $10^{-5}$--$5\times10^{-3}$ & $10^{6}$--$10^{12}$ & $1$--$10^{7}$ \\
		\bottomrule
	\end{tabular}
	\tablecomments{$^{\rm a}$Isotropic-equivalent; the true luminosity is lower by the beaming factor. $L_{\rm peak}$, durations, and $\nH$ are indicative orders of magnitude; $D$ is the characteristic source--cloud distance (the YSO range corresponds to $1$--$10^{3}$~au). $\Lambda=\tau_{\rm source}/\tau_{\rm gas}$ (Eq.~\ref{eq:lambda}) takes the bright-phase duration as $\tau_{\rm source}$ and evaluates $\tau_{\rm gas}$ (Eq.~\ref{eq:taudamp_gas}) for $a=0.1\,\mu$m grains at the quoted $\nH$ and representative gas temperatures: $\Lambda\gg1$ permits slow RAT alignment/disruption, whereas $\Lambda<1$ requires the fast channel. CSM = circumstellar medium; NLR = narrow-line region.}
\end{table*}

\section{Summary}
\label{sec:summary}

We have introduced {\it TransRAT}, a self-consistent framework for modeling the time-dependent response of dust to an arbitrary transient light curve, $L(t_r)$. It jointly evolves grain heating, magnetic properties, rotational dynamics, alignment, and disruption, and predicts the resulting extinction, emission, and polarization. We applied {\it TransRAT} to a Type~IIP supernova illuminating a homogeneous one-zone dense cloud at several distances. Our principal results are as follows:

\begin{enumerate}
	
	\item \textit{The {\it TransRAT} framework.} We solve the rotational dynamics of grains under time-dependent RATs, gas damping, and infrared damping to determine the evolving critical sizes for fast alignment and RAT disruption. Combined with the low-$J$/high-$J$ populations and their respective alignment axes, these quantities define the time-dependent aligned grain population, size distribution, and Stokes parameters.
	
	\item \textit{RAT-D, grain-size distribution, and extinction.} RAT-D opens a disruption band around $a\sim\bar{\lambda}/2\sim0.1$--$0.2,\mu{\rm m}$ that broadens with time and transfers mass toward small grains and nanodust. This produces a non-monotonic evolution of $R_V$, while the 2175~\AA\ feature persists because PAHs are unaffected by RAT-D.
	
	\item \textit{Fast alignment and polarization signatures.} As the transient brightens, fast RAT alignment extends to smaller grains, producing polarization flares in extinction and thermal emission and shifting $\lambda_{\max}$ blueward. In nearby clouds, subsequent RAT-D causes a polarization dip and can expose the low-$J$, $\kRAT$-aligned population, rotating the polarization angle by the projected $\Bv$--$\kv$ separation. As the transient fades, Larmor precession restores $\BRAT$ alignment, producing an angle reverberation whose recovery timescale depends on grain magnetism.
	
	\item \textit{Physical memory and fossil imprints.} Dust relaxes through a hierarchy of timescales: alignment-axis memory persists until Larmor precession dominates again, fast-alignment memory decays over $\sim1$--$10,\tau_{\rm gas}$, and disruption memory can survive for $\gtrsim100$~kyr until grain growth replenishes large grains. Consequently, enhanced polarization, blueshifted $\lambda_{\max}$, and altered $R_V$ can persist long after the transient fades, forming \emph{fossil imprints} and, in resolved clouds, an extended fossil region trailing the IR-bright shell.
	
	\item \textit{Probing past irradiation.} Present-day dust observables can therefore encode past irradiation rather than only the current radiation field. A discrepancy $U_{\rm align}\gg U_{\rm current}$ can diagnose fossil alignment, while an anomalously small $a_{\rm max}$ can indicate past RAT-D. The super-Serkowski polarization toward the local superbubble shell, requiring $U_{\rm align}\sim10^{2}$--$10^{4}$, may provide such a fossil record of past supernova irradiation.
	
\end{enumerate}

These results establish the physical basis of the key observables identified in \paperi---the polarization flare--dip cycle, blueward drift of $\lambda_{\max}$, $\BRAT!\to!\kRAT$ polarization-angle rotation, and $\kRAT!\to!\BRAT$ reverberation---while providing the full theoretical formulation, numerical validation, and complementary extinction predictions. Because {\it TransRAT} accepts an arbitrary light curve, it can be applied directly to SNe~Ia, novae, GRB afterglows, TDEs, and other cosmic transients (Table~\ref{tab:transients}). Together, the two papers establish the methodological foundation of time-domain dust astrophysics and demonstrate that dust extinction and polarization can serve both as a real-time probe of grain physics under transient irradiation and as a fossil probe of the irradiation history of the ISM.

%==========================================================================
\begin{acknowledgments}
We thank Alex Lazarian and Le Ngoc Tram for insightful discussions. This work was supported by the major research project (No. 2026183200) from Korea Astronomy and Space Science Institute (KASI) funded by the Ministry of Science and ICT (MSIT). This work was partially funded by a grant from the Simons Foundation (SFI-MPS-T-Institutes-00021894, HTN). We thank the ICISE staff for excellent support and DAP (VNSC) for hospitality. Claude \citep{claude} was used to assist with code development and ChatGPT \citep{gpt} with polishing the manuscript.
\end{acknowledgments}

\software{Python, \texttt{DustPOL\_py} \citep{Tram.2023}, ChatGPT \citep{gpt}, Claude \citep{claude}.}

%==========================================================================
%\section*{Software Availability}
%==========================================================================

%==========================================================================
\appendix
% Ensure that appendix equation anchors remain unique when hyperref is active.
\renewcommand{\theHequation}{\Alph{section}.\arabic{equation}}
%==========================================================================
%--------------------------------------------------------------------------
\section{Numerical Solution for Time-Dependent Grain Angular Velocity}
\subsection{Numerical solution of equation of rotational motion}
\label{app:diag_heat}
\label{app:spinup}

The critical size of alignment and disruption as well as the alignment axis of dust grains due to RATs depend on the grain angular velocity. Here, we describe the numerical solution of equation of rotational motion to find $\Omega$ as a function of time.

The equation of rotational motion (Eq.~\ref{eq:omegaEoM}) can be written in relaxation form as
\bea
d\Omega = \frac{\Gamma_{\rm RAT}(t_{r})}{I_{\rm grain}}dt_{r} - \Omega \frac{dt_{r}}{\tau_{\rm damp}},\label{eq:domega_dt}
\ena
which becomes
\bea
d\Omega + \Omega d\tau = \overline{\Omega}(\tau)d\tau,
\ena
where $d\tau\equiv dt_{r}/\tau_{\rm damp}(t_{r})$ is the dimensionless time increment, and
\bea
\overline{\Omega}(\tau)=\frac{\tau_{\rm damp}\,\Gamma_{\rm RAT}(\tau)}{I_{\rm grain}},\label{eq:Source}
\ena
is the instantaneous terminal angular velocity for fixed $\Gamma_{\rm RAT}$ and $\tau_{\rm damp}$.

\begin{comment}
For $F_{\rm IR}\ll 1$, $\tau_{\rm damp}=\tau_{\rm gas}$ is constant. Multiplying both sides by $e^{\tau}$ gives
\bea
e^{\tau}\left(d\Omega + \Omega d\tau\right)= e^{\tau}\overline{\Omega}(\tau)d\tau ,\\
d\left(e^{\tau}\Omega\right)=e^{\tau}\overline{\Omega}(\tau)d\tau
\ena
Integrating both sides from $\tau'=0$ to $\tau$ yields the solution given in Eq.~(\ref{eq:omega_time}).

For a radiation source of constant luminosity (e.g., massive stars or YMSCs), $\Gamma_{\rm RAT}$ is constant, and the solution reduces to
\begin{eqnarray}
	\Omega(\tau)=\Omega_{\rm RAT}\left[1-\exp\left(-\tau\right)\right],\label{eq:omega_const}
\end{eqnarray}
where
\begin{eqnarray}
	\Omega_{\rm RAT}=\frac{\Gamma_{\rm RAT}\tau_{\rm damp}}{I_{\rm grain}}\label{eq:omega_RAT0}
\end{eqnarray}
is the terminal angular velocity at $t\gg \tau_{\rm damp}$, the maximum rotation rate spun up by RATs.
\end{comment}
For a time-varying source, both $\Gamma_{\rm RAT}$ and $\tau_{\rm damp}$ evolve. Let $\Delta t_{r}=t_{r,i+1}-t_{r,i}$ be the time step and assume that during this timestep, the luminosity is quasi-static. Thus, the grain angular velocity can be evaluated
\begin{eqnarray}
\Omega_{i+1}&&=\;\Omega_{{\rm term},i}\;+\;\bigl(\Omega_i-\Omega_{{\rm term},i}\bigr)\,e^{-\Delta t_{r}/\tau_{{\rm damp},i}},\\
\qquad \Omega_{{\rm term},i}&&=\frac{\Gamma_{{\rm RAT},i}\,\tau_{{\rm damp},i}}{I_{\rm grain}},
\label{eq:omegaUpdate}
\end{eqnarray}
where the timestep $\Delta t$ is described in Table \ref{tab:fid} (see more in Sec \ref{sec:timeresol}).

\subsection{Characteristic Dynamical Timescales and Temporal Resolution}
\label{sec:timeresol}

\begin{figure*}
	\centering
	\begin{overpic}[width=0.99\textwidth]{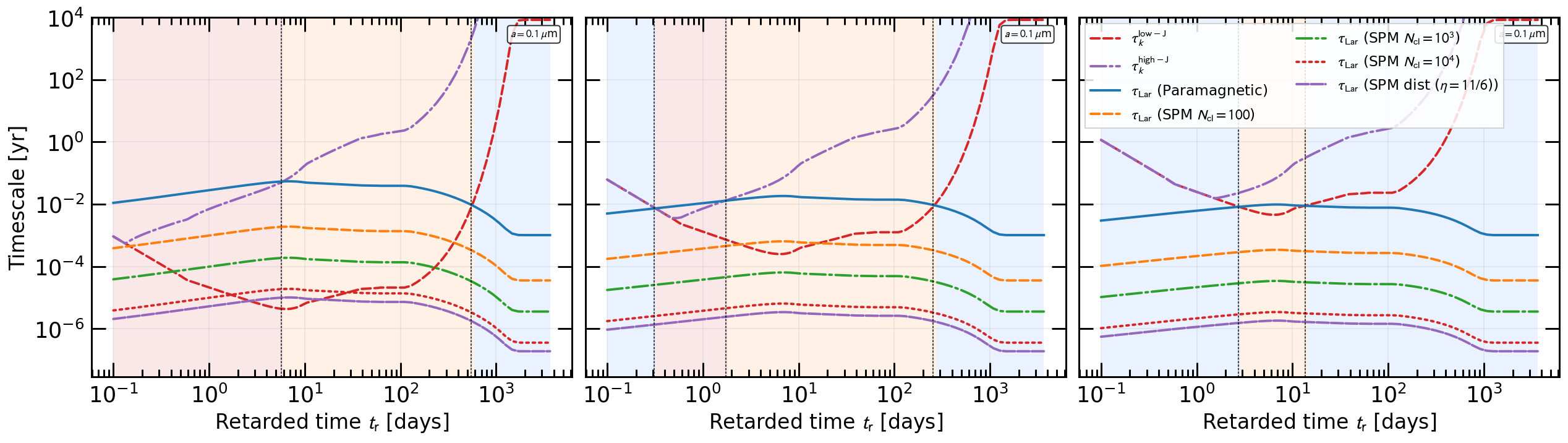}
		\put(7.5,52){\textbf{(a)}}\put(46,52){\textbf{(b)}}\put(72.5,52){\textbf{(c)}}
	\end{overpic}
	\caption{Radiative and Larmor precession timescales for a typical grain size of $a=0.1\mu$m for the  three cloud distances. Red and purple curves show $\tkRAT$ for low-$J$ and high-$J$ grains, respectively; the remaining curves show $\tLar$ for PM and the four SPM models. The faster precession sets the alignment axis (Eq.~\ref{eq:brat_krat}): $\kRAT$ is possible where $\tkRAT<\tLar$, most readily for low-$J$ PM grains, whereas rapid Larmor precession allows strongly magnetic SPM grains to retain $\BRAT$ alignment.}
	\label{fig:timescales}
\end{figure*}

Figure~\ref{fig:timescales} compares the radiative precession and Larmor precession timescales for a typical size $a=0.1\mu$m of different magnetic properties. Radiative precession for low-$J$ aligned grains is increasingly faster than high-$J$ ones. For clouds close to the SN ($D=0.1$ pc), low-$J$ aligned grains can switch from $\BRAT$ to $\kRAT$ alignment regime from $10-300$ d (marked by orange area in the figure) and returns to $\BRAT$ afterward due to the decrease of the SN radiation. In contrast, grains at high-$J$ generally remain in the $\BRAT$ regime because their higher rotation rates increase the characteristic $\kRAT$ precession time, $\tkRAT^{\rm high\text{-}J}$. For $D=1$ and 5 pc, only low-J PM grains can experience $\kRAT$ alignment, and SPM grains follow $\BRAT$ alignment regardless of their rotational state.

These dynamical timescales also set the temporal resolution required to follow changes in the alignment axis. The quasi-static treatment of the $\BRAT$ vs. $\kRAT$ alignment state is valid only when the time step $\Delta t$ exceeds the shortest relevant precession timescale,
\begin{equation}
	\tau_{\rm prec,min}=\min(\tkRAT^{\rm low-J},\tkRAT^{\rm high-J},\tLar).
\end{equation}

When this condition is not satisfied, the system does not have sufficient time to establish the instantaneous equilibrium alignment state; we therefore retain the alignment state from the last temporally resolved epoch. Our temporal grid initially consists of $N_T=70$ logarithmically spaced epochs and is then adaptively refined around the onset of RAT-D and the B$\rightarrow$$\kRAT$ transition at each distance. This refinement typically adds $\sim20$--$80$ epochs in intervals where the critical grain sizes and alignment states evolve most rapidly, allowing the transient response of the grain population to be resolved without unnecessarily increasing the temporal resolution during slowly evolving phases.

%--------------------------------------------------------------------------
\section{Modified Grain Size Distribution under Partial RAT-D}
\label{app:gsd}

Sections~\ref{sec:components}--\ref{sec:mass_weights} define the six active components (A, B, C, $D_h$, $D_l$, and F) and their mass weights (Eqs.~\ref{eq:Mdef}--\ref{eq:wE_def}). Here we specify the irreversible disruption boundary, the normalization after mass redistribution, and the adopted fragment-size distributions.

\subsection{Irreversible disruption and transferred mass}

Because grain regrowth timescales far exceed the observation window, disruption is irreversible. RAT-D removes grains only inside the bounded band $[\adisr^{\min},\adisr^{\max}]$, whose effective edges $\adisr^{\min,\rm eff}(t_{r})$ and $\adisr^{\max,\rm eff}(t_{r})$ accumulate as the running extrema of the instantaneous bands over the preceding radiation history (Eq.~\ref{eq:adisrirrev}). Throughout this appendix, $\adisr\equiv\adisr^{\min,\rm eff}(t_r)$ denotes the lower effective boundary. The transferred dust-mass fraction is $w_E(t_r)$ (Eq.~\ref{eq:wE_def}), which is a diagnostic of mass redistribution.

\subsection{Component normalization after redistribution}

For radiative transfer, each component $X$ is evaluated over its active interval $[a_{\rm lo},a_{\rm hi})$ using a reference distribution normalized to $M_{\rm tot}$. For the fast-aligned component, the corresponding amplitude satisfies
\begin{eqnarray}
C_{\rm hi}(t_{r})\,\mathcal{M}\!\bigl(\afast(t_{r}),\,a_{\rm align}^{\rm fast,top}(t_{r})\bigr)
= C_0\,\mathcal{M}(a_{\min},a_{\max}),
\label{eq:Chi_norm}
\end{eqnarray}
where $C_{0}$ is the amplitude of the initial MRN distribution, $C_{\rm hi}(t_{r})$ is the renormalized amplitude of the fast-aligned component, $a_{\rm align}^{\rm fast,top}=\min(\aali^{\rm ISRF},\adisr)$, and $\mathcal{M}(a_1,a_2)=2(a_2^{1/2}-a_1^{1/2})$ is the initial MRN mass measure (Eq.~\ref{eq:Mdef}). This is only a reference normalization, and multiplication by the post-redistribution weight $w_X$ (Eq.~\ref{eq:weights}) yields the physical mass $M_X=w_XM_{\rm tot}$. The effective grain size distribution is therefore
\begin{eqnarray}
\left(\frac{dn}{da}\right)_{\!\rm eff} =
\sum_{X\in\{A,B,C,D,F\}} w_X
\left(\frac{dn}{da}\right)_{\!X}^{\rm norm},
\label{eq:gsd_eff}
\end{eqnarray}
where $(dn/da)_X^{\rm norm}$ is restricted to the interval occupied by component $X$ and normalized to $M_{\rm tot}$. The weights then ensure that the sum contains exactly the conserved mass. Fragmentation changes both the component distributions and their weights before the extinction, emission, and polarization integrals are evaluated. Component~D occupies the two intact intervals $[\aali^{\rm ISRF},\adisr^{\min})$ and $[\adisr^{\max},a_{\max}]$, whereas component~F contains the surviving low-$J$ grains inside the disruption band.

\subsection{Fragment size distribution models}
\label{app:fragmodels}

We consider three prescriptions for the fragment distribution. Each redistributes the same transferred mass, $M_{\rm frag}=\fDisrJ\,\mathcal{M}_E$ (Eq.~\ref{eq:Mfrag}), supplied only by disrupted high-$J$ grains within the band. The low-$J$ survivors (component~F) and intact grains above $\adisr^{\max}$ are not fragment sources. The fragments enter the effective GSD, $(dn/da)_{\rm eff}$, and hence all extinction, thermal emission, and polarization integrals; $M_{\rm frag}$ fixes their normalization but is not retained as a separate additive component.

\paragraph{Two-component prescription.}
The pre-transient GSD below the disruption band consists of the Astrodust log-normal nanodust component and the MRN power law, joined at $a_{\rm junc}=5$~nm:
\begin{eqnarray}
\left(\frac{dn}{da}\right)_{\rm nano} &=& \frac{B_{\rm nano}}{a}\,
  \exp\!\left[-\frac{\ln^{2}(a/a_{0})}{2\sigma^{2}}\right],
  \quad a\le a_{\rm junc},
\label{eq:gsd_nano}\\
\left(\frac{dn}{da}\right)_{\rm MRN} &=& C_{\rm MRN}\,a^{q},
  \quad a_{\rm junc}\le a\le a_{\max},
\label{eq:gsd_mrn}
\end{eqnarray}
where $B_{\rm nano}$ and $C_{\rm MRN}$ are the pre-transient amplitudes and $a_{0}$, $\sigma$ are the parameters of the Astrodust nanograin distribution \citep{DraineHensley.2023}. In the fiducial model, the transferred mass $M_{\rm frag}$ is split between these two sub-components: a fraction $f_{\rm nano}$ enters the log-normal component, and the remainder $1-f_{\rm nano}$ enters the surviving MRN range $[a_{\rm junc},\adisr]$ (the disruption efficiency $\fDisrJ$ is already contained in $M_{\rm frag}$). Because the functional forms in Equations~\eqref{eq:gsd_nano}--\eqref{eq:gsd_mrn} are held fixed, the injection amounts to rescaling the amplitudes,
\begin{eqnarray}
B_{\rm nano} \to \zeta_{\rm nano}(t_{r})\,B_{\rm nano},\qquad
C_{\rm MRN} \to \zeta_{\rm MRN}(t_{r})\,C_{\rm MRN} \quad (a\le\adisr),
\label{eq:ratd_amp}
\end{eqnarray}
with the time dependent renormalization factors
\begin{eqnarray}
\zeta_{\rm nano}(t_{r}) &=& 1 + f_{\rm nano}
  \frac{C_{\rm MRN}\,M_{\rm frag}(t_{r})}{M_{\rm nano}},
\label{eq:ratd_renorm_nano}\\
\zeta_{\rm MRN}(t_{r})  &=& 1 + (1-f_{\rm nano})
  \frac{M_{\rm frag}(t_{r})}{\mathcal{M}(a_{\rm junc},\adisr)},
\label{eq:ratd_renorm_mrn}
\end{eqnarray}
with fiducial $f_{\rm nano}=0.2$. Here $M_{\rm frag}$ is expressed per unit MRN amplitude $C_0$, $C_{\rm MRN}$ converts it to the physical nanodust mass scale, and $M_{\rm nano}$ is the pre-transient nanodust mass obtained by integrating Equation~\eqref{eq:gsd_nano}. The rescaled components enter the effective distribution $(dn/da)_{\rm eff}$ (Eq.~\ref{eq:gsd_eff}) and hence all observables. Because the two fractions sum to unity and the source is restricted to the disrupted high-$J$ mass, Equations~\eqref{eq:ratd_renorm_nano}--\eqref{eq:ratd_renorm_mrn} conserve mass exactly.

Both factors exceed unity once the disruption band opens: $\zeta_{\rm nano}$ rises because disrupted large grains feed the small-grain reservoir, while $\zeta_{\rm MRN}$ rises because the fragment MRN population adds to the sub-band range. Their magnitude follows directly from the MRN mass measure: in terms of the transferred fraction $w_E=M_{\rm frag}/M_{\rm tot}$ (Eq.~\ref{eq:wE_def}), $\zeta_{\rm nano}-1=f_{\rm nano}\,w_E\,(M_{\rm tot}/M_{\rm nano})$ and $\zeta_{\rm MRN}-1=(1-f_{\rm nano})\,w_E\,M_{\rm tot}/\mathcal{M}(a_{\rm junc},\adisr)$. For the fiducial $f_{\rm nano}=0.2$ and $\fDisrJ=\fhif=0.25$, the band at the onset of RAT-D ($[0.1,0.2]\,\mu$m) gives $w_E\approx0.05$ and $\zeta_{\rm nano}\approx\zeta_{\rm MRN}\approx1.1$, whereas the fully developed band reached at $D\lesssim0.1$~pc ($[0.02,0.5]\,\mu$m) gives $w_E\approx0.2$, $\zeta_{\rm MRN}\approx2.6$, and $\zeta_{\rm nano}\approx1.4$--1.8 for a nanodust mass fraction $M_{\rm nano}/M_{\rm tot}\approx0.05$--0.1. At $D\gtrsim1$~pc, the band remains narrow, and both factors stay within $\sim$10\% of unity, consistent with the modest nanodust enhancement visible in Figure~\ref{fig:gsd_time}. These factors set the amplitudes of the enhanced nanodust and sub-band populations in Figure~\ref{fig:gsd_time}.%, and the nanodust enhancement $\zeta_{\rm nano}$ powers the transient spinning dust emission presented in Paper~III.

%\begin{comment}
	content...
\paragraph{Uniform-cascade prescription: Wyatt (2011) model.}
The transferred mass $M_{\rm frag}$ is redistributed into a pure MRN power law below $\adisr$ with a uniform disruption rate $K_{\rm disr}={\rm const}$. The net distribution is $dn/da = (C_{\rm MRN}+C_{\rm frag})\,a^{q}$ for $a\le\adisr$, where $C_{\rm frag}$ is set by mass conservation of the disrupted grain population, $C_{\rm frag}\,\mathcal{M}(a_{\min},\adisr)=C_{0}\,M_{\rm frag}$, following \cite{Wyatt.2011}.

\paragraph{Size-dependent cascade prescription: Smoluchowski model.}

The disruption rate varies as $K_{\rm disr}(a)\propto(a/\adisr)^{s_{\rm disr}}$, reflecting the scalings of the RAT-driven terminal angular velocity, $\Omega_{\rm RAT}\propto a^{1/2}$, and of the critical disruption velocity, $\Omega_{\rm disr}\propto a^{-1}$ (Eq.~\ref{eq:omegadisr}). We adopt $s_{\rm disr}=3/2$, and the limit $s_{\rm disr}=0$ recovers the uniform-cascade prescription. In both cases, $C_{\rm frag}$ is fixed by normalization to $M_{\rm frag}$.
%\end{comment}

\section{Mass Conservation, Alignment-State Decomposition, and Validation}
\label{app:bookkeeping}

\begin{figure*}[ht]
	\centering
	\includegraphics[width=\textwidth]{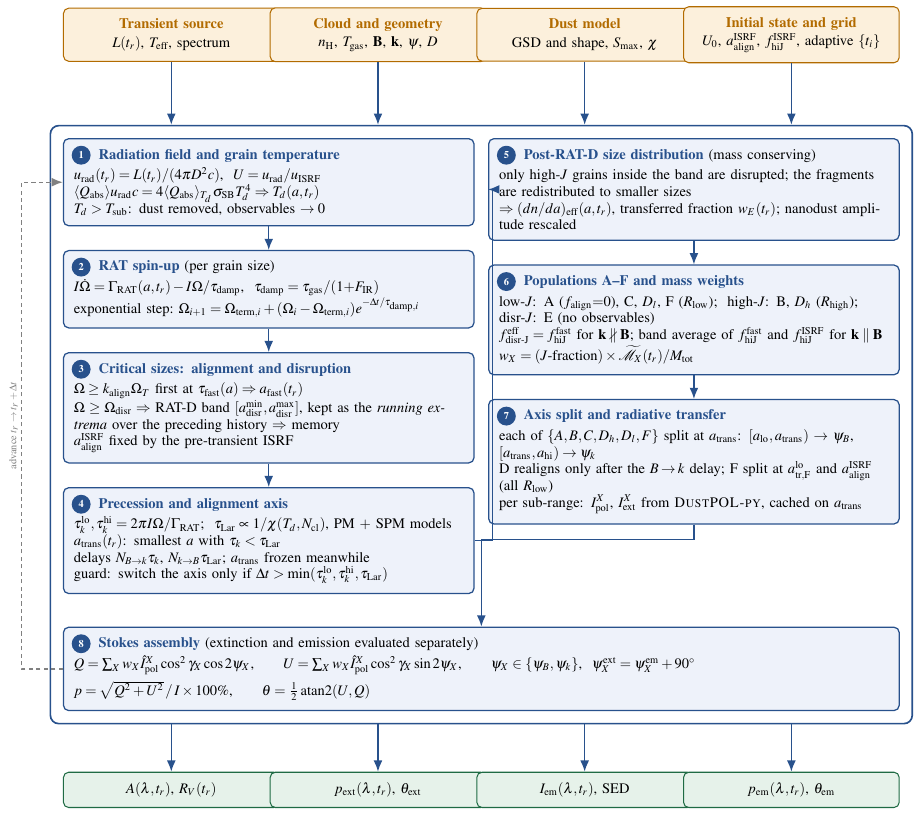}
	\caption{Flow of the {\it TransRAT} calculation. \textit{Top:} The transient light curve, the cloud and magnetic geometry, the grain model, and the pre-transient ISRF state and time grid define the physical conditions at each source--cloud distance $D$. \textit{Center:} At every epoch the model updates $U$ and $T_d$ (1) and integrates RAT spin-up (2), which fixes the critical sizes $\afast$ and the RAT-D band (3, accumulated irreversibly) and, through the competing precession timescales, the alignment-axis boundary $a_{\rm trans}$ (4). Disruption then reshapes the grain size distribution at fixed total mass (5), the size--$J$ populations A--F and their mass weights are reassigned (6), and each component is split at $a_{\rm trans}$ into its $\BRAT$ and $\kRAT$ sub-ranges for the radiative transfer calls (7) before the Stokes parameters are assembled (8); the dashed arrow denotes propagation to the next epoch. \textit{Bottom:} The principal outputs are the extinction curve and $R_V$, polarized extinction and thermal emission with their position angles, and the dust-emission SED.}
	\label{fig:pipeline}
\end{figure*}

Figure~\ref{fig:pipeline} summarizes the sequence of the {\it TransRAT} calculation from physical inputs to observables. Here we specify how the grain populations of Sections~\ref{sec:mass_weights}--\ref{sec:dsplit} are combined without double counting and test the resulting calculation in the analytically tractable parallel geometry (Figure~\ref{fig:pipeline}, center and bottom rows).

\subsection{Weight normalization and mass-budget closure}
\label{app:massbudget}

\begin{figure*}[ht]
\centering
\begin{overpic}[width=\textwidth]{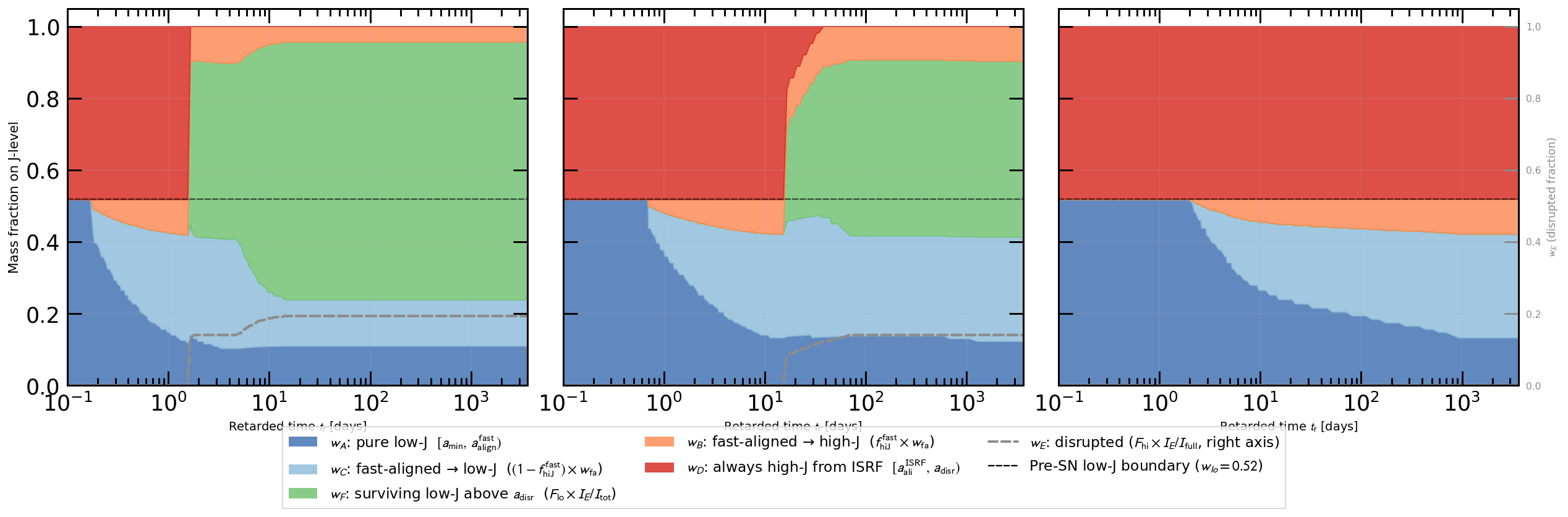}
	\put(7.5,25){\textbf{(a)}}\put(40.5,25){\textbf{(b)}}\put(72.5,25){\textbf{(c)}}
\end{overpic}
\caption{Mass-budget closure (Eq.~\ref{eq:massbudget}) for the $\psi=0$ validation model at different cloud distances ($D=0.1, 1,5$ pc). Colored bands show the post-redistribution mass fractions $w_A$ (unaligned low-$J$), $w_C$ (fast-aligned $\rightarrow$ low-$J$), $w_F$ (band survivors), $w_B$ (fast-aligned $\rightarrow$ high-$J$), and $w_D=w_{D_h}+w_{D_l}$ (ISRF-aligned); their sum is unity at every epoch. The gray dashed curve $w_E=M_{\rm frag}/M_{\rm tot}$ is the fraction transferred from disrupted grains and already redistributed among the colored components, not an additional reservoir. The horizontal dashed line marks the pre-transient low-$J$ fraction. At large $D$ (bottom row), RAT-D is inactive ($w_E=0$) and only the alignment partition evolves.}
\label{fig:massbudget_fractions}
\end{figure*}

Figure~\ref{fig:massbudget_fractions} shows the mass fractions of different components for the $\psi=0$ validation calculation at three source--cloud distances. At every epoch and distance, $w_A+w_B+w_C+w_D+w_F=1$ (Eq.~\ref{eq:weights}, with $w_D=w_{D_h}+w_{D_l}$). The dashed $w_E$ curve records the fraction of the conserved mass transferred from disrupted grains and is not an additional component.

%Each \texttt{DustPOL} sub-range call internally re-normalises the MRN distribution to the full dust mass within its own interval (amplitude fixed by the dust-to-gas mass ratio).
The mass contribution of each component to Stokes parameters is evaluated from a reference distribution normalized to $M_{\rm tot}$. Multiplication by $w_X$ (Eq.~\ref{eq:weights}) assigns the physical component mass and makes the contributions additive, as expressed by Equation~\eqref{eq:gsd_eff}. We also test Equation~\eqref{eq:massbudget} by integrating the complete post-redistribution distribution and evaluating the six component masses $M_X=w_XM_{\rm tot}$ at 400 logarithmically spaced epochs over $t_r=0.5$--600~d (a dedicated test grid, finer than the adaptive production grid). Four cases are considered: $\psi=0$ (mass-weighted $\fDisrJ$, Equation~\ref{eq:fdisr_eff_parallel}) and $\psi\neq0$ ($\fDisrJ=\fhif$), each with the disruption band either entirely above $\aali^{\rm ISRF}$ or sweeping below it. In every case, the component budget closes to machine precision, with a maximum deviation $\max_{t_r}|(M_A+M_B+M_C+M_{D_h}+M_{D_l}+M_F)/M_{\rm tot}-1| = 2.2\times10^{-16}$.

Note that $w_E=M_{\rm frag}/M_{\rm tot}$ is retained only as the diagnostic fraction of the original mass that has been transferred to fragment sizes. It is already contained in the A--F masses through $(dn/da)_{\rm eff}$ and is therefore not included in this sum.

\subsection{Axis-dependent population decomposition}
In the sums of Equations~\eqref{eq:QUext} and \eqref{eq:QU}, $X$ runs over all active $\BRAT$ and $\kRAT$ intervals. For $\bk\|\bB$, these are $\{A,B,C,D_{\rm hi},D_{\rm lo},F_{\rm BC,B},F_{\rm D,B}\}$. For $\bk\nparallel\bB$, each of B, C, $D_{\rm hi}$, and $D_{\rm lo}$ is divided at its B/$\kRAT$ boundary, while F yields $\{F_{\rm BC,B},F_{\rm D,B},F_{\rm BC,K},F_{\rm D,K}\}$. Including the two surviving D intervals, the general case contains at most 18 integration intervals per epoch.

Component~F is split at $a_{\rm tr,F}^{\rm lo}$ (the low-$J$ B/$\kRAT$ axis boundary) into grains aligned along $\Bv$ and along $\kv$; within each axis, the mass integral is further partitioned at $\aali^{\rm ISRF}$, but all four intervals carry the same low-$J$ efficiency $\Rlow$.

When the D population straddles the high-$J$ B/$\kRAT$ transition (Sec.~\ref{sec:dsplit}), grains in $D_h$ and $D_l$ are divided at $a_{\rm tr,D}^{\rm hi}$:
\begin{itemize}
\item $\mathbf{D_{hB}}$ (both $J$-levels, $\BRAT$ retained): $[\aali^{\rm ISRF},\,a_{\rm tr,D}^{\rm hi})$ ---
      no axis change; $D_h$ and $D_l$ keep $\Rhigh$ and $\Rlow$, i.e.\ combined $f_{\rm align}^{\rm eff}=\fHiISRFeff+\fLoISRFeff$.
\item $\mathbf{D_{hK}}$: $[a_{\rm tr,D}^{\rm hi},\,\adisr^{\min})$ ---
      fraction $\fhif$ fast-realigns to high-$J$ at $\kv$
      ($f_{\rm align}=\Rhigh$; the $\fhif$ fraction is carried by $w_{D_{hK}}$).
\item $\mathbf{D_{lB}}$/$\mathbf{D_{lK}}$: remainder $(1-\fhif)$ of the
      $\kRAT$ sub-range drops to low-$J$, split further at the low-$J$
      transition $a_{\rm tr,D}^{\rm lo}$ ($f_{\rm align}=\Rlow$; the $1-\fhif$ fraction is carried by $w_{D_{lB}}+w_{D_{lK}}$).
\end{itemize}

The mass weights are
\begin{eqnarray}
\begin{aligned}
w_{D_{hB}} &= \frac{\widetilde{\mathcal{M}}(\aali^{\rm ISRF},\,a_{\rm tr,D}^{\rm hi};t_r)}{M_{\rm tot}},\\
w_{D_{hK}} &= \fhif\,\frac{\widetilde{\mathcal{M}}(a_{\rm tr,D}^{\rm hi},\,\adisr^{\min};t_r)}{M_{\rm tot}},\\
w_{D_{lB}}+w_{D_{lK}} &= (1-\fhif)\,\frac{\widetilde{\mathcal{M}}(a_{\rm tr,D}^{\rm hi},\,\adisr^{\min};t_r)}{M_{\rm tot}},
\end{aligned}
\label{eq:Dsplit}
\end{eqnarray}
and analogously for the above-band interval $[\adisr^{\max},a_{\max}]$ with its own $a_{\rm tr,D}^{\rm hi}$, $a_{\rm tr,D}^{\rm lo}$; the two sets of sub-range weights are added.

If $\kv\parallel\Bv$, the $\kRAT$ condition has not been met, or the switching delay has not elapsed, the full D range remains in $\BRAT$ and $w_{D_{hB}}=\widetilde{\mathcal{M}}_{D}/M_{\rm tot}=w_{D_h}+w_{D_l}$.

\subsection{Parallel-geometry validation (\texorpdfstring{$\bk\parallel\bB$}{k parallel B})}
\label{app:kpB_validation}
%==========================================================================
Here we discuss the special geometry $\bk\|\bB$ ($\psi=0$) which provides a benchmark for testing the alignment axis switching because the $\BRAT$ and $\kRAT$ axes have identical sky projections. In this special case, the transient radiation only redistributes grains among the three $J$-levels: grains pre-aligned at high-$J$ by the ISRF (component $D_h$) retain the fraction $\fHiISRF$ and are disrupted with efficiency $\fHiISRF$; grains at low-$J$ attractors (component $D_l$) remain trapped there by the enhanced transient RATs \citep{Hoang.2025}; and previously unaligned grains ($a<\aali^{\rm ISRF}$) acquire fast alignment with a fraction $\fhif$ captured at high-$J$ and disruption efficiency $\fhif$ (see Eq.~\ref{eq:fdisr_eff_parallel}).

%The polarization angle must therefore remain constant while the polarization degree varies due to time dependent alignment and disruption as discussed in Section \ref{sec:TransRATmodel}. 

\begin{figure}
		\centering
	\begin{overpic}[width=0.99\textwidth]{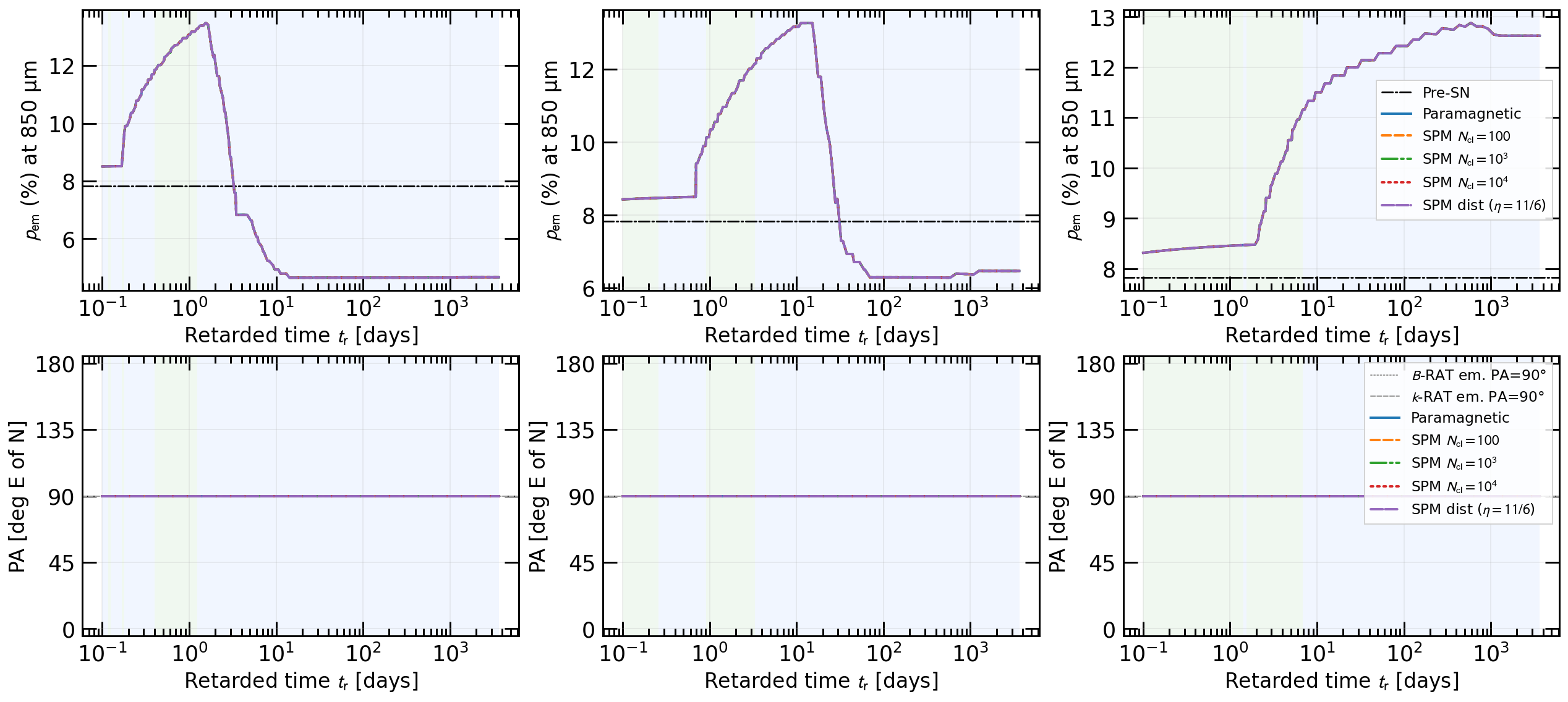}
		\put(17.5,40.5){\textbf{(a) D=0.1 pc}}\put(50.5,40.5){\textbf{(b) D=1 pc}}\put(72.5,40.5){\textbf{(c) D=5 pc}}
	\end{overpic}
	\caption{Validation of the thermal emission polarization calculation for $\bk\parallel\bB$. Upper panels show $p_{\rm em}(850\,\mu{\rm m})$ and lower panels show the position angle for the three modeled distances, ordered from $D=0.1, 1, 5$ pc across the two image blocks, and for the five PM/SPM susceptibility models. The polarization fraction responds to fast alignment and RAT-D, whereas the position angle remains at the common $\BRAT$/$\kRAT$ value, as required when the radiation and magnetic field directions are parallel.}
	\label{fig:pem_mag_Bk0}
\end{figure}

Figures \ref{fig:pem_mag_Bk0} shows the time dependent polarization fraction and position angle for the different source-cloud distances. The polarization fraction varies with time---a flare followed by a RAT-D-driven dip in nearby clouds and a gradual rise to saturation at $D\ge5$~pc, while the position angle remains constant, as expected from the fixed alignment axis.

\bibliographystyle{aasjournalv7}
\bibliography{references}

\end{document}